\documentclass[preprint,review,12pt,authoryear]{elsarticle}
\usepackage{multirow}
\usepackage{mathtools}
\usepackage[utf8]{inputenc}
\usepackage[margin=1 in]{geometry}
\usepackage{multirow}
\usepackage{mathtools}
\usepackage{siunitx}
\usepackage{empheq}
\usepackage{amsmath}
\usepackage{cases}
\usepackage{enumitem}
\newcommand\norm[1]{\left\lVert#1\right\rVert}
\usepackage[usenames, dvipsnames]{color}
\definecolor{UW}{RGB}{64, 38, 96}

\date{August 2021}

\begin{document}
\begin{titlepage}

\clearpage\thispagestyle{empty}



\noindent

\hrulefill

\begin{figure}[h!]

\centering

\includegraphics[width=1.5 in]{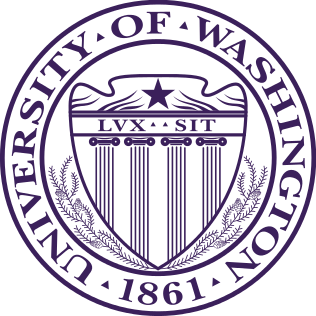}

\end{figure}

\begin{center}

{\color{UW}{

{\bf A\&A Program in Structures} \\ [0.1in]

William E. Boeing Department of Aeronautics and Astronautics \\ [0.1in]

University of Washington \\ [0.1in]

Seattle, Washington 98195, USA

}

}

\end{center} 

\hrulefill \\ \vskip 2mm

\vskip 0.5in

\begin{center}

{\large {\bf Report E21-8: \\Stress Distributions and Notch Stress Intensity Factors in Multimaterial V-notches Under Antiplane Shear and Torsion}}\\[0.5in]
{\large {\sc Marco Salviato}}\\[0.45in]

%



%

\end{center}

\noindent {\footnotesize {{\em Submitted to International Journal of Solids and Structures \hfill November 2021} }}

\end{titlepage}

\begin{frontmatter}


\cortext[cor1]{Corresponding Author, \ead{salviato@aa.washington.edu}}

\title{Report E21-8: \\Stress Distributions and Notch Stress Intensity Factors in Multimaterial V-notches Under Antiplane Shear and Torsion}


\author[addressa]{Marco Salviato\corref{cor1}}

\address[addressa]{William E. Boeing Department of Aeronautics and Astronautics, University of Washington, Seattle, Washington 98195, USA}

\begin{abstract}
\linespread{1}\selectfont
This study investigates how the insertion of multimaterial circular regions embracing the tip of a finite V-notch can be used to reduce the Notch Stress Intensity Factors (NSIFs) in structures subjected to antiplane shear or torsion. Towards this goal, this work presents a novel theoretical framework to calculate stress distributions and NSIFs in closed-form. Thanks to the new solution, it is shown that by tuning multimaterial region radii and elastic properties it is possible to significantly reduce the NSIFs and stress concentrations at the material interfaces.

To investigate whether the proposed multimaterial system translates into increased structural capacity even in the presence of significant nonlinear deformations, computational simulations were conducted using nonlinear hyperelastic-damage and elasto-plastic-damage models. The preliminary results show increases of structural capacity up to $46\%$ and of nominal strain at failure of up to $86\%$ at the expenses of only a $8\%$ reduction in structural stiffness. 

It is expected that a similar approach can be extended to other loading conditions (e.g. mode I and mode II, and fatigue) and that even larger gains can be obtained by performing thorough optimization studies.  

\end{abstract}


\end{frontmatter}

\section{Introduction}
The advent of multimaterial additive manufacturing has given researchers and designers the opportunity to explore unprecedented ways to increase the damage tolerance of structural components \citep{bandyopadhyay2018additive,rafiee2020multi}. \cite{ubaid2018strength}, for instance, investigated multimaterial jetting additive manufacturing to realize strength and performance enhancement of multilayered materials by spatial tailoring of adherend compliance and morphology. Compared to the baseline homogeneous system, they were able to obtain an increase of strength by $20\%$, toughness by $48\%$, and strain at failure by $18\%$. 

\cite{lin2014tunability} leveraged multimaterial additive manufacturing to explore the mechanical behavior of suture interfaces inspired by the intricate, hierarchical designs of ammonites. They showed that proper combinations of soft and hard materials along with proper selection of the order of hierarchy of the interface can lead to significant stiffness, tensile strength, and toughness compared to conventional interfaces. 

\cite{suksangpanya2018crack} used multimaterial additive manufacturing to reproduce the Bouligand structure present in the dactyl club of the smashing mantis shrimp. Leveraging experiments on three-point bend specimens and semi-analytical modeling, they showed that, thanks to changes in local fracture mode and increases in crack surface area, the initiation fracture toughness can be increased almost twofold while the fracture toughness at catastrophic failure can be increased of an order of magnitude. \cite{zaheri2018revealing} showed that similar microstructures are the secrets for the outstanding stiffness and toughness of the cuticle of the figeater beetle (\textit{Cotinis mutabilis}). 

\cite{raney2018rotational} developed a novel rotational 3D printing method that enables spatially controlled orientation of short fibers in polymer matrices by varying the nozzle rotation speed relative to the printing speed. Using this technology, they fabricated carbon fiber-epoxy composites composed of volume elements with defined fiber arrangements. By tailoring the fiber orientation in select areas, the authors were able to demonstrate the possibility of increasing damage tolerance and capacity of several structural components.  

\cite{martin2015designing} proposed a new additive manufacturing technology called ``3D magnetic printing" that is capable of printing dense ceramic/polymer composites in which the direction of the ceramic-reinforcing particles can be finely tuned. Thanks to this new method, the authors explored the mechanics of complex bioinspired reinforcement architectures showing the possibility of steering cracks using controlled mesostructures and increase damage tolerance.

Leveraging 3D printing of continuous carbon fiber composites \cite{sugiyama20203d} investigated the optimization of curved fiber trajectories to realize variable fiber volume fraction and stiffness composites to increase the capacity of composite structures. They were able to show that proper selection of the fiber paths can increase the ratio between the structural capacity and its overall weight of almost $60\%$ compared to traditional designs. Using a novel isogeometric computational framework \cite{suzuki2021isogeometric} showed that the stress concentration factor of notched additively manufactured composites can be reduced to almost half without affecting the structural stiffness.

\begin{figure}[!ht]
\center
\includegraphics[width=1\textwidth]{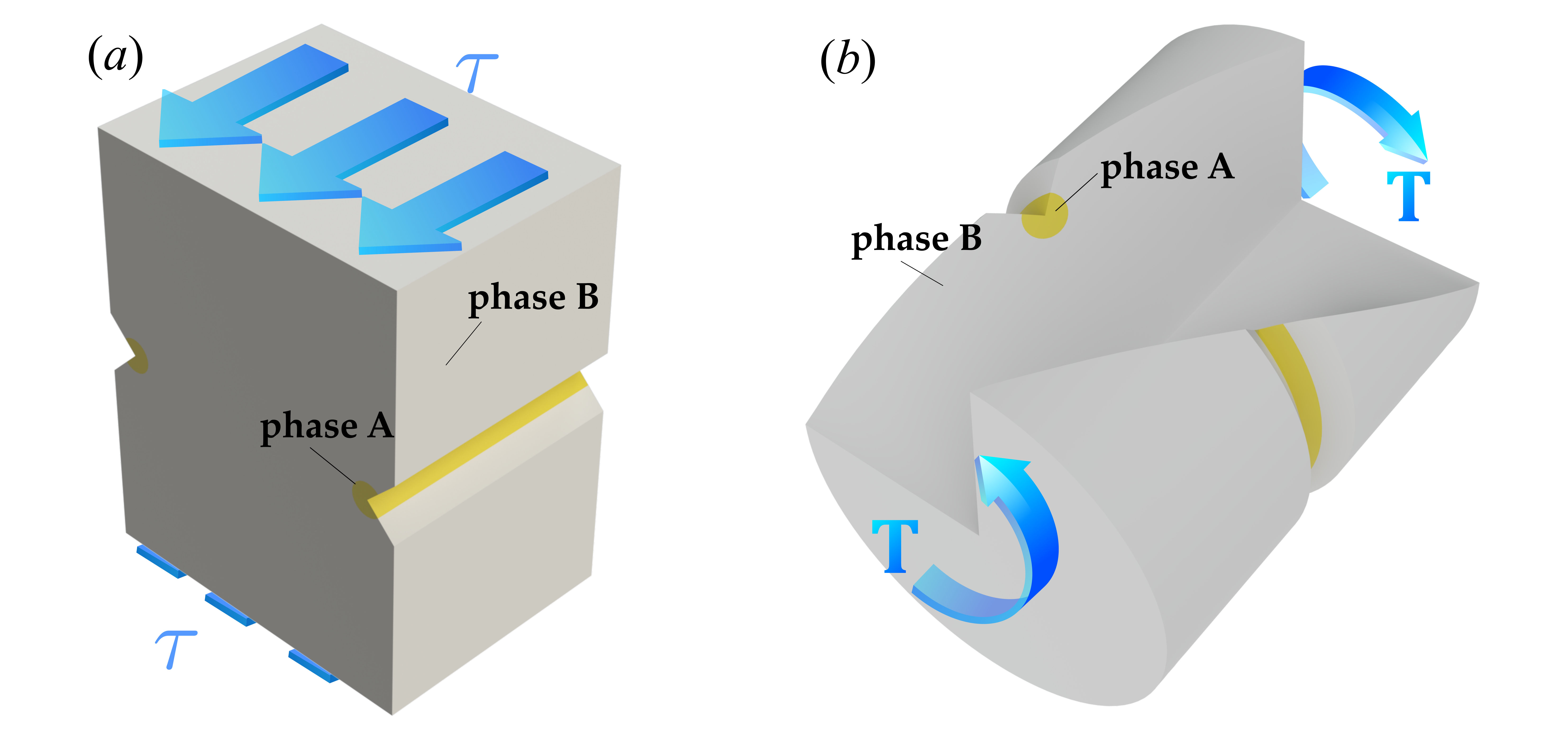}
\caption{Examples of the systems investigated in this work: (a) component under antiplane shear and (b) three-quarter section view of a circular shaft under torsion. Both systems are weakened by a V-notch featuring a circular region made of a different material embracing the tip.}
\label{Load case}
\end{figure}

Inspired by the foregoing recent advances in additive manufacturing, this work aims at exploring the use of multimaterial configurations to increase the capacity and toughness of V-notched structural components under antiplane shear or torsion. As shown in Figs. (\ref{Load case}a,b), the idea is to investigate the addition at the tip of the notch of multiple circular regions made of different material combinations to reduce the Notch Stress Intensity Factor (NSIF) \citep{Gross72} and promote higher energy dissipation by plastic deformations. 

In pursuit of this goal, the problem is initially investigated within the framework of Linear Elastic Fracture Mechanics (LEFM) to determine the effects of the material properties and notch configuration on the NSIF. Recently, significant studies have investigated the stress field and stress intensity factors in multimaterial structures under antiplane shear and torsion. Leveraging a complex potential approach \cite{Zappa19Sal} developed closed form and approximate solutions to describe the antiplane shear stress distribution in orthotropic plates with lateral blunt notches. They later extended this approach for the case of in-plane shear loading \citep{PastrelloSal21}. 

\cite{Sal18} and \cite{Zappa19Sal2} investigated the mode III stress distribution ahead of cracks initiated at sharp and blunt notch tips. They were able to provide closed form solutions for the calculation of the Stress Intensity Factors (SIF) for a number of notch configurations. 

However, notwithstanding the efforts devoted in the past decades, a theoretical framework for the calculation of the displacement and stress distributions for multimaterial V-notches as the ones shown in Fig. (\ref{Load case}a,b) is still elusive. This is the topic of the first part of the present study in which, combining conformal mapping and complex potentials, a new general method for calculating the stress distribution in such multimaterial systems in closed form is presented. Thanks to the proposed theoretical framework, the present study investigates the effects of the material properties and region radii on the NSIF providing clear guidelines for the mitigation of the stress intensity at the notch and the stress concentration at the material interfaces. 

Finally, in the second part of the study, advanced elastoplastic-damage and hyperelastic-damage models are implemented to investigate the efficacy of the proposed multimaterial system even in the presence of significant inelastic strains which are likely to occur under the investigated loading conditions. The results confirm that the concept of multimaterial notch blunting can lead to significant reductions of the NSIF, which translates into higher structural capacity and toughness. 

The article is organized as follows. Section 2 presents a novel general theoretical framework for the analytical solution of the stress field in multimaterial domains subject to antiplane shear and torsion. Section 3 discusses the application of the approach to the case of a bimaterial system. It provides closed form sulutions for stresses, NSIFs, and displacement as function of the geometrical configuration of the notch and the elastic properties of the materials. Section 4 presents the application of the theoretical framework to the case of a trimaterial configuration. Several contour plots are presented to discuss the complex influence on the NSIF of the radii of the regions and their elastic properties. Section 5 discusses the implementation of elastoplastic-damage and hyperelastic-damage models for the nonlinear simulation of torsion in bimaterial system composed of vulcanized rubber and epoxy. The models allow to capture the plastic deformation close to the notch tip and enable the evaluation of the effectiveness of the proposed multimaterial system in the presence of realistic deformations. The manuscript ends with Section 6 where a thorough analysis of the main conclusions of this work is presented.

\section{Unified solution for the stress fields in antiplane shear and torsion problems in multi-material domains}

\subsection{Governing equations in isotropic and homogeneous media}\label{intro}
\begin{figure}[!ht]
\center
\includegraphics[width=0.88\textwidth]{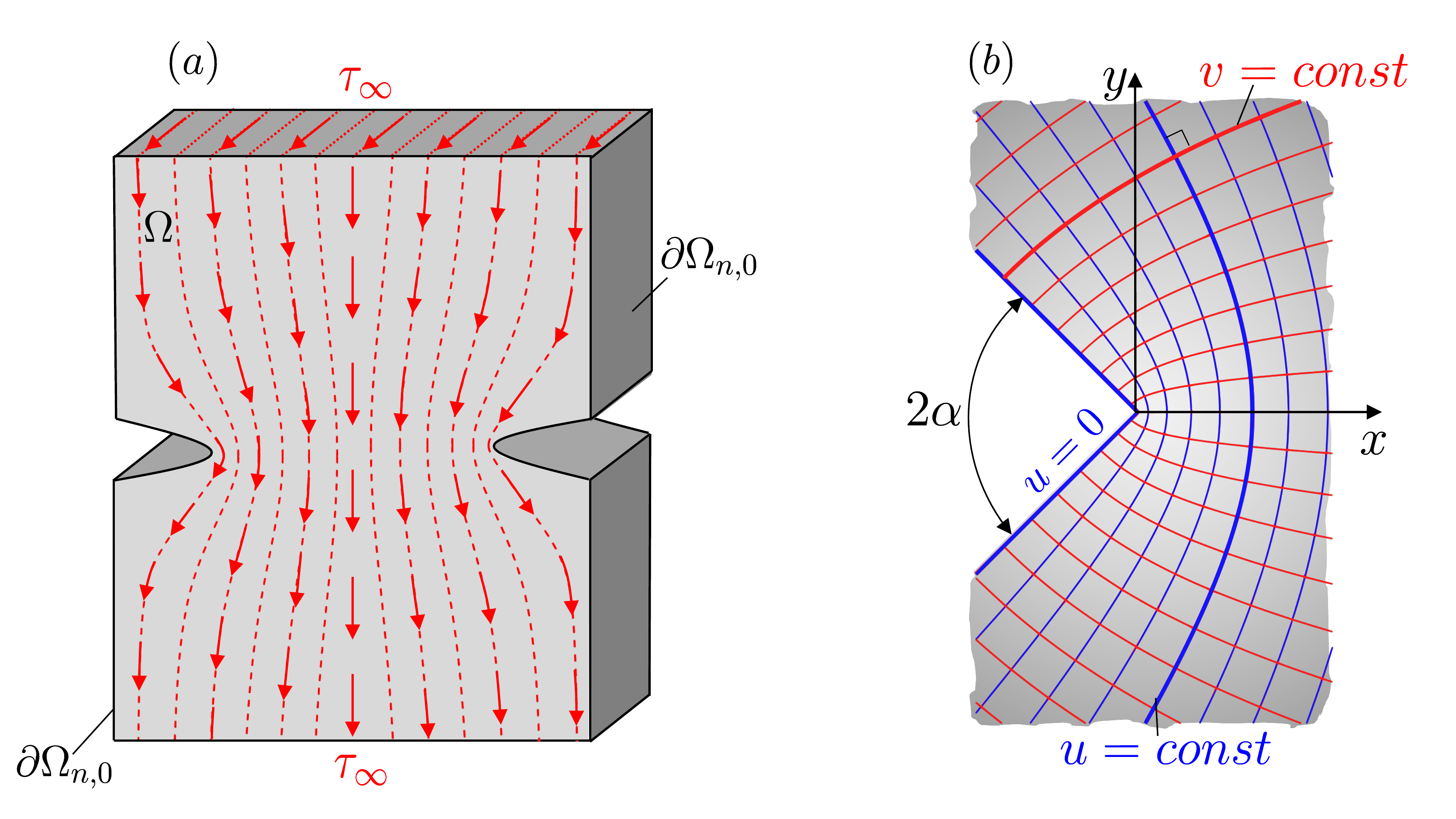}
\caption{(a) 2D domain $\Omega\cup\partial\Omega$ subjected to a remote antiplane shear stress $\tau_{\infty}$ and (b) example of the conformal transformation $z=\xi^q$ with $q=2\left(\pi-\alpha\right)/\pi$ and $\xi=u+iv$ typically used for the description of deep hyperbolic notches \citep{Neuber58a, Neuber58b}. }
\label{mapping}
\end{figure}

Let us consider first a semi-infinite 2D domain $\Omega \cup \partial \Omega$ made of a isotropic and homogeneous material featuring a linear elastic behavior and assume that the domain is subjected to a remote antiplane shear stress $\tau_{\infty}$ (Fig. \ref{mapping}a). Considering the Cartesian coordinate system $\left(x,y,z\right)$ defined in Figure \ref{mapping}b, the equilibrium equation assuming the absence of body forces can be written as follows:
\begin{equation}\label{equilibrium}
    \frac{\partial \tau_{zx}}{\partial x}+\frac{\partial \tau_{zy}}{\partial y}=0
\end{equation}
where $\tau_{zi}(i=x,y)=$ shear stress components in $x-$ and $y-$ directions. The shear stress components are linked to the engineering shear strains through Hooke's law: $\tau_{zi}=1/G\gamma_{zi}(i=x,y)$ where $G$ represents the elastic shear modulus of the material. Introducing the kinematic relationships, the stresses can be written as a function of the displacement in the z-direction, $w$: $\tau_{zx}=G\partial w/\partial x$ and $\tau_{zy}=G\partial w/\partial y$. Then, if one substitutes the foregoing expressions into Eq. \ref{equilibrium}, the governing equation in terms of the displacement becomes a two-dimensional Laplace equation:
\begin{equation}\label{laplacian}
    \nabla^2w=0
\end{equation}
where $\nabla^2=\partial^2/\partial x^2+\partial^2/\partial y^2=$ the laplacian operator. 

The solution of Eq. \ref{laplacian} in the two-dimensional domain $\Omega \cup \partial \Omega$ in the presence of Dirichlet or von Neumann boundary conditions on $\partial \Omega$ can be particularly cumbersome except for cases in which the domain is simple. To overcome this problem, it is convenient to leverage a conformal mapping to transform the complex domain in one for which it is easier to apply the boundary conditions. Conformal maps can be described by complex analytic functions \citep{fisher1999complex, brown2009complex} so that the change of coordinates takes the form: $z=z\left(\xi\right)$ with $z=x+iy$ and $\xi=u+iv$. It can be shown that the transformation is always bijective and satisfies the Cauchy-Riemann (C-R) conditions $\partial u/\partial x=\partial v/\partial y$, $\partial u/\partial y=-\partial v/\partial x$ in $\Omega$. Since the Laplace equation is conformally invariant, the advantage of using a conformal map is that the governing equation in the new coordinates takes the same form as the one in Cartesian coordinates:
\begin{equation}
    \frac{\partial^2 w}{\partial u^2}+\frac{\partial^2 w}{\partial v^2}=0
\end{equation}
while making it easier to apply the boundary conditions \citep{fisher1999complex, brown2009complex}. The curvilinear stress components in the new coordinate system can be calculated as follows:
\begin{equation}\label{stress}
    \tau_{zu}=\frac{G}{h}\frac{\partial w}{\partial u}, \quad ~ \tau_{zv}=\frac{G}{h}\frac{\partial w}{\partial v}
\end{equation}
where $h=h_u=\sqrt{\left(\partial x/\partial u\right)^2+\left(\partial y/\partial u\right)^2 }=h_v=\sqrt{\left(\partial x/\partial v\right)^2+\left(\partial y/\partial v\right)^2 }$ are the metric coefficients of the transformation \citep{sokolnikoff1956mathematical}.

Salviato and Zappalorto \citep{Sal16a} showed that a significant simplification to the solution can be provided by considering conformal maps in which the condition $u=u_0$ or $v=v_0$ (with $u_0$ and $v_0$ being real constants) describes the portion of boundary, $\partial \Omega_{n,0}$, where the shear stress normal to the boundary is null: $\tau_{zn}=0$. 

For the case in which $u=u_0$ describes $\partial \Omega_{n,0}$, Salviato and Zappalorto \citep{Sal16a} showed that:
\begin{equation}
    w=Av
\end{equation}
and the curvilinear stress components can be calculated directly from the analytic function describing the conformal map using Eqs \ref{stress}a,b:
\begin{equation}
    \tau_{zu}=0, \qquad~\quad \tau_{zv}=\frac{\psi}{\norm{z'\left(\xi\right)}}
\end{equation}
where $\norm{z'\left(\xi\right)}=h$ is the magnitude of the first derivative of the conformal map and $\psi=AG$ is a constant to be determined by imposing the remote stress conditions. Furthermore, Salviato and Zappalorto \citep{Sal16a} noted an immediate link between the Cartesian stress components and the conformal map. In fact, the stresses can be derived directly by taking the first derivative of the conformal transformation as follows \citep{Sal16a}:
\begin{equation}
    \tau_{zy}+i\tau_{zx}=\psi\frac{\mbox{d}\xi\left(z\right)}{\mbox{d}z}
\end{equation}
which means that the problem of finding the stress components comes down to simply identifying the proper conformal map and differentiate it. 

For the case in which $v=v_0$ describes $\partial \Omega_{n,0}$:
\begin{equation}\label{displacement}
    w=Au
\end{equation}
and the curvilinear stress components can be calculated as follows:
\begin{equation}
    \tau_{zu}=\frac{\psi}{\norm{z'\left(\xi\right)}}, \qquad~\quad \tau_{zv}=0
\end{equation}
Then, the expression for the Cartesian stress components takes the following form \citep{Sal16a}:
\begin{equation}\label{cartstresses}
    \tau_{zx}-i\tau_{zy}=\psi\frac{\mbox{d}\xi\left(z\right)}{\mbox{d}z}
\end{equation}
The mathematical proof in support of the foregoing equations can be found in \citep{Sal16a} or \citep{Sal19a} where a similar framework was used to solve problems in electrostatics. In the latter case, the shear stresses $\tau_{zx}$ and $\tau_{zy}$ must be substituted by the components of current density in the $x-$ and $y-$ directions while one must use the electric potential in lieu of the displacement $w$. 

\begin{figure}[!ht]
\center
\includegraphics[width=0.48\textwidth]{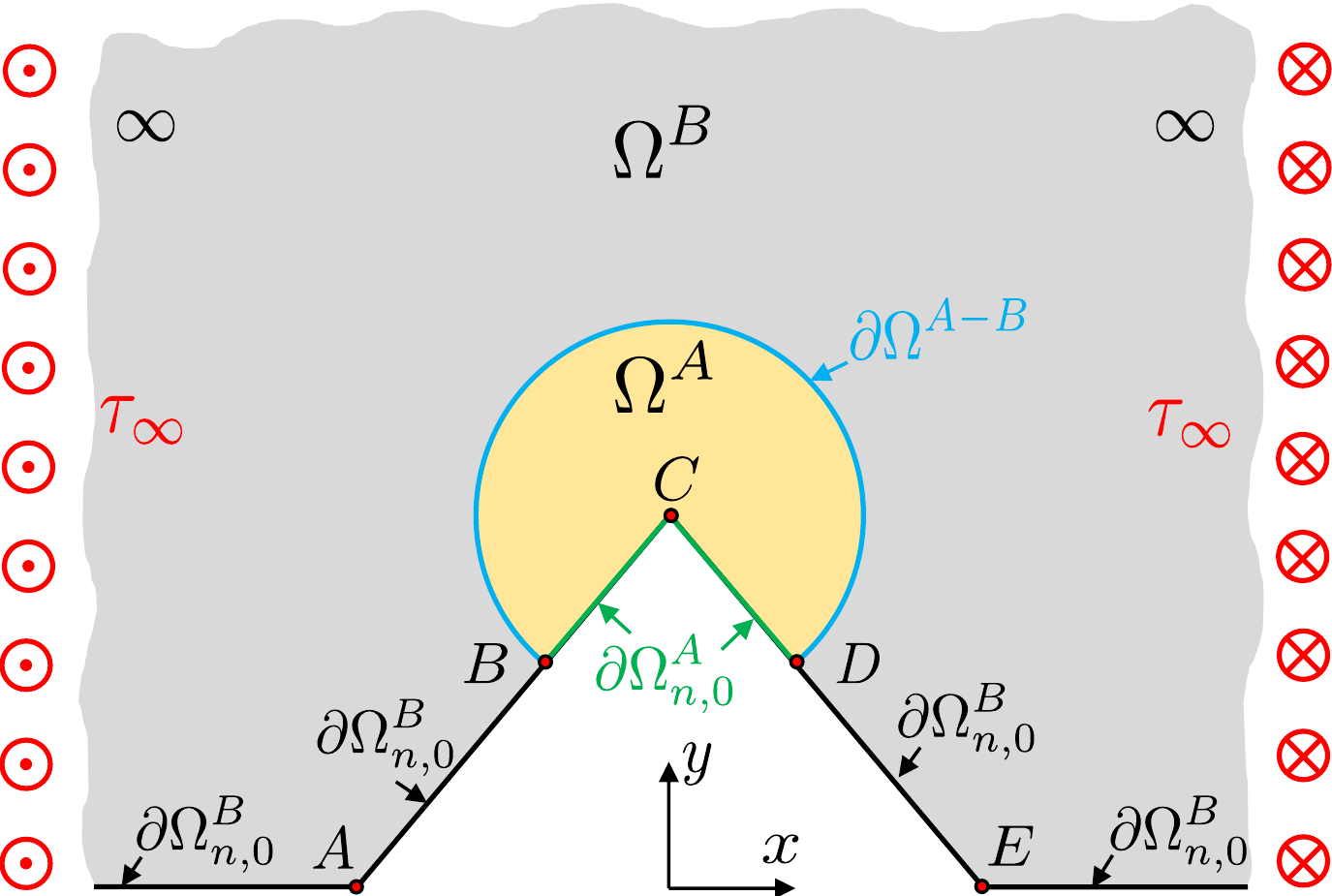}
\caption{Example of semi-infinite domain $\Omega=\Omega^A \cup \Omega^B$ with a finite notch featuring one or more regions of different materials surrounding the tip. The $\tau_{zn}=0$ condition is applied on $\Omega_{n,0}^A$ and $\Omega_{n,0}^B$ for regions A and B respectively. $\Omega^{A-B}$ represents the interface between regions A and B.}
\label{multidomain}
\end{figure}

\subsection{Governing equations for notches featuring multi-material regions embracing the tip}
We seek the solution for the antiplane shear stress field for notches featuring multi-material regions embracing the tip (Fig. \ref{multidomain}). With reference to Fig. \ref{multidomain}, let regions $\Omega^A$ and $\Omega^B$ be made of different linear elastic materials featuring shear elastic moduli $G_A$ and $G_B$ respectively. Let $\partial\Omega^A_{n,0}$ and $\partial\Omega^B_{n,0}$ be the portions of boundary subjected to no stresses for regions $\Omega^A$ and $\Omega^B$ respectively while $\partial \Omega^{A-B}$ represents the interface boundary between the two regions. Let also $w_A$ and $w_B$ be the displacement components in the $z-$direction in regions $\Omega^A$ and $\Omega^B$ respectively.  

Taken singularly, each region can be considered as homogeneous. Hence, the same governing equation for homogeneous materials, Eq. (\ref{laplacian}), can be applied to each region separately. Accordingly, the system of equations to be solved reads:
\begin{numcases}{}
  \nabla^2 w_A = 0, & $\mbox{for } (x,y) \in \Omega^A$\label{lapla}\\
  \label{BCa}\tau^A_{zn} = 0, & $\mbox{for } (x,y) \in \partial \Omega^A_{n,0}$\\
  \nabla^2 w_B =0, & $\mbox{for } (x,y) \in \Omega^B$\\
  \label{BCb}\tau^B_{zn} = 0, & $\mbox{for } (x,y) \in \partial \Omega^B_{n,0}$\\
  \label{BCc}\tau^A_{zn} = \tau^B_{zn}, & $\mbox{for } (x,y) \in \partial \Omega^{A-B}$\\
  \label{compatibility}w_A = w_B, & $\mbox{for } (x,y) \in \partial \Omega^{A-B}$
\end{numcases}
where Eqs. (\ref{BCa}), (\ref{BCb}), and (\ref{BCc}) represent equilibrium conditions on the boundaries and Eq. (\ref{compatibility}) guarantees the compatibility of the displacement field. This latter condition, along with Eq. (\ref{BCc}), is necessary since $\Omega=\Omega^A\cup\Omega^B$ is not homogeneous. 

The solution of the foregoing system of equations can be generalized and significantly simplified by taking advantage of proper conformal maps as described next.

\subsection{General solution framework}\label{framework}
Let us consider two conformal mappings $z=z\left(\xi_v\right)$ and $z=z\left(\xi\right)$ with $\xi_v=u_v+iv_v$ and $\xi=u+iv$. The two maps are defined so that the condition $v_v=v_{v,0}$ describes $\partial\Omega^A_{n,0}\cup\partial\Omega^B_{n,0}$ while $v=v_0$ describes $\partial\Omega^B_{n,0}\cup\partial\Omega^{A-B}$ (see Fig. \ref{multidomain}). Furthermore, the complex function $\xi_v=p\left(\xi\right)$ defining the relation between the two curvilinear variables is an analytic function. Taking advantage of the proposed conformal maps, the system of governing equations (\ref{lapla})-(\ref{compatibility}) can be significantly simplified as follows:
\begin{numcases}{}
  \frac{\partial^2 w_A}{\partial u_v^2}+\frac{\partial^2 w_A}{\partial v_v^2} = 0, & $\mbox{for } \xi_v \in \left(-\infty,\infty\right)\times\left[v_{v,0},\infty\right)$\label{prima}\\
  \frac{\partial w_A}{\partial v_v} = 0, & \mbox{for } $v_{v}=v_{v,0}$\label{seconda}\\
  \frac{\partial^2 w_B}{\partial u^2}+\frac{\partial^2 w_B}{\partial v^2} = 0, & $\mbox{for } \xi \in \left(-\infty,\infty\right)\times\left[v_{0},\infty\right)$\label{terza}\\
  \frac{\partial w_B}{\partial v} = 0, & \mbox{for } $v=v_{0}$\label{quarta}\\
G_A\frac{\partial \omega_A}{\partial v} = G_B\frac{\partial w_B}{\partial v}, & $\mbox{for } \xi \in \partial \Omega^{A-B}$\label{equi}\\
  \omega_A = w_B, & $\mbox{for } \xi \in \partial \Omega^{A-B}$\label{ultima}
\end{numcases}
where $\omega_A\left(u,v\right)=w_A\left[u_v\left(u,v\right),v_v\left(u,v\right)\right]$ represents the displacement in region $\Omega^A$ written as a function of the curvilinear coordinates $u,v$.

\begin{figure}[!ht]
\center
\includegraphics[width=1\textwidth]{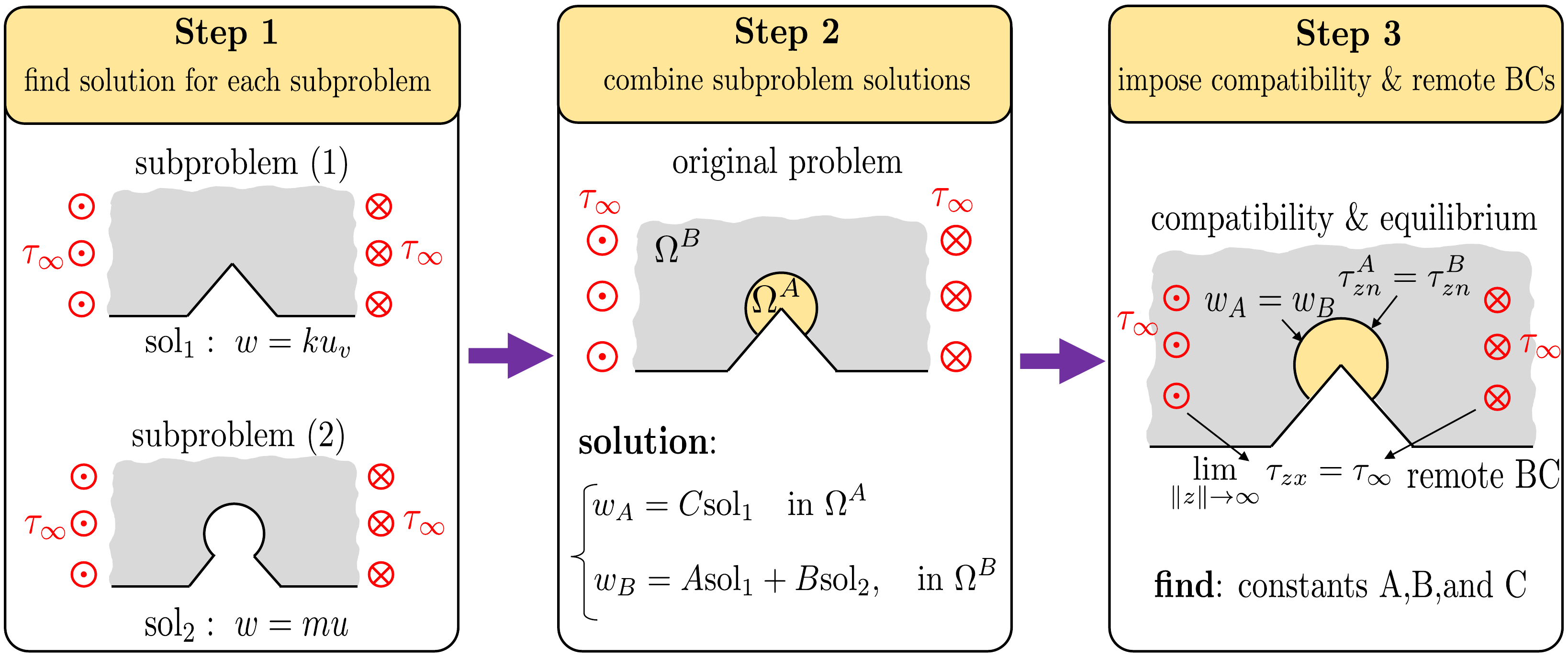}
\caption{Flowchart summarizing the three-step strategy for calculating the stress distribution. Step 1 requires the calculation of the stresses for the subproblems defined by the transformations $z=z\left(\xi_v\right)$ and $z=z\left(\xi\right)$. In Step 2, the general solution of the original problem is obtained as a combination of the subproblem solutions. Finally, in Step 3, the final solution is uniquely defined by imposing the equilibrium and compatibility conditions on the interface, and the remote conditions.  }
\label{flowchart}
\end{figure}

For the solution of equations (\ref{prima})-(\ref{ultima}) one can note that an easy way to satisfy compatibility, Eq. (\ref{ultima}), and Eqs. (\ref{prima})-(\ref{quarta}) at the same time would be to extend the solution for $\Omega^A$ to $\Omega^B$ (subproblem 1 in Fig. \ref{flowchart}). In fact, let $w_A=w_A\left(u_v,v_v\right)$ satisfy Eqs. (\ref{prima}) and (\ref{seconda}). Then, as composed function of two harmonic functions, $w_B=\omega_A\left(u,v\right)=w_A\left[u_v\left(u,v\right),v_v\left(u,v\right)\right]$ would be harmonic in $\xi$. Considering the definitions of $\xi_v$ and $\xi$, this means that $w_B$ would also satisfy both Eq. (\ref{terza}) and (\ref{quarta}). However, a simple extension would not be enough to fulfill the equilibrium condition, Eq. (\ref{equi}), on the interface $\Omega^{A-B}$ which also depends on the elastic properties of each region. To solve this problem, the solution for the displacement in $\Omega^B$ must be augmented by an additional term. Such term should still satisfy the equilibrium condition in $\partial \Omega^B_{n,0}$. A natural choice is to use $Bu$ since it automatically satisfies the condition that $\partial (Bu)/\partial v=0$ in $\partial \Omega^B_{n,0}$. It is interesting to note that, following the derivations presented in Section \ref{intro}, $Bu$ would be the solution for the displacement field in region B if region A were eliminated and the same remote boundary conditions were applied (subproblem 2 in Fig. \ref{flowchart}). 

Following Eq. (\ref{displacement}), it is easy to derive the expression for the displacement $w$ in $\Omega=\Omega^A\cup\Omega^B$:
\begin{numcases}{w=}
     w_A\left(u_v,v_v\right) = Cu_v,\quad  &\quad~\quad$\mbox{for }\xi_v\in\Omega^A\quad~\quad$ \label{w1}\\ 
     w_B\left(u,v\right) = A\omega_A\left(u,v\right)+Bu=Au_v\left(u,v\right)+Bu,\quad &\quad~\quad$\mbox{for }\xi\in\Omega^B\quad~\quad$ \label{w2}
\end{numcases}
where $A,B,C$ are real constants to be determined by imposing the remote stress, and the equilibrium and compatibility conditions on the interface. Finally, leveraging Eq. (\ref{cartstresses}), the Cartesian components of the stresses can be calculated by taking the real and imaginary parts of the following equation:

\begin{numcases}{\tau_{zx}-i\tau_{zy}=}
     CG_A\frac{\mbox{d}\xi_v\left(z\right)}{\mbox{d}z},\quad  &\quad~\quad$\mbox{for }\xi_v\in\Omega^A\quad~\quad$\label{stress1}\\
     G_B\left[A\frac{\mbox{d}\xi_v\left(z\right)}{\mbox{d}z}+B\frac{\mbox{d}\xi\left(z\right)}{\mbox{d}z}\right],\quad &$\quad~\quad\mbox{for }\xi\in\Omega^B\quad~\quad$\label{stress2}
\end{numcases}\label{multistresses}

In summary, the displacement and stress distributions in multi-material domains as the one shown in Fig. \ref{multidomain} can be obtained by performing the following three-step procedure (Fig. \ref{flowchart}):
\begin{enumerate}[leftmargin=2cm]
    \item[\textbf{Step 1}: ] Using the analytical approach described in Section \ref{intro}, find the solutions for the displacement and stress distributions underlain by the transformations $z=z\left(\xi_v\right)$ and $z=z\left(\xi\right)$ (subproblems 1 and 2 in Fig. \ref{flowchart}).
    \item[\textbf{Step 2}: ] Find the general solution for the displacement of the original problem, defined except for the constants $A,B,$ and $C$, by combining the solutions of the subproblems, Eqs. (\ref{prima})-(\ref{ultima}) (see Fig \ref{flowchart}, Step 2). Then, calculate the stress distribution leveraging Eqs. (\ref{stress1}) and (\ref{stress2}). 
    \item[\textbf{Step 3}: ] Find the values of the constants $A,B,$ and $C$ by imposing compatibility and equilibrium along the interface $\partial\Omega^{A-B}$, and by imposing the remote stress condition: $\lim_{\|z\|\rightarrow \infty}\tau_{zx}=\tau_{\infty}$ (Fig \ref{flowchart}, Step 3).
\end{enumerate}

The application of the general framework to the particular case of a multi-material V-notch as the one shown in Fig. (\ref{multidomain}) requires the solutions of two subproblems: the \textit{finite V-notch} and the \textit{finite V-notch with a circular end hole} (see Fig. \ref{flowchart}, Step 1). While the solution for the displacement and stress distributions for the finite V-notch has already been presented in \citep{Sal16a}, the solution of the finite V-notch with a circular end hole is still elusive. This solution is presented in the next sections for the first time along with a review of the solution for the finite V-notch.

\begin{figure}[!ht]
\center
\includegraphics[width=1\textwidth]{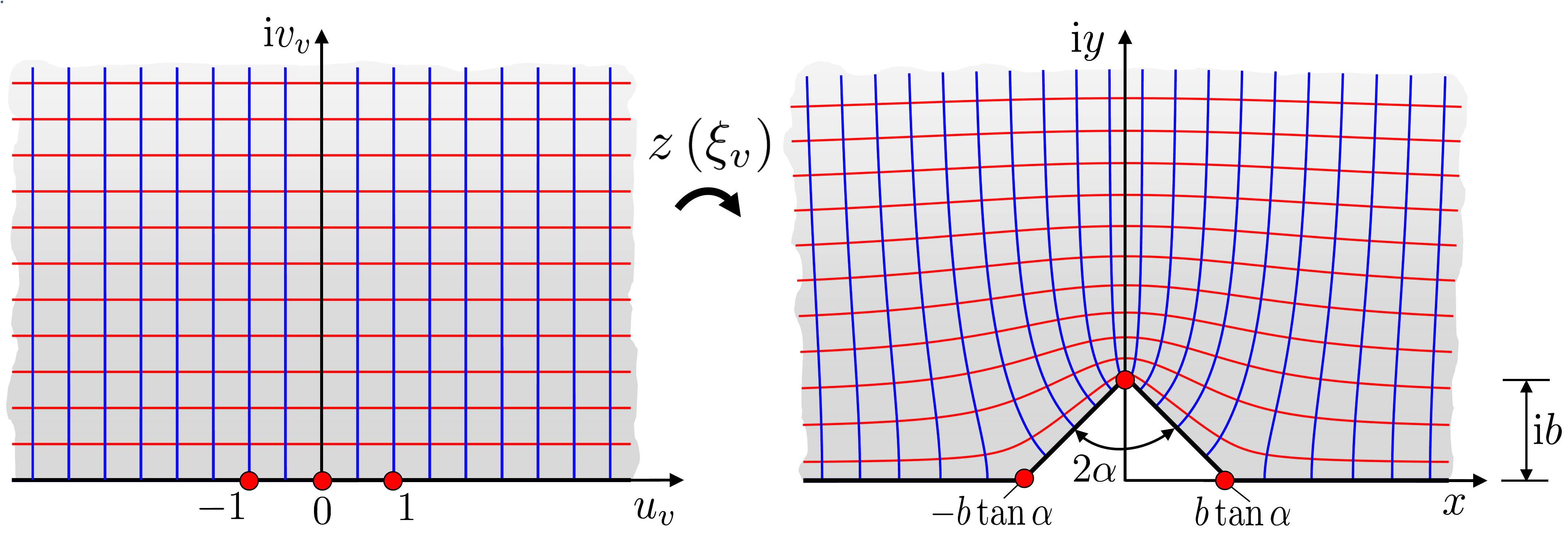}
\caption{Conformal map, Eq. (\ref{Eq.V}), transforming the upper half plane, $v_v\ge 0$ into a semi-infinite domain weakened by a finite V-notch of depth $b$ and opening angle $2\alpha$.}
\label{V-conformal}
\end{figure}

\subsection{Finite V-notch}
As discussed in the previous sections, the solution for the displacement and shear stress distributions for the finite V-notch can be easily obtained from Eq. (\ref{cartstresses}) provided that one knows the conformal map $Z=Z\left(\xi_v\right)$ transforming the upper half-plane ($v_v\ge 0$) into a semi-infinite plane featuring a finite V-notch (Fig. \ref{V-conformal}).
With reference to the coordinate system shown in Fig. (\ref{V-conformal}), such mapping was derived in \citep{Sal16a} leveraging a Schwarz-Christoffel transformation \citep{Driscoll02}:
\begin{equation}\label{Eq.V}
    Z\left(\xi\right)=\frac{A_v\pi\exp\left[i\left(\alpha-\frac{\pi}{2}\right)\right]}{2\left(\pi-\alpha\right)}\xi_v^{2\left(1-\frac{\alpha}{\pi}\right)} H\left(\alpha,\xi_v\right)+ib 
\end{equation}
where $b$ represents the depth of the V-notch, $2\alpha \in \left[0,\pi\right]$ is the notch opening angle, and the boundary featuring zero shear stress components in the normal direction, $\partial\Omega_{n,0}$, is defined by the condition $v_v=0$ (Fig. \ref{V-conformal}). Furthermore,
\begin{equation}\label{notchconstant}
    A_v=\frac{b\sqrt{\pi}}{\cos\alpha\Gamma\left(1-\alpha/\pi\right)\Gamma\left(1/2+\alpha/\pi\right)}
\end{equation}
is a constant depending on the notch depth $b$ and notch opening angle $\alpha$ \citep{Sal16a} with $\Gamma(t)=\int_0^{\infty}x^{t-1}\exp(-x)\mbox{d}x$ being the gamma function \citep{abramowitz1964handbook}. Finally:
\begin{equation}
    H\left(\alpha,\xi_v\right)={_2F_1}\left(\frac{1}{2}-\frac{\alpha}{\pi},1-\frac{\alpha}{\pi},2-\frac{\alpha}{\pi},\xi_v^2\right)
\end{equation}
where $_2F_1(a,b;c;z)=\sum_{k=0}^{\infty}(a)_k(b)_k/(c)_kz^k/k!$ is the Gaussian hypergeometric function \citep{Andrews99,Yoshida13}.

Starting from the conformal map, Eq. (\ref{Eq.V}), the Cartesian stress components written as a function of the curvilinear coordinates can be calculated taking advantage of Eq. (\ref{cartstresses}):
\begin{equation}\label{vcart}
    \tau_{zx}-i\tau_{zy}=\psi\left[\frac{\mbox{d}Z\left(\xi_v\right)}{\mbox{d}\xi_v}\right]^{-1}=\psi\frac{\left(\xi_v^2-1\right)^{1/2-\alpha/\pi}}{\xi_v^{1-2\alpha/\pi}}
\end{equation}
from which one can note that the remote boundary condition gives $\lim_{\|\xi_v\|\rightarrow \infty}\left(\tau_{zx}-i\tau_{zy}\right)=\psi=\tau_{\infty}$. Then, the equations for the stress components can be rewritten as a function of the remote stress $\tau_{\infty}$ by extracting the real and imaginary parts of Eq. (\ref{vcart}) as follows:
\begin{equation}\label{stressv1}
    \tau_{zx}\left(u_v,v_v\right)=\tau_{\infty}\frac{\left[\left(u_v^2-v_v^2-1\right)^2+4u_v^2v_v^2\right]^{1/4-\alpha/2\pi}}{\left(u_v^2+v_v^2\right)^{1/2-\alpha/\pi}}\cos{\eta}
\end{equation}
\begin{equation}\label{stressv2}
    \tau_{zy}\left(u_v,v_v\right)=\tau_{\infty}\frac{\left[\left(u_v^2-v_v^2-1\right)^2+4u_v^2v_v^2\right]^{1/4-\alpha/2\pi}}{\left(u_v^2+v_v^2\right)^{1/2-\alpha/\pi}}\sin{\eta}
\end{equation}
where $\eta=\left(\alpha/\pi-1/2\right)\arg{\left(u_v^2-v_v^2-1+2iu_vv_v\right)}+\left(1-2\alpha/\pi\right)\arg{\left(u_v+iv_v\right)}$. It is worth mentioning that, leveraging Eqs. (\ref{stressv1}) and (\ref{stressv2}), the polar components can be calculated easily: $\tau_{zr}=\cos{\theta}\tau_{zx}+\sin{\theta}\tau_{zy}$ and $\tau_{z\theta}=\cos{\theta}\tau_{zy}-\sin{\theta}\tau_{zx}$.

\begin{figure}[!ht]
\center
\includegraphics[width=1\textwidth]{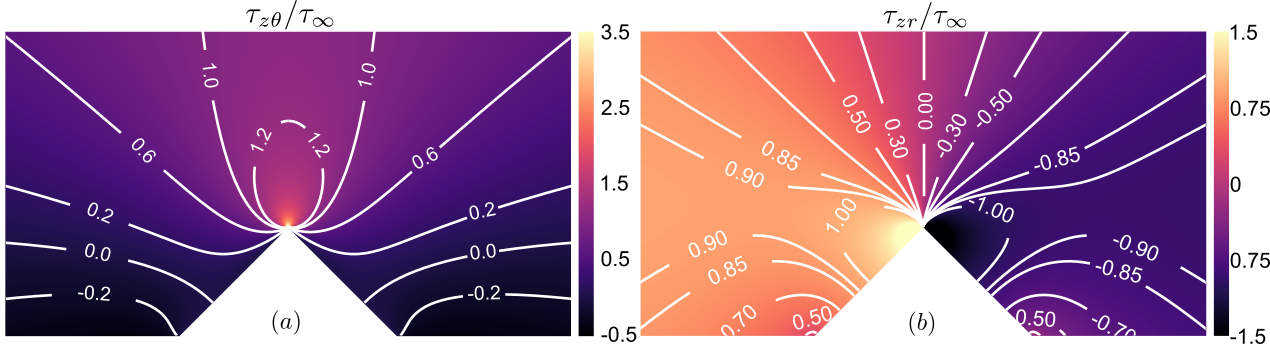}
\caption{Normalized stress distribution for a finite V-notch of depth $b=5$ mm and opening angle $2\alpha=90^{\circ}$ subjected to a remote stress $\lim_{z\rightarrow \infty}\tau_{zx}=-\tau_{\infty}$: (a) $\tau_{z\theta}/\tau_{\infty}$ and (b) $\tau_{zr}/\tau_{\infty}$.}
\label{stressesVnotch}
\end{figure}

Contour plots of the shear stress components calculated by means of Eqs. (\ref{stressv1}) and (\ref{stressv2}) are shown in Fig. \ref{stressesVnotch}.

\subsubsection{Near-tip stress field}\label{finitevneartip}
The stress distribution described by Eqs. (\ref{stressv1}) and (\ref{stressv2}) is written as a function of the curvilinear coordinates $u_v$ and $v_v$. However, for practical uses, it would be convenient to express the stress components as a function of a Cartesian coordinate system centered at the tip, $(x,y)$, or a polar coordinate system $(r,\theta)$. This would require the inversion of the conformal map expressed in Eq. (\ref{Eq.V}): $\xi_v\left(x,y\right)=Z^{-1}\left(x,y\right)$. However, considering the complexity of the transformation, a closed-form solution would not be attainable. A possible way to overcome this issue is to acknowledge that the failure behavior of the structure will mostly depend on the stress distribution close to the notch tip rather than away from it. Furthermore, for the calculation of the Notch Stress Intensity Factor (NSIF), only the near-tip stress field is of relevance. Accordingly, one can focus on the calculation of the stress components in a region sufficiently close to the tip where significant simplifications can be introduced. Salviato and Zappalorto \citep{Sal16a} showed that by expanding Eq. (\ref{Eq.V}) in Laurent series around $\xi_v=0$, which is the condition describing the location of the notch tip, one gets:
\begin{equation}
    \frac{Z-ib}{\xi_v^{1-2\alpha/\pi}}=\lim_{\xi_v\rightarrow 0}\left[\frac{Z' \left(\xi_v\right)}{\xi_v^{1-2\alpha/\pi}}-\frac{1-2\alpha/\pi}{2\left(1-\alpha/\pi\right)}\frac{Z' \left(\xi_v\right)}{\xi_v^{1-2\alpha/\pi}}\right]\xi_v+\mathcal{O}\left(\xi_v^2\right)
\end{equation}
and introducing the change of variables $z=Z-ib$, it is possible to obtain the following approximation for the relation between Cartesian and curvilinear coordinates close to the notch tip \citep{Sal16a}:
\begin{equation}\label{vz}
    \xi_v\approx \exp\left[\frac{i\left(\pi/2-\alpha\right)}{2\left(1-\alpha/\pi\right)}\right]\left[\frac{2\left(1-\alpha/\pi\right)}{A_v}\right]^{1/2\left(1-\alpha/\pi\right)}z^{1/2\left(1-\alpha/\pi\right)}
\end{equation}
Finally, substituting Eq. (\ref{vz}) into Eq. (\ref{vcart}) makes it possible to calculate the stresses near the tip as a function of Cartesian coordinates:
\begin{equation}\label{vcartstress}
    \tau_{zx}-i\tau_{zy}=\tau_{\infty}\exp\left[i\frac{\pi\left(q-1\right)}{2q}\right]\left(\frac{A_v}{q}\right)^{1/q-1}z^{1/q-1}
\end{equation}
where the equality $2\alpha=\pi\left(2-q\right)$ has been used. Using the foregoing equation, the Notch Stress Intensity Factor (NSIF) \citep{Gross72} can be easily determined leveraging its definition $K_3=\sqrt{2\pi}\lim_{y\rightarrow 0}\tau_{zx}\left(0,y\right)y^{1-1/q}$ which leads to the following expression \citep{Sal16a}:
\begin{equation}
    K_3=\tau_{\infty}b^{1-1/q}k_3
\end{equation}
where:
\begin{equation}\label{kappa}
    k_3=\sqrt{2\pi}\left[\frac{\sqrt{\pi}}{q\cos\alpha\Gamma\left(1-\frac{\alpha}{\pi}\right)\Gamma\left(\frac{1}{2}+\frac{\alpha}{\pi}\right)}\right]^{1-1/q}
\end{equation}
is a dimensionless function depending solely on the notch opening angle.
After calculating the NSIF and extracting the real and imaginary parts of Eq. (\ref{vcartstress}), the near-tip Cartesian stress components can be written as follows:
\begin{equation}
    \tau_{zx}=\frac{K_3}{\sqrt{2\pi}\left(x^2+y^2\right)^{\frac{1}{2}\left(1-\frac{1}{q}\right)}}\cos{\left[\left(1-\frac{1}{q}\right)\left(\arctan{\frac{y}{x}}-\frac{\pi}{2}\right)\right]}
\end{equation}
\begin{equation}
    \tau_{zy}=\frac{K_3}{\sqrt{2\pi}\left(x^2+y^2\right)^{\frac{1}{2}\left(1-\frac{1}{q}\right)}}\sin{\left[\left(1-\frac{1}{q}\right)\left(\arctan{\frac{y}{x}}-\frac{\pi}{2}\right)\right]}
\end{equation}
or, in polar coordinates:
\begin{equation}\label{tvbi1}
    \tau_{zx}=\frac{K_3}{\sqrt{2\pi}r^{1-\frac{1}{q}}}\cos{\left[\left(1-\frac{1}{q}\right)\left(\theta-\frac{\pi}{2}\right)\right]}
\end{equation}
\begin{equation}\label{tvbi2}
    \tau_{zy}=\frac{K_3}{\sqrt{2\pi}r^{1-\frac{1}{q}}}\sin{\left[\left(1-\frac{1}{q}\right)\left(\theta-\frac{\pi}{2}\right)\right]}
\end{equation}
Finally, leveraging the relationship $\tau_{zr}-i\tau_{z\theta}=\exp\left(i\theta\right)\left(\tau_{zx}-i\tau_{zy}\right)$, one can derive the equations for the polar stress components:
\begin{equation}\label{neartiptheta1}
    \tau_{zr}=\frac{K_3}{\sqrt{2\pi}r^{1-\frac{1}{q}}}\cos{\left[\frac{1}{q}\left(\frac{\pi}{2}-\theta\right)-\frac{\pi}{2}\right]}
\end{equation}
\begin{equation}\label{neartiptheta}
    \tau_{z\theta}=\frac{K_3}{\sqrt{2\pi}r^{1-\frac{1}{q}}}\sin{\left[\frac{1}{q}\left(\frac{\pi}{2}-\theta\right)-\frac{\pi}{2}\right]}
\end{equation}

\begin{figure}[!ht]
\center
\includegraphics[width=0.5\textwidth]{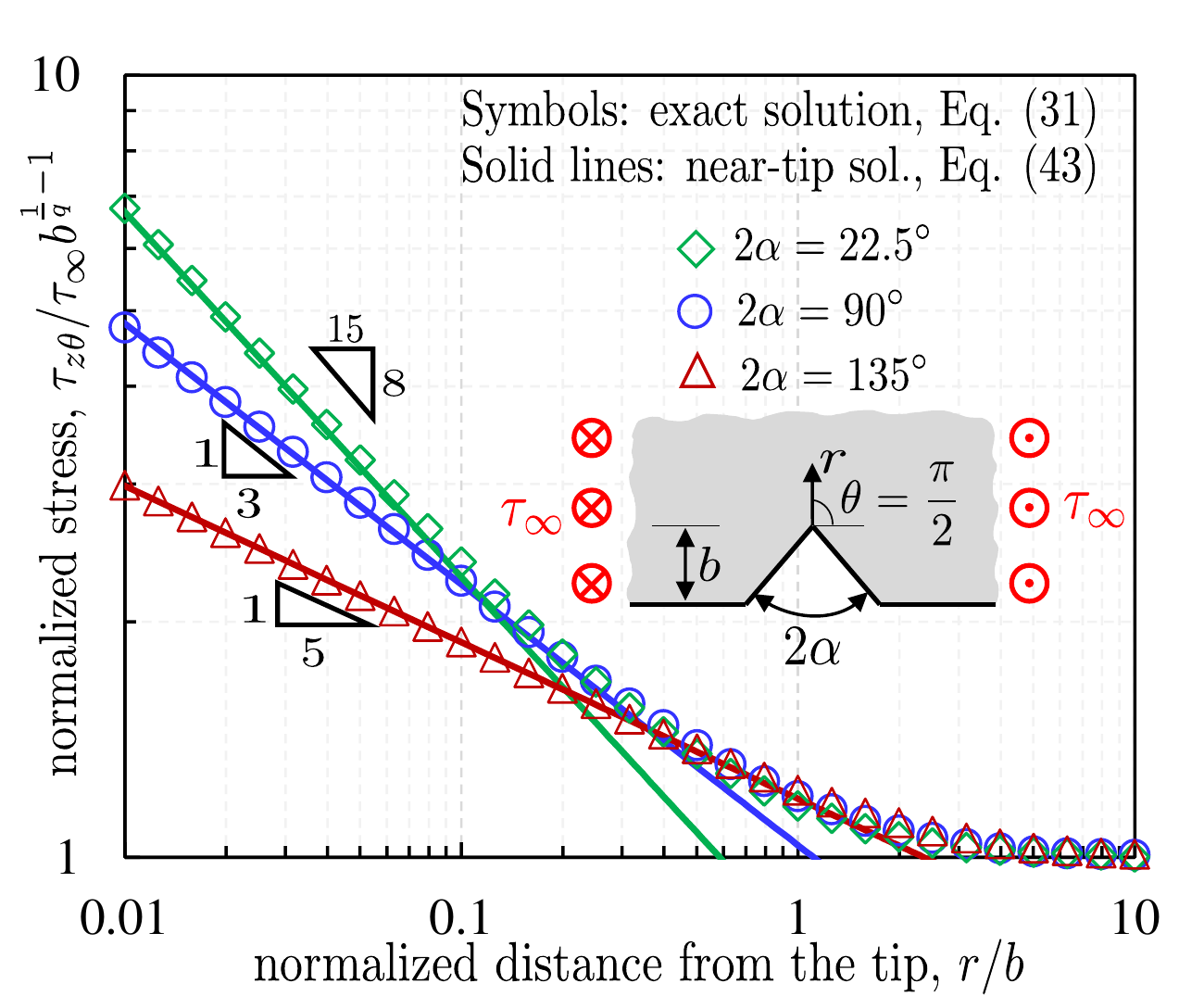}
\caption{Double-logarithmic plot showing the normalized stress, $\tau_{z\theta}/\tau_{\infty}b^{\frac{1}{q}-1}$, as a function of the normalized distance from the notch tip, $r/b$ along the bisector ($\theta=\pi/2$).}
\label{Figure_vstresses}
\end{figure}

Fig. \ref{Figure_vstresses} shows the normalized stress, $\|\tau_{z\theta}\|/\tau_{\infty}b^{\frac{1}{q}-1}$, as a function of the normalized distance from the notch tip along the bisector, $r/b$, in double-logarithmic scale for various notch opening angles. The symbols represent the exact solution calculated numerically by means of Eq. (\ref{stressv1}) while the solid lines show the stresses calculated using the near-tip solution, Eq. (\ref{neartiptheta}). As can be noted, the stresses feature a singularity of order $1-1/q$ with $q=2-2\alpha/\pi$ for $r\rightarrow 0$ as abundantly reported in previous literature \Citep{Neuber58a,Neuber58b,Zappa08}. The strength of the singularity increases with decreasing values of the notch opening angle, $2\alpha$ and takes the maximum value of $1/2$ for the case of a crack, $2\alpha=0$. Another interesting observation that can be drawn from Fig. (\ref{Figure_vstresses}) is that the near-tip solution, Eq. (\ref{neartiptheta}), provides a  remarkably good approximation of the exact solution for distances up to one tenth of the notch depth, $b$. Considering that for brittle materials the size of the Fracture Process Zone (FPZ) is generally considerably smaller than typical notch depths, the use of the near-tip solution for the application of failure criteria is more than justified. The same holds true for quasibrittle media featuring a finite FPZ provided that Irwin's characteristic length, $l_{ch}=G_{IIIc}G/f_s^2$ \citep{Irwin58} with $G=$ shear modulus, $G_{IIIc}=$ mode-III fracture energy, and $f_s=$ shear strength, is smaller than $b/10$ \citep{Baz97, Sal21a}. Of course, for all the other cases one can always use the exact solution, Eq. (\ref{stressv1}), which captures the transition from the singular field to the remote stress $\tau_{\infty}$ (Fig. \ref{Figure_vstresses}).

\begin{figure}[!ht]
\center
\includegraphics[width=0.5\textwidth]{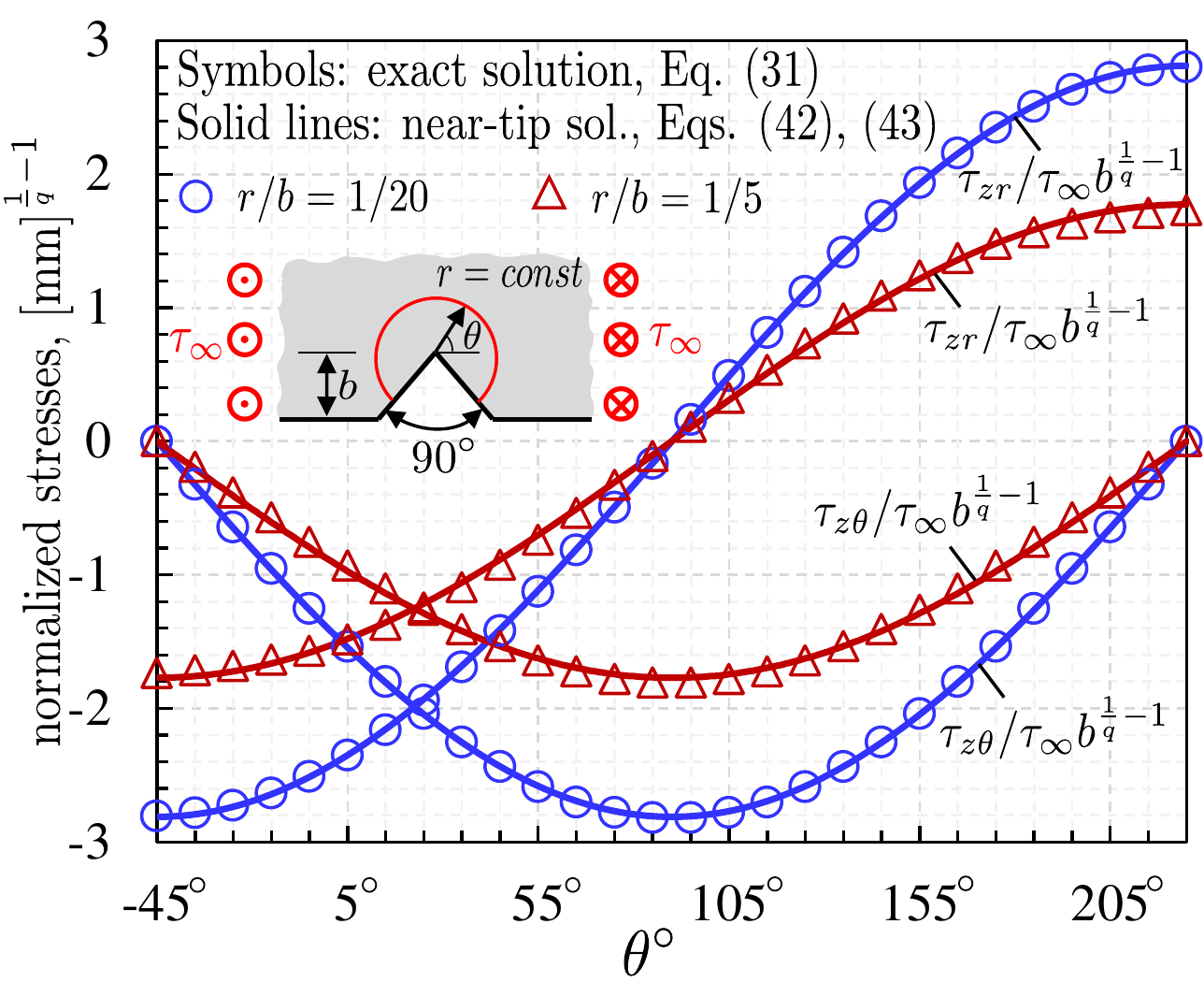}
\caption{Normalized stress distributions $\tau_{z\theta}/\tau_{\infty}b^{\frac{1}{q}-1}$ and $\tau_{zr}/\tau_{\infty}b^{\frac{1}{q}-1}$ along circular paths of radii $r=1/20b$ and $r=1/5b$ centered at the tip and embracing the notch. The notch opening angle, $2\alpha$, is equal to $90^{\circ}$ in both cases.}
\label{v_stress-Curve}
\end{figure}

Fig. \ref{v_stress-Curve} shows the normalized stresses $\tau_{z\theta}/\tau_{\infty}b^{\frac{1}{q}-1}$ and $\tau_{zr}/\tau_{\infty}b^{\frac{1}{q}-1}$ along circular paths of radii $r=1/20b$ and $r=1/5b$ centered at the tip and embracing the notch for an opening angle $2\alpha=90^{\circ}$. Again, the symbols represent the exact solution calculated numerically by means of Eq. (\ref{stressv1}) while the solid lines show the stresses calculated using the near-tip solution, Eqs. (\ref{neartiptheta1}) and (\ref{neartiptheta}). It is interesting to note that the near-tip solution provides a very good approximation of the stresses for both the radii.

\begin{figure}[!ht]
\center
\includegraphics[width=1\textwidth]{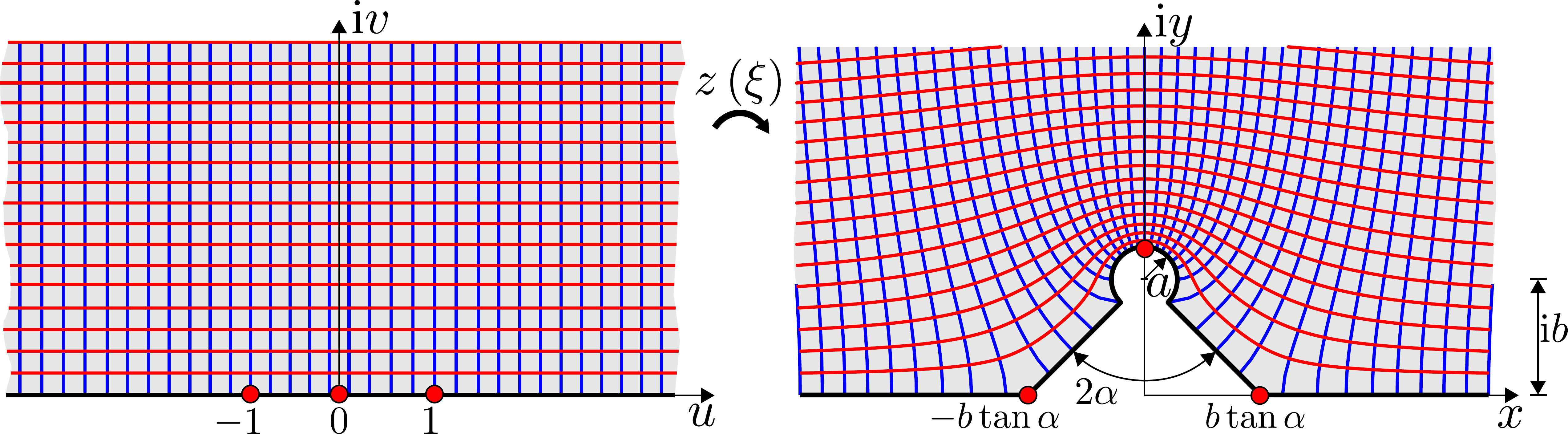}
\caption{Schematic representation of the conformal map provided by equation (\ref{keyhole_map}), transforming the upper half-plane $v\ge 0$, into a semi-infinite domain featuring a finite V-notch with circular end hole of depth $b$, radius $a$, and opening $2\alpha$}
\label{Keyhole_map}
\end{figure}

\subsection{Finite V-notch with circular end hole}
The mapping for the finite V-notch with a circular end hole can be obtained introducing the following transformation on the V-notch curvilinear coordinates:
\begin{equation}\label{v-key}
    \xi_v=u_v+iv_v=p\left(\xi,t\right)=\frac{i}{2}\left(-i\xi+\sqrt{4t^2-\xi^2}\right)
\end{equation}
where $\xi_v$ represents the curvilinear coordinates of the conformal map describing a finite V-notch of depth $b$ (see Eq. \ref{Eq.V}), and $\xi$ represents the curvilinear coordinates of the finite V-notch of depth $b$ and a final circular end hole of radius $a$ as shown in Fig. \ref{Keyhole_map}. By introducing the change of coordinates in Eq. (\ref{Eq.V}), the following expression can be found for the finite V-notch with a circular end hole:
\begin{equation}\label{keyhole_map}
\begin{split}
    Z\left(\xi\right)=\frac{A_v\pi\exp\left[i\left(\alpha-\frac{\pi}{2}\right)\right]}{2\left(\pi-\alpha\right)}\left[\frac{i}{2}\left(-i\xi+\sqrt{4t^2-\xi^2}\right)\right]^{2-\frac{2\alpha}{\pi}} \chi\left(\alpha,t,\xi\right)+ib
\end{split}
\end{equation}
where:
\begin{equation}
    \chi \left(\alpha,t,\xi\right)=H\left[\alpha,p\left(\xi,t\right)\right]={_2F_1}\left\{\frac{1}{2}-\frac{\alpha}{\pi},1-\frac{\alpha}{\pi},2-\frac{\alpha}{\pi},\left[\frac{i}{2}\left(-i\xi+\sqrt{4t^2-\xi^2}\right)\right]^2\right\}
\end{equation}
and $t$ is a parameter depending on the ratio between the notch radius and depth, $a/b$, while the constant $A_v$ was defined in Eq. (\ref{notchconstant}). Figure \ref{Keyhole_map} shows how Eq. (\ref{keyhole_map}) transforms the upper half-plane into the desired semi-infinite plane weakened by a V-notch with a circular end hole.

\begin{figure}[!ht]
\center
\includegraphics[width=1\textwidth]{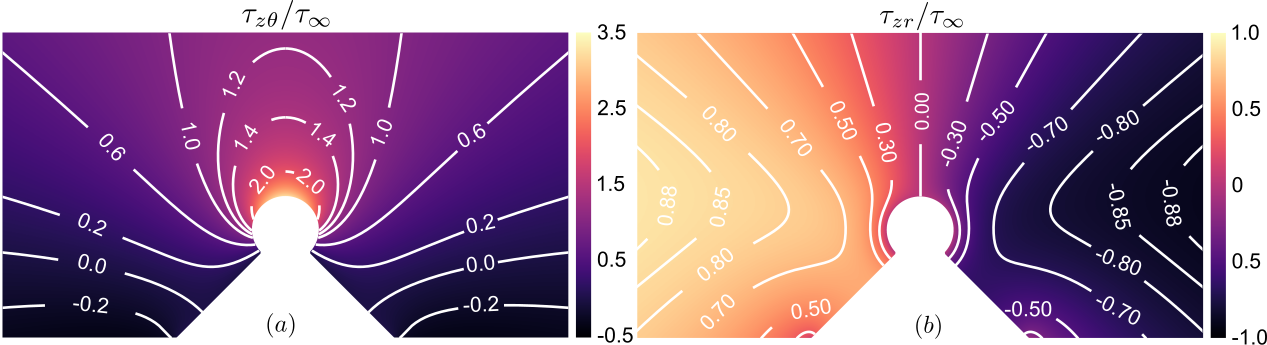}
\caption{Normalized stress distribution for a finite V-notch of depth $b=5$ mm and opening angle $2\alpha=90^{\circ}$ with circular end hole of radius $a=0.3b$ subjected to a remote stress $\lim_{z\rightarrow \infty}\tau_{zx}=-\tau_{\infty}$: (a) $\tau_{z\theta}/\tau_{\infty}$ and (b) $\tau_{zr}/\tau_{\infty}$.}
\label{stresses}
\end{figure}

Once the conformal mapping of the notch is known, the stresses can be calculated by extracting the real and imaginary parts from the following simple expression:
\begin{equation}\label{keyhole stresses}
     \tau_{zx}-i\tau_{zy}=\psi\frac{\mbox{d}\xi\left(z\right)}{\mbox{d}z}=\psi\left(\frac{\mbox{d}Z\left(\xi\right)}{\mbox{d}\xi}\right)^{-1}=\tau_{\infty}g\left(\xi\right)
\end{equation}
where:
\begin{equation}\label{g_function}
   g\left(\xi,t\right)= \frac{2^{\frac{1}{2}-\frac{\alpha}{\pi}}\sqrt{4t^2-\xi^2}\left(\xi+i\sqrt{4t^2-\xi^2}\right)^{\frac{2\alpha}{\pi}}}{\left[-2\left(1+t^2\right)+\xi\left(\xi+i\sqrt{4t^2-\xi^2}\right)\right]^{-\frac{1}{2}+\frac{\alpha}{\pi}}\left[2it^2+\xi\left(-i\xi+\sqrt{4t^2-\xi^2}\right)\right]}
\end{equation}
and $\tau_{\infty}=\lim_{\|\xi\| \rightarrow 0}\tau_{zx}-i\tau_{zy}$ is the applied remote stress. Eq. (\ref{keyhole stresses}) provides the expression of the stresses as a function of curvilinear coordinates in the whole domain. Contour plots of the shear stress components calculated by means of Eq. (\ref{keyhole stresses}) are shown in Figs. \ref{stresses}a,b.

\subsubsection{Maximum shear stress at the notch tip}
Owed to the complexity of Eqs. (\ref{keyhole_map}) and (\ref{keyhole stresses}), providing explicit expressions for the stresses as a function of Cartesian coordinates can be particularly cumbersome. However, simple relationships can be easily found to describe the stress fields close to the tip of the notch as it was done in Section \ref{finitevneartip} for the finite V-notch. By expanding Eq. (\ref{keyhole_map}) in Laurent series around $\xi=0$, which defines the location of the notch tip, one gets:
\begin{equation}
    Z\left(\xi\right)-ib=Z\left(0\right)-ib-\frac{A_vt^{1-2\alpha/\pi}}{2}\left(1+t^2\right)^{-1/2+\alpha/\pi}\xi+\mathcal{O}\left(\xi^2\right)
\end{equation}
Introducing the change of coordinates $z=Z-ib$, recalling that $Z(0)=i(a+b)$, and retaining only the linear terms of the expansion one gets:
\begin{equation}
    \xi\approx \frac{2\left(z-ia\right)}{A_vt^{1-2\alpha/\pi}}\left(1+t^2\right)^{1/2-\alpha/\pi}
\end{equation}

\begin{table}[ht]
	\caption{Coefficients $c_i$ of Eq. (\ref{tap}) as a function of the notch opening angle.}
	\centering
	\begin{tabular}{ccccc}
		\hline\hline
		Opening angle, $2\alpha$ & \textbf{$c_3$} & \textbf{$c_2$} & \textbf{$c_1$} & \textbf{$c_0$} \\
		\hline
		$0^{\circ}$ & $5.71 \times 10^{-1}$& $-1.37\times 10^0$ & $2.28\times 10^0$ & $2.52\times 10^{-1}$  \\
		$22.5^{\circ}$ & $5.14\times 10^{-1}$ & $-1.24\times 10^0$ & $2.09\times 10^0$ & $2.07\times 10^{-1}$ \\
		$45^{\circ}$ & $4.47\times 10^{-1}$ & $-1.09\times 10^0$ & $1.87\times 10^0$ & $1.62\times 10^{-1}$ \\
		$90^{\circ}$ & $2.78\times 10^{-1}$ & $-6.99\times 10^{-1}$ & $1.33\times 10^0$ & $7.88\times 10^{-2}$ \\
		$120^{\circ}$ & $1.49\times 10^{-1}$ & $-3.90\times 10^{-1}$ & $8.80\times 10^{-1}$ & $3.50\times 10^{-2}$ \\
		$135^{\circ}$ & $8.91\times 10^{-2}$ & $-2.40\times 10^{-1}$ & $6.38\times 10^{-1}$ & $1.90\times 10^{-2}$ \\
		\hline\hline
	\end{tabular}\label{coefficients}
\end{table}

\begin{figure}[!ht]
\center
\includegraphics[width=1\textwidth]{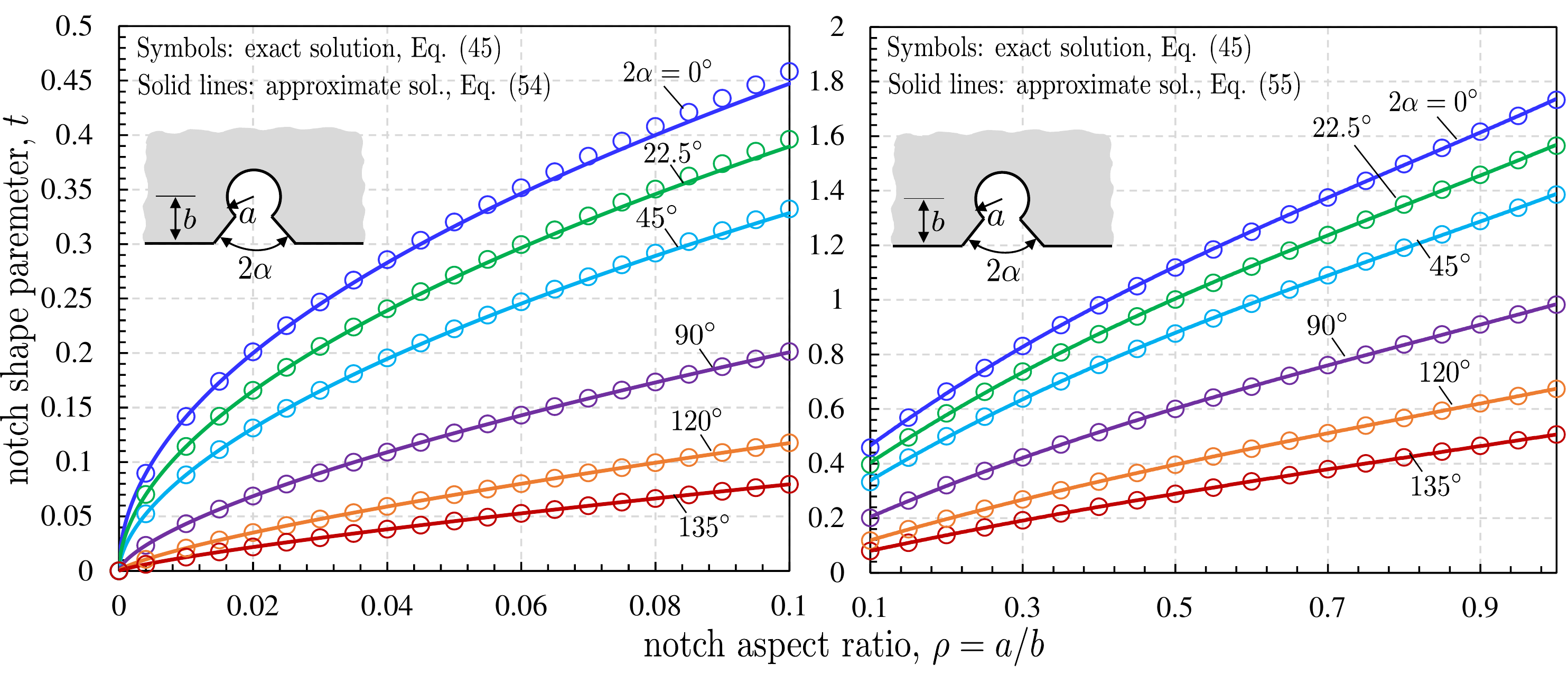}
\caption{Notch shape parameter $t$ used in Eq. (\ref{keyhole_map}) as a function of the ratio between notch depth and radius, $\rho=a/b$}
\label{t_evolution}
\end{figure}

The same procedure can be used to approximate Eq. (\ref{g_function}) giving the following expression:
\begin{equation}\label{g_approx}
   g\left(\xi,t\right)\approx \frac{2}{t^{1-2\alpha/\pi}}\left(1+t^2\right)^{1/2-\alpha/\pi}+i\frac{t^{2\alpha/\pi-2}\left[\pi\left(2+t^2\right)-2\alpha\right]}{\left(1+t^2\right)^{1/2+\alpha/\pi}\pi}\xi
\end{equation}
which, by setting the condition for the notch tip ($\xi=0$) and combining it with Eq. (\ref{keyhole stresses}), gives the expression for the maximum shear stress:
\begin{equation}\label{t_max}
\tau_{zx}^{\max}=2\tau_{\infty}\left[\frac{\sqrt{1+t^2}}{t}\right]^{1-2\alpha/\pi}
\end{equation}
This simple expression, presented for the first time in this work, shows that the maximum shear stress depends on the remote stress $\tau_{\infty}$, the notch opening angle $2\alpha$, and the notch radius/depth ratio $a/b$ through the parameter $t$. It is difficult to express the dependence of $t$ on the ratio $a/b$ in closed-form however, a very accurate approximation for $a/b\in\left[0,0.1\right]$ can be achieved by imposing the condition for the notch tip, $\xi=0$, and expanding Eq. (\ref{keyhole_map}) in Taylor series around $t=0$:
\begin{equation}
Z\left(0,t\right)= i\frac{\pi A_v }{2\left(\pi-\alpha \right)}t^{\frac{2\left(1-\alpha\right)}{\pi}}+\mathcal{O}\left(t^{4-\frac{2\alpha}{\pi}}\right)
\end{equation}
Then, after retaining only the first term of the equation, imposing the condition that $Z=i\left(a+b\right)$, and rearranging, one can find a direct relationship between the parameter $t$ and the shape ratio $a/b$:
\begin{equation}\label{tsmall}
    t\approx\left[\frac{2\left(\pi-\alpha\right)}{\pi A_v/b}\rho\right]^{\frac{\pi}{2\pi-2\alpha}}\qquad \mbox{for}~\rho=\frac{a}{b}\in \left[0,0.1\right]
\end{equation}
For $a/b \in \left(0.1,1\right]$, the following polynomial approximation can be used:
\begin{equation}\label{tap}
    t\approx c_3\rho^3+c_2\rho^2+c_1\rho+c_0 \qquad \mbox{for}~\rho=\frac{a}{b}\in \left(0.1,1\right]
\end{equation}
This expression was obtained by best fitting of numerical results obtained from Eq. (\ref{keyhole_map}). Table \ref{coefficients} provides a summary of the coefficients $c_i$ used in Eq. (\ref{tap}) for various notch opening angles while Figs. (\ref{t_evolution}a,b) show the relation between the parameter $t$ and $\rho=a/b$. The symbols represent the exact solution solved numerically by means of Eq. (\ref{keyhole_map}) while the solid lines show the excellent approximation provided by Eqs. (\ref{tsmall}) and (\ref{tap}).

\begin{figure}[!ht]
\center
\includegraphics[width=0.5\textwidth]{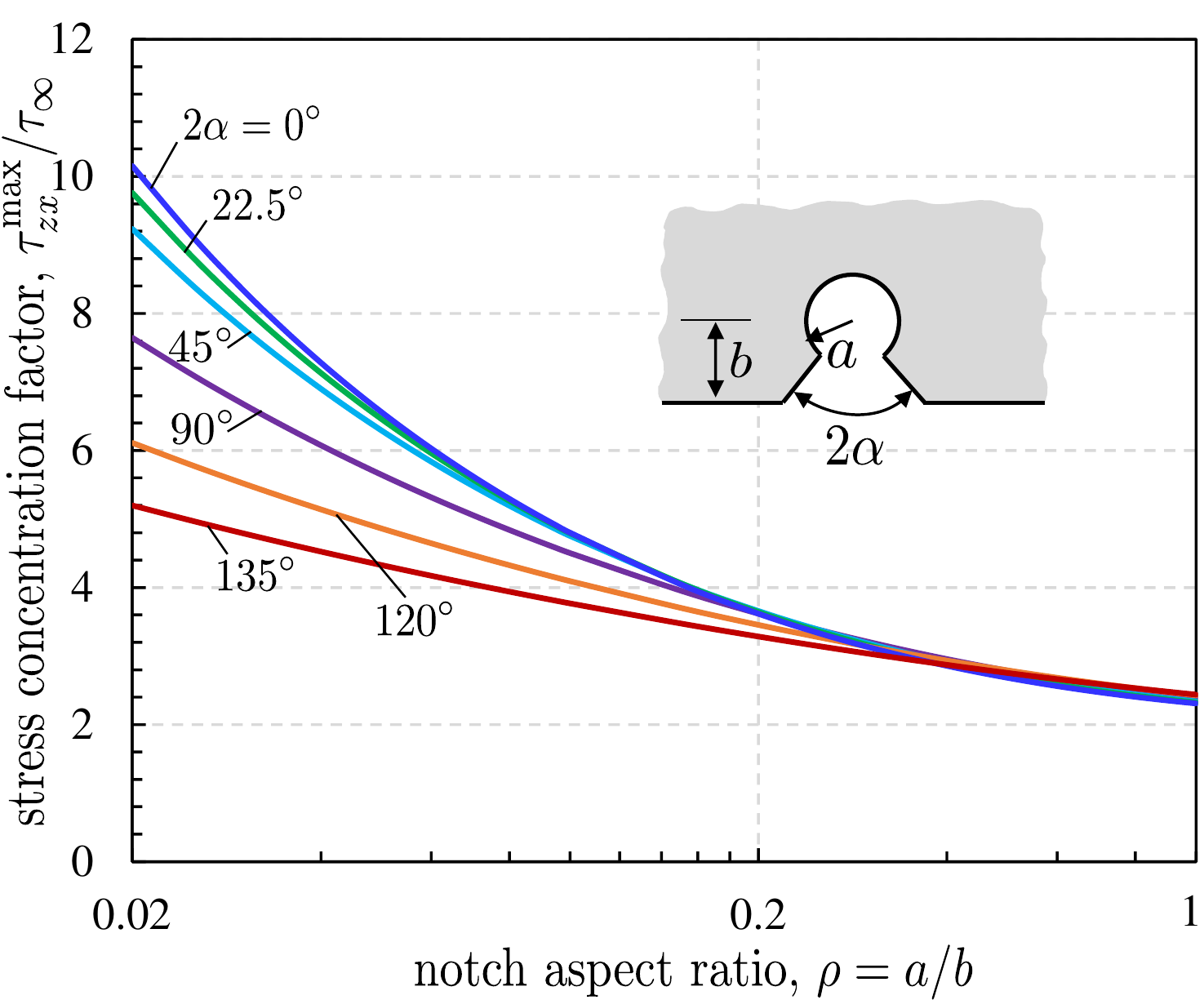}
\caption{Stress concentration factor for a finite V-notch with a circular end hole as a function of the ratio between the notch depth and radius, $\rho=a/b$, calculated leveraging Eqs. (\ref{t_max}), (\ref{tsmall}), and (\ref{tap}).}
\label{stress_concent}
\end{figure}

Thanks to Eqs. (\ref{t_max}), (\ref{tsmall}), and (\ref{tap}), it is possible to calculate the stress concentration factor, $\tau_{zx}^{max}/\tau_{\infty}$, as a function of the ratio between the notch depth and radius, $\rho=a/b$ for various notch opening angles, $2\alpha$, as shown in Fig. (\ref{stress_concent}). As can be noted from the figure, $\rho$ has a significant effect on the maximum stress regardless of the notch opening angle. In particular, a smaller $\rho$ leads to higher stress concentrations with the limit case $\rho \rightarrow 0$ leading to a stress singularity. On the other hand, it is interesting to note that the dependence of the stress concentration factor on the notch opening angle is more complex. In the range $\rho \in (0,0.2]$ the opening angle has a significant effect on the stress concentration, lower $2\alpha$ leading to higher concentrations. However, for $\rho>0.2$, the effect of the notch opening angle becomes negligible, with all the curves converging to one master curve regardless of the value of $2\alpha$.

\subsubsection{Near-tip stress field}

After the maximum stress is found, the stress field close to the notch tip can be found using the following conformal map describing an infinite V-notch with an end hole:
\begin{equation}
    z=i\left[\frac{1}{2}\left(-i\xi+\sqrt{4a^{2/q}-\xi^2}\right)\right]^q
\end{equation}
where $q=2(\pi-\alpha)/\pi$. This transformation is obtained by introducing the transformation defined in Eq. (\ref{v-key}) into the conformal map $z=i\xi^q$ which describes a deep hyperbolic notch of opening $2\alpha$ \citep{Neuber58a,Neuber58b}. Since only the stress field close of the tip is of interest and the stress concentration factor is known from the previous calculations, the effects of the finiteness of the V-notch portion can be neglected. This makes it possible to express the curvilinear coordinate $\xi$ as a function of $z$:
\begin{equation}
    \xi=i\left[-a^{2/q}(-iz)^{-1/q}+\left(-iz\right)^{1/q}\right]
\end{equation}
Now, the stresses can be easily calculated in polar coordinates as follows:
\begin{equation}
    \tau_{zr}-i\tau_{z\theta}=\psi\exp\left(i\theta\right)\frac{\mbox{d}\xi\left(z\right)}{\mbox{d}z}=i\frac{\psi}{q} \exp\left(i\theta\right)\left[a^{2/q}(-i)^{-1/q}z^{-1-1/q}+\left(-i\right)^{1/q}z^{-1+1/q}\right]
\end{equation}
which, after a few algebraic manipulations, can be re-written as:
\begin{equation}
  \tau_{zr}-i\tau_{z\theta}=\frac{i}{q}r^{1-1/q}\left\{\left(\frac{a}{r}\right)^{2/q}\exp\left[-i\left(\theta-\frac{\pi}{2}\right)\right]+\exp\left[i\left(\theta-\frac{\pi}{2}\right)\right]\right\}
\end{equation}
After extracting the real and imaginary parts, the stresses take the following expressions:
\begin{equation}
  \tau_{zr}=\frac{\psi r^{1/q-1}}{q}\left[\left(\frac{a}{r}\right)^{2/q}\sin\left(\theta-\frac{\pi}{2}\right)-\sin\left(\theta-\frac{\pi}{2}\right)\right]
\end{equation}
\begin{equation}
  \tau_{z\theta}=\frac{\psi r^{1/q-1}}{q}\left[\left(\frac{a}{r}\right)^{2/q}\cos\left(\theta-\frac{\pi}{2}\right)+\cos\left(\theta-\frac{\pi}{2}\right)\right]
\end{equation}
By setting $\theta=\pi/2$ and $r=a$ it is possible to find the relation between the constant $\psi$ and the maximum shear stress:
\begin{equation}
  \tau_{z\theta}^{max}=\frac{2\psi a^{1-1/q}}{q}
\end{equation}
Finally, the stress equations can be rewritten as a function of the maximum stress:
\begin{equation}\label{key1}
  \tau_{zr}=\frac{\tau_{z\theta}^{max} }{2}\left(\frac{r}{a}\right)^{1/q-1}\left[\left(\frac{a}{r}\right)^{2/q}\sin\left(\theta-\frac{\pi}{2}\right)-\sin\left(\theta-\frac{\pi}{2}\right)\right]
\end{equation}
\begin{equation}\label{key2}
  \tau_{z\theta}=\frac{\tau_{z\theta}^{max} }{2}\left(\frac{r}{a}\right)^{1/q-1}\left[\left(\frac{a}{r}\right)^{2/q}\cos\left(\theta-\frac{\pi}{2}\right)+\cos\left(\theta-\frac{\pi}{2}\right)\right]
\end{equation}
where $\tau_{z\theta}^{max}$ was derived from the full-field stress distribution, Eq. (\ref{keyhole stresses}), in the previous sections and can be calculated leveraging Eqs. (\ref{tsmall}) and (\ref{tap}). 

It is worth noting that Eqs. (\ref{key1}) and (\ref{key2}) agree with the equations proposed by Zappalorto and Lazzarin \citep{Zappa11a} for the analysis of deep V-notches with circular end holes. However, the solution presented in this work is more comprehensive since it can be applied to both finite and deep notches and it accounts for the effects of the depth of the notch and the radius of the end hole. The latter condition is necessary to be able to calculate the maximum stress $\tau_{z\theta}^{max}$ without relying on numerical approaches such as the Finite Element Method (FEM).

\begin{figure}[!ht]
\center
\includegraphics[width=0.5\textwidth]{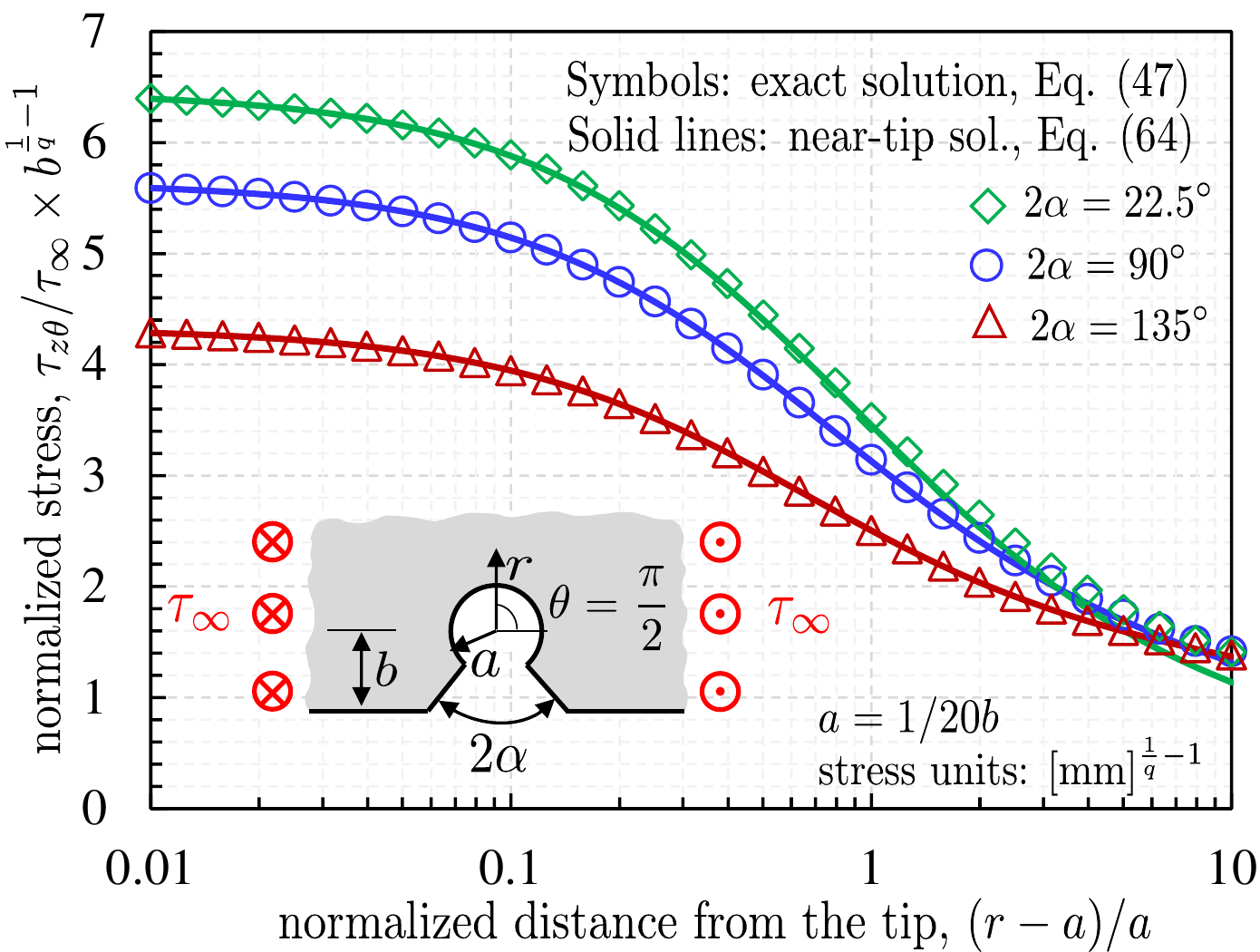}
\caption{Normalized stress, $\tau_{z\theta}/\tau_{\infty}b^{\frac{1}{q}-1}$, as a function of the normalized distance from the notch tip along the bisector, $(r-a)/a$ for $a=1/20b$.}
\label{key_bisector}
\end{figure}

\begin{figure}[!ht]
\center
\includegraphics[width=0.5\textwidth]{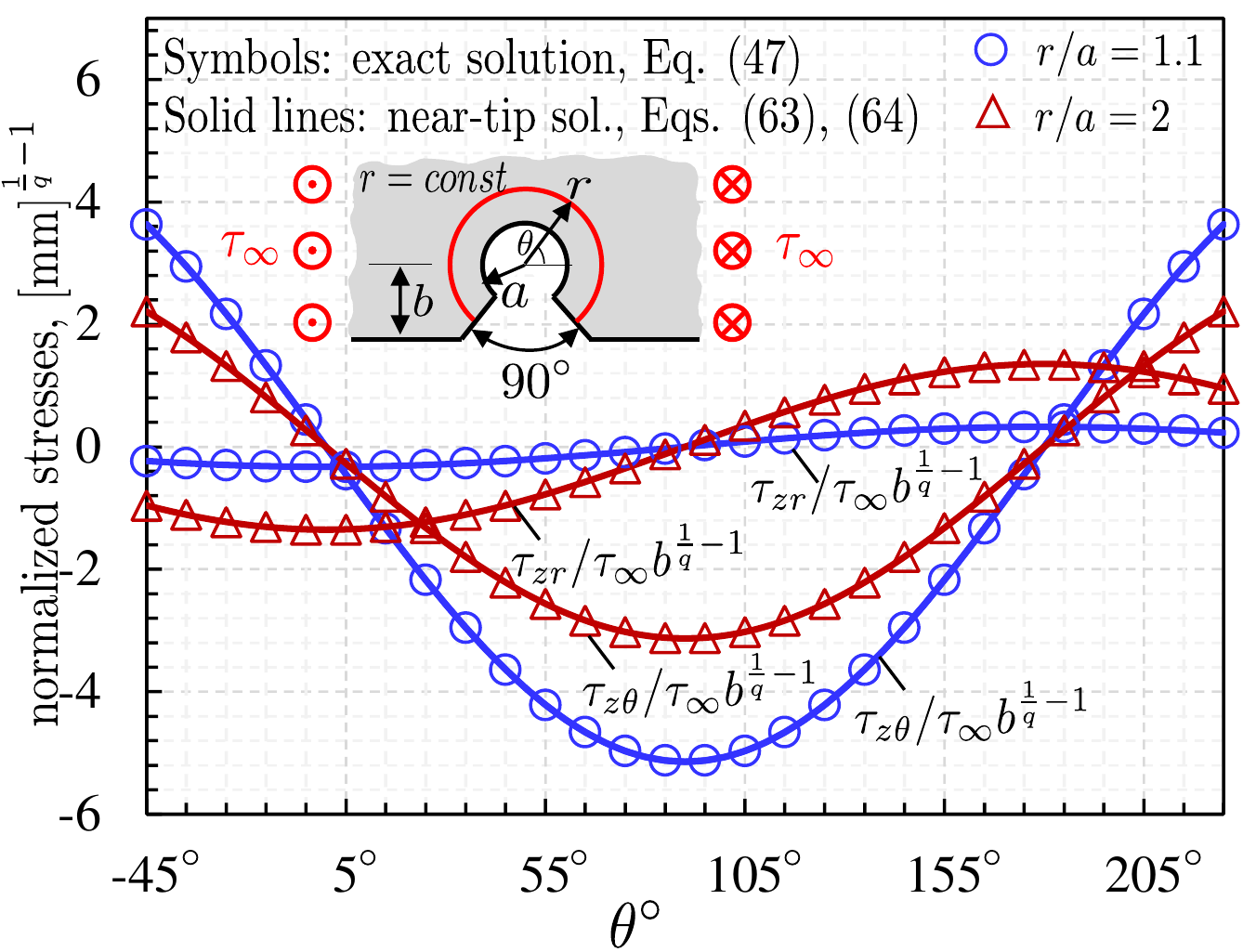}
\caption{Normalized stresses, $\tau_{z\theta}/\tau_{\infty}b^{\frac{1}{q}-1}$ and $\tau_{zr}/\tau_{\infty}b^{\frac{1}{q}-1}$, along circumferential paths embracing a V-notch with a circular end hole of radius $a=1/20b$.}
\label{keyhole_circle}
\end{figure}

The stress distributions calculated by means of the exact and near-tip solutions are shown in Figs \ref{key_bisector} and \ref{keyhole_circle}. Fig. \ref{key_bisector} shows the normalized stress, $\|\tau_{z\theta}\|/\tau_{\infty}b^{\frac{1}{q}-1}$, as a function of the normalized distance from the notch tip along the bisector, $(r-a)/a$ in semi-logarithmic scale for various notch opening angles and $a=1/20b$. The symbols represent the exact solution calculated numerically by means of Eq. (\ref{keyhole stresses}) while the solid lines show the stresses calculated using the near-tip solution, Eq. (\ref{key2}). As can be noted, the near-tip solution, Eq. (\ref{key2}), provides a remarkably good approximation of the exact solution.

Fig. \ref{keyhole_circle} shows the normalized stresses $\tau_{z\theta}/\tau_{\infty}b^{\frac{1}{q}-1}$ and $\tau_{zr}/\tau_{\infty}b^{\frac{1}{q}-1}$ along circular paths of radii $r=1.1a$ and $r=2a$ embracing the notch for an opening angle $2\alpha=90^{\circ}$. Again, the symbols represent the exact solution calculated numerically by means of Eq. (\ref{keyhole stresses}) while the solid lines show the stresses calculated using the near-tip solution, Eqs. (\ref{key1}) and (\ref{key2}). It is interesting to note that the near-tip solution provides a very good approximation of the stresses for both the radii.

\section{V-notch with circular region of a different material embracing the tip}
The previous sections analyzed the displacement and stress distributions for the case of a V-notch and a V-notch with a circular end hole. As explained in Section \ref{framework}, these distributions are required to find the solution for a V-notch featuring a circular region of a different material embracing the tip (Fig. \ref{multidomain}). With these solutions in hand, the next step is to characterize the constants $A,B,$ and $C$ utilized in Eqs. (\ref{w1}) and (\ref{w2}) to calculate the displacements in regions $\Omega^A$ and $\Omega^B$. This can be accomplished by imposing the equilibrium and compatibility conditions on the interface, Eqs. (\ref{equi}) and (\ref{ultima}), and the remote condition $\lim_{\xi\rightarrow \infty}\tau_{zx}=\tau_{\infty}$. Towards this goal, it is convenient to note that $v=0$ on the interface $\Omega^{A-B}$ so that the following relation between the coordinates $\xi_v$ and $\xi$ holds:
\begin{equation}
    u_v+iv_v=\frac{u}{2}+\frac{i}{2}\sqrt{4t^2-u}
\end{equation}
which means that at the interface, where $r=a$, the following simple relation between the curvilinear coordinates applies: $u=2u_v$. Substituting this result into Eqs. (\ref{equi}) and (\ref{ultima}) and performing a few algebraic manipulations gives the following system of equations:
\begin{numcases}{}
     G_AC = G_BA \label{equig}\\ 
     C = A+2B \label{compg}
\end{numcases}
which leads to $C=2B/\left(1-G_A/G_B\right)$ and $A=G_A/G_B\left[2B/\left(1-G_A/G_B\right)\right]$. To determine the value of the constant $B$, one can impose the remote condition:
\begin{equation}\label{remoteg}
    \tau_{\infty}=\lim_{\xi \rightarrow \infty}\frac{G_B}{h}\frac{\partial w_B}{\partial u}=G_B\left(A+B\right)
\end{equation}
Then, after combining Eqs. (\ref{equig}), (\ref{compg}), and (\ref{remoteg}) it is finally possible to get the following expressions:
\begin{equation}\label{A}
    A=\frac{2G_A\tau_{\infty}}{G_B^2\left(1+G_A/G_B\right)} 
\end{equation}
\begin{equation}\label{B}
    B=\frac{\tau_{\infty}}{G_B}\frac{\left(1-G_A/G_B\right)}{\left(1+G_A/G_B\right)} 
\end{equation}
\begin{equation}\label{C}
    C=\frac{2\tau_{\infty}}{G_B\left(1+G_A/G_B\right)} 
\end{equation}
Substituting into Eqs. (\ref{w1}) and (\ref{w2}) the displacement can be calculated as follows:
\begin{numcases}{w=}
     \frac{2\tau_{\infty}}{G_B\left(1+G_A/G_B\right)}u_v\quad  &$\quad~\quad\mbox{for }\xi_v\in\Omega^A\quad~\quad$ \label{w1g}\\ 
     \frac{2G_A\tau_{\infty}}{G_B^2\left(1+G_A/G_B\right)}u_v\left(u,v\right)+\frac{\tau_{\infty}}{G_B}\frac{\left(1-G_A/G_B\right)}{\left(1+G_A/G_B\right)}u\quad &$\quad~\quad\mbox{for }\xi\in\Omega^B\quad~\quad$ \label{w2g}
\end{numcases}
showing that the displacement depends on the domain geometry through the curvilinear coordinates $u_v$, $u$, and $v$, the remote stress $\tau_{\infty}$, and the shear moduli $G_A$ and $G_B$.

After deriving the equation for the displacement, the stresses in region $\Omega^A$ can be calculated using the following relation:
\begin{equation}\label{v_stresses}
    \tau_{zx}-i\tau_{zy}=G_AC\frac{\mbox{d}\xi_v\left(z\right)}{\mbox{d}z}=G_AC\left(\frac{\mbox{d}Z\left(\xi_v\right)}{\mbox{d}\xi_v}\right)^{-1}=\frac{2\tau_{\infty}G_A/G_B}{\left(1+G_A/G_B\right)}\frac{\left(\xi_v^2-1\right)^{1/2-\alpha/\pi}}{\xi_v^{1-2\alpha/\pi}}
\end{equation}
By extracting the real and imaginary parts of the equation, the stresses can be written as follows:
\begin{equation}\label{stressv1bi}
    \tau_{zx}\left(u_v,v_v\right)=\frac{2\tau_{\infty}G_A/G_B}{\left(1+G_A/G_B\right)}\frac{\left[\left(u_v^2-v_v^2-1\right)^2+4u_v^2v_v^2\right]^{1/4-\alpha/2\pi}}{\left(u_v^2+v_v^2\right)^{1/2-\alpha/\pi}}\cos{\eta}
\end{equation}
\begin{equation}\label{stressv2bi}
    \tau_{zy}\left(u_v,v_v\right)=\frac{2\tau_{\infty}G_A/G_B}{\left(1+G_A/G_B\right)}\frac{\left[\left(u_v^2-v_v^2-1\right)^2+4u_v^2v_v^2\right]^{1/4-\alpha/2\pi}}{\left(u_v^2+v_v^2\right)^{1/2-\alpha/\pi}}\sin{\eta}
\end{equation}
where $\eta=\left(\alpha/\pi-1/2\right)\arg{\left(u_v^2-v_v^2-1+2iu_vv_v\right)}+\left(1-2\alpha/\pi\right)\arg{\left(u_v+iv_v\right)}$. It is worth mentioning that Eqs. (\ref{stressv1bi}) and (\ref{stressv2bi}) differ from the expressions for the homogeneous case, Eqs. (\ref{stressv1}) and (\ref{stressv2}), only for the factor $2G_A/[G_B(1+G_A/G_B)]$. It is interesting to note that for $G_A/G_B \rightarrow 1$, Eqs. (\ref{stressv1bi}) and (\ref{stressv2bi}) tend to the homogeneous case. On the other hand, for the case in which region $\Omega^A$ is significantly stiffer than $\Omega^B$, $G_A/G_B \rightarrow \infty$, the multiplying factor tends to $2$. This means that, in such scenario, the stresses in $\Omega^A$ tend to be two times larger than the homogeneous case for the same remote stress $\tau_{\infty}$. As expected, the case in which region $\Omega^A$ is significantly more compliant than $\Omega^B$, $G_A/G_B \rightarrow 0$, leads to negligible stresses.

For region $\Omega^B$, the stresses can be calculated starting from the following equation:
\begin{equation}
    \tau_{zx}-i\tau_{zy}=G_B\left[A\frac{\mbox{d}\xi_v\left(z\right)}{\mbox{d}z}+B\frac{\mbox{d}\xi\left(z\right)}{\mbox{d}z}\right]=G_B\left[A\left(\frac{\mbox{d}Z\left(\xi_v\right)}{\mbox{d}\xi_v}\right)^{-1}+B\left(\frac{\mbox{d}Z\left(\xi\right)}{\mbox{d}\xi}\right)^{-1}\right]
\end{equation}
which, after inserting Eqs. (\ref{A})-(\ref{C}), gives:
\begin{equation}\label{keyb}
    \tau_{zx}-i\tau_{zy}=\frac{2\tau_{\infty}G_A/G_B}{1+G_A/G_B}f\left(\xi_v\right)+\tau_{\infty}\frac{1-G_A/G_B}{1+G_A/G_B}g\left(\xi,t\right)
\end{equation}
where:
\begin{equation}\label{f_function}
   f\left(\xi_v\right)= \frac{\left(\xi_v^2-1\right)^{1/2-\alpha/\pi}}{\xi_v^{1-2\alpha/\pi}}
\end{equation}
and $g\left(\xi,t\right)$ was defined in Eq. (\ref{g_function}). By taking the real and imaginary parts of the equation, it is possible to extract the stress components as a function of the curvilinear coordinates $\xi_v$ and $\xi$. In regard to the stresses, it is interesting to note that for $G_A/G_B \rightarrow 0$ the stresses in Eq. (\ref{keyb}) tend to the ones of the homogeneous case of the V-notch with a circular end hole, Eq. (\ref{keyhole stresses}). Of course, this is expected since in this case region $\Omega^A$ acts as an empty space. Furthermore, $G_A/G_B \rightarrow 1$ gives the homogeneous case of a finite V-notch, Eq. (\ref{vcart}). Finally, for $G_A/G_B \rightarrow \infty$ the stresses depend on the contribution of both the domains $\Omega^A$ and $\Omega^B$: $\tau_{zx}-i\tau_{zy}= \tau_{\infty}\left[2f\left(\xi_v\right)-g\left(\xi\right)\right]$. 

\begin{figure}[!ht]
\center
\includegraphics[width=1\textwidth]{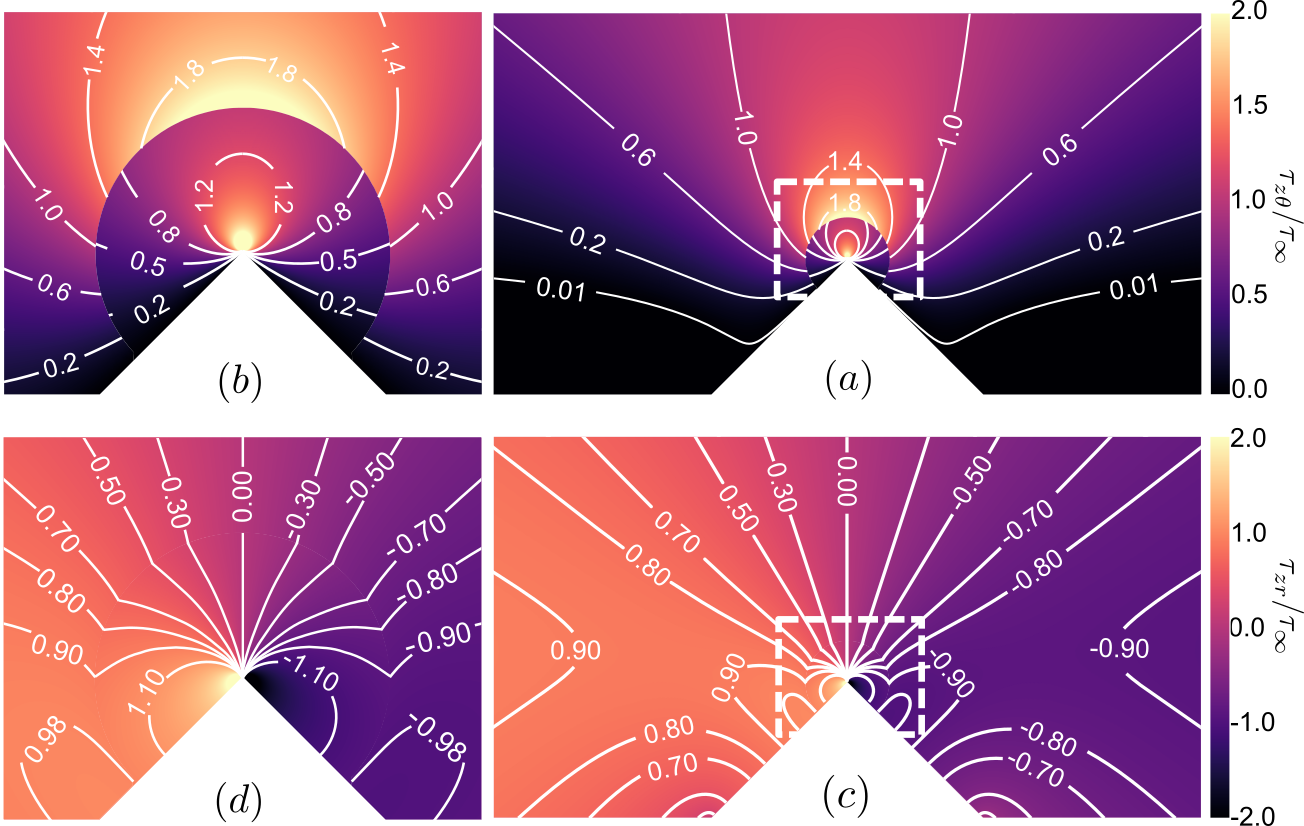}
\caption{Normalized stress distribution for a finite V-notch of depth $b=5$ mm and opening angle $2\alpha=90^{\circ}$ with circular region made of different material of radius $a=0.3b$ embracing the tip. The notch is subjected to a remote stress $\lim_{z\rightarrow \infty}\tau_{zx}=-\tau_{\infty}$. The material inside the circular region $\Omega^A$ features a shear modulus $G_A=1500$ MPa while the outer region, $\Omega^B$, has a shear modulus $G_B=3000$ MPa: (a) $\tau_{z\theta}/\tau_{\infty}$, (b) magnified view, (c) $\tau_{zr}/\tau_{\infty}$, and (d) magnified view. Note the discontinuity on $\tau_{z\theta}/\tau_{\infty}$ on the interface between $\Omega^A$ and $\Omega^B$ while $\tau_{zr}/\tau_{\infty}$ is perfectly continuous for equilibrium.}
\label{stressesbi}
\end{figure}

Contour plots of the shear stresses calculated by means of Eqs. (\ref{v_stresses}) and (\ref{keyb}) for $b=5$ mm, $a=0.3b$, $2\alpha=90^{\circ}$, and $G_A/G_B=1/2$ are shown in Figs. (\ref{stressesbi}a-d). As can be noted, the difference in material properties leads to a significant discontinuity in the tangential component of the stress (Fig. \ref{stressesbi}a,b). On the other hand, the radial component of the stress is continuous throughout the entire domain for equilibrium as shown by Figs (\ref{stressesbi}c,d).

\subsection{Near-tip stress field}
The stress field near the tip of the V-notch can be calculated leveraging the equation for the near-tip stress field of the homogeneous case, Eq. (\ref{vcartstress}). Noting that for the multi-material case the stresses in region $\Omega^A$ differ from the homogeneous case only for a multiplying factor, one can easily obtain the following expression:
\begin{equation}
   \tau_{zx}-i\tau_{zy}=\frac{2\tau_{\infty}G_A/G_B}{\left(1+G_A/G_B\right)}\exp\left[\frac{\pi\left(q-1\right)}{2q}i\right]\left(\frac{A_v}{q}\right)^{1/q-1}z^{1/q-1}
\end{equation}
Based on the foregoing equation, the NSIF \citep{Gross72} can be easily determined leveraging its definition:
\begin{equation}
\begin{split}
   K_3=\sqrt{2\pi}\lim_{y\rightarrow 0}\tau_{zx}y^{1-1/q}&=\frac{2\tau_{\infty}G_A/G_B}{\left(1+G_A/G_B\right)}\sqrt{2\pi}\left(\frac{A_v}{q}\right)^{1-1/q} \\
   & = \frac{2\tau_{\infty}G_A/G_B}{\left(1+G_A/G_B\right)}\sqrt{2\pi}\left[\frac{b\sqrt{\pi}/\cos\alpha}{q \Gamma\left(1-\alpha/\pi\right)\Gamma\left(1/2+\alpha/\pi\right)}\right]^{1-1/q}
   \end{split}
\end{equation}
or, in a more compact form:
\begin{equation}\label{NSIFbi}
   K_3=\tau_{\infty} b^{1-1/q} k_3\left(\alpha, G_A/G_B\right)
   \end{equation}
where:
\begin{equation}\label{k3bi}
   k_3\left(\alpha, G_A/G_B\right)=\frac{2\sqrt{2\pi}G_A/G_B}{\left(1+G_A/G_B\right)}\left[\frac{\sqrt{\pi}/\cos\alpha}{q \Gamma\left(1-\alpha/\pi\right)\Gamma\left(1/2+\alpha/\pi\right)}\right]^{1-1/q}
   \end{equation}
is a dimensionless function that, different from the homogeneous case, depends not only on the opening angle $2\alpha$ but also the ratio between the shear moduli $G_A/G_B$. It is interesting to note that the dependence of $K_3$ on the notch depth, $b$, features the same exponent $1-1/q$ as the homogeneous case which only depends on the notch opening angle. 

\begin{figure}[!ht]
\center
\includegraphics[width=0.5\textwidth]{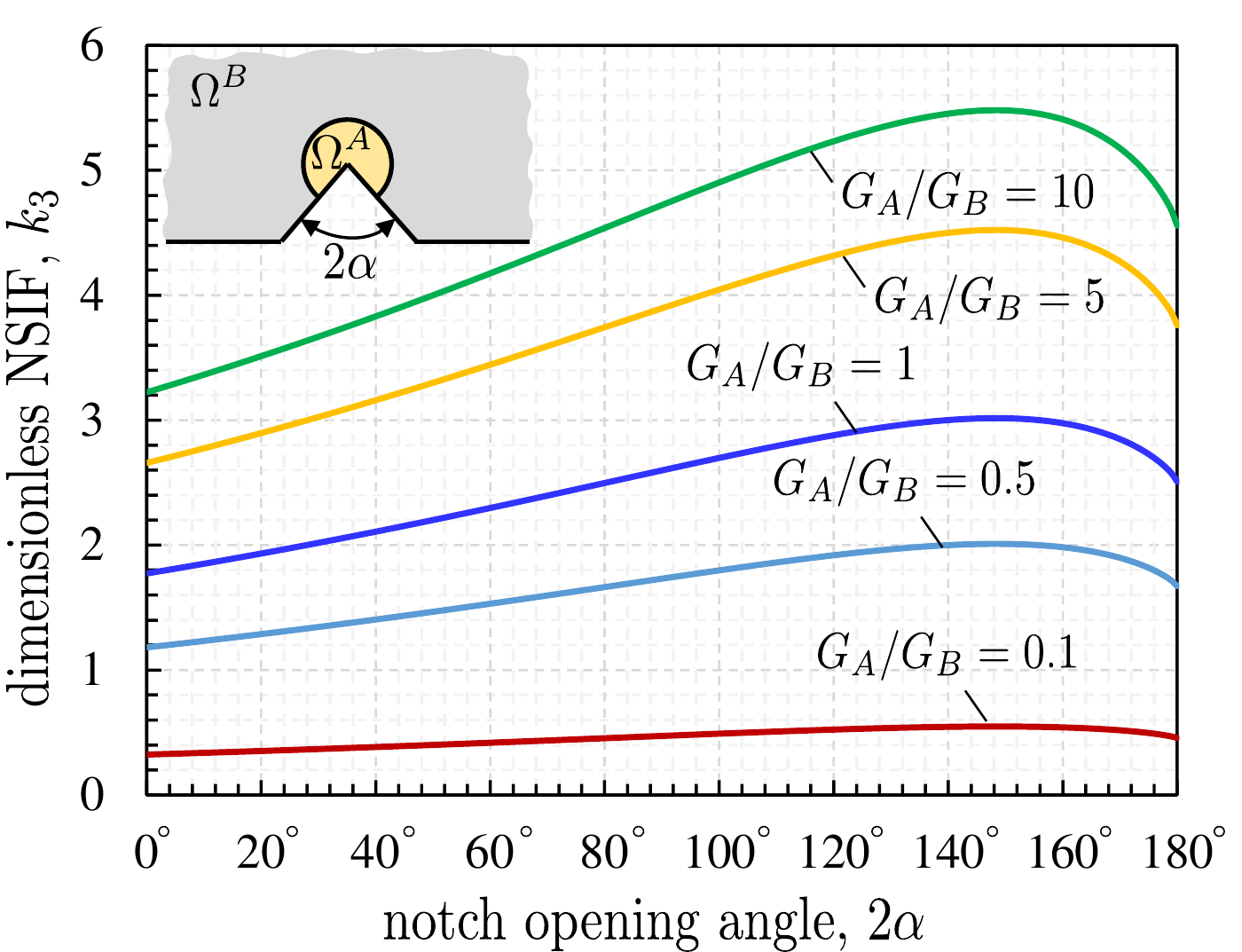}
\caption{Evolution of the dimensionless Notch Stress Intensity Factor (NSIF), Eq. (\ref{k3bi}), as a function of the notch opening angle for various ratios $G_A/G_B$. Note that, for a given opening angle, stiffer materials in region $\Omega^A$ lead to higher dimensionless NSIFs. }
\label{NSIF_bimaterial}
\end{figure}

Fig. (\ref{NSIF_bimaterial}) shows the dimensionless Notch Stress Intensity Factor $k_3\left(\alpha,G_A/G_B\right)$ as a function of the notch opening angle for various sets of elastic properties. As can be noted, the elastic properties have a significant effect on the NSIF, regardless of the opening angle. In particular, the figure shows that increasing the ratio $G_A/G_B$ leads to higher values of the NSIF. This result is particularly interesting since it means that it is possible to reduce the NSIF of the notch by finely controlling the material introduced in the circular region embracing the tip. 

After calculating the NSIF leveraging Eqs. (\ref{NSIFbi}) and (\ref{k3bi}), the Cartesian and polar stress distributions can be calculated leveraging Eqs. (\ref{tvbi1}), (\ref{tvbi2}) and (\ref{neartiptheta1}), (\ref{neartiptheta}) respectively.

\subsection{Maximum stress in $\Omega^B$}
In addition to the NSIF, it is important to investigate how the maximum stress in $\Omega^B$ is affected by the material and geometrical configurations. Towards this goal, it is useful to note that $v=u=u_v=0$ at the point of maximum stress. Considering Eq. (\ref{v-key}), this means that $v_v=t$ in the location of maximum stress. Inserting this result into Eq. (\ref{keyb}) and extracting the real part of the equation leads to the following expression for the maximum stress:
\begin{equation}\label{taumaxbi}
    \tau_{zx}^{\max}=\frac{2\tau_{\infty}}{1+G_A/G_B}\left(\frac{\sqrt{1+t^2}}{t}\right)^{1-\frac{2\alpha}{\pi}}
\end{equation}
It is interesting to note that the foregoing equation differs from the homogeneous case, Eq. (\ref{t_max}), only for the multiplying factor $1/(1+G_A/G_B)$. As expected, when $G_A/G_B \rightarrow 0$, the stress concentration tends to the homogeneous case of V-notch with a circular end hole. For the case in which the material in $\Omega^A$ is very stiff compared to the one in $\Omega^B$, $G_A/G_B \rightarrow \infty$, the maximum stress in $\Omega^B$ becomes negligible. 

\begin{figure}[!ht]
\center
\includegraphics[width=0.5\textwidth]{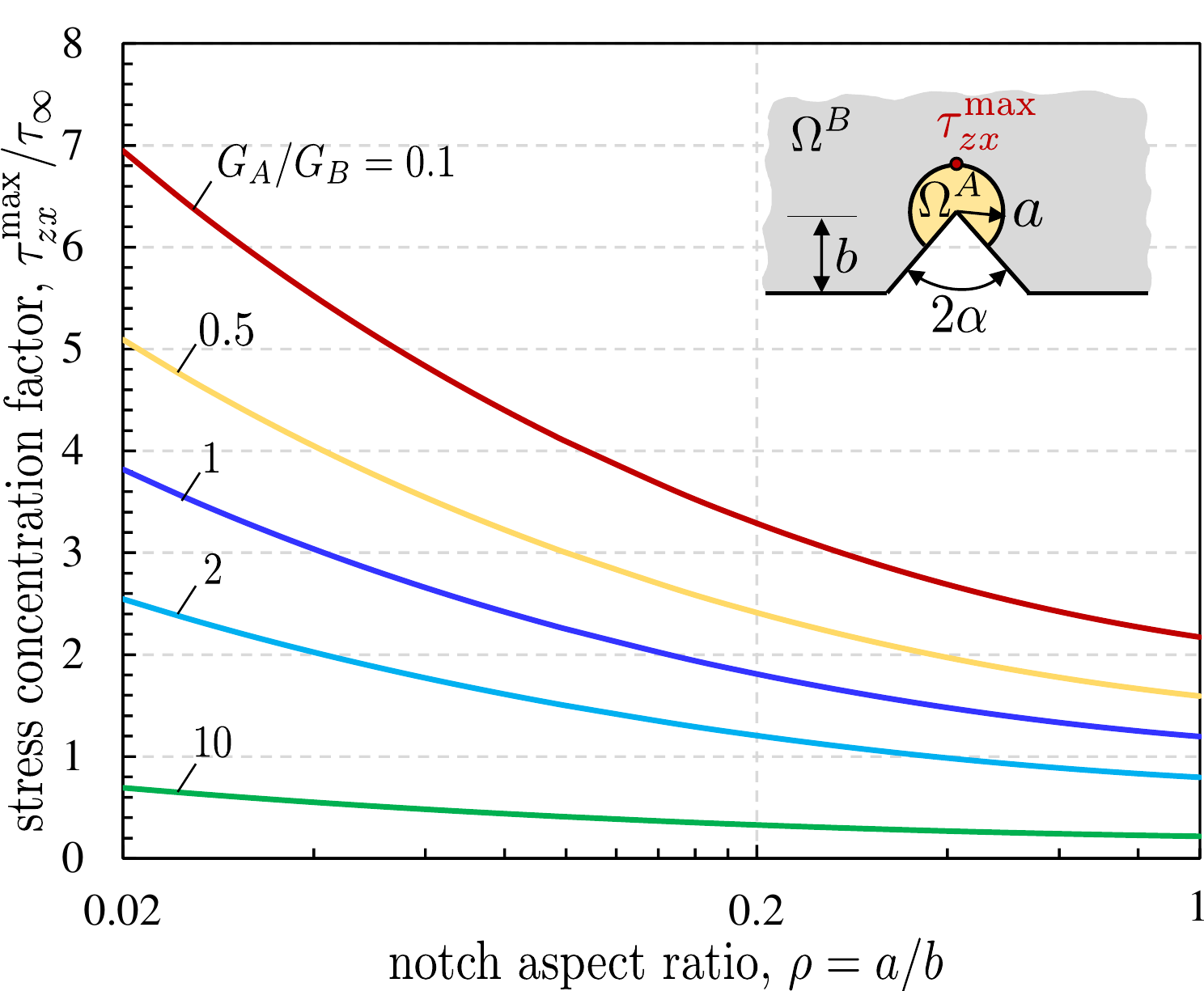}
\caption{Stress concentration factor in region $\Omega^B$ calculated by means of Eq. (\ref{taumaxbi}) as a function of the notch aspect ratio, $\rho=a/b$, for various sets of elastic properties. The notch opening angle is $2\alpha=90^{\circ}$. Note that, for a given $\rho$, lower $G_A/G_B$ ratios lead to higher concentration factors.}
\label{stress_con_bi}
\end{figure}

It is noteworthy that the maximum stress in $\Omega^B$ decreases for increasing values of $G_A/G_B$ which is an opposite trend compared to the evolution of the NSIF. This means that particular care must be devoted to the selection of the combination of materials in $\Omega^A$ and $\Omega^B$. Selecting a very soft material in $\Omega^A$ does reduce the NSIF but it does so at the expenses of the stress concentration in $\Omega^B$, which increases. Ultimately the best combination is the one that leads to the maximum structural capacity by striking the right balance between NSIF reduction and stress concentration increase. 

A summary of these observations is provided in Fig. (\ref{stress_con_bi}) which shows the evolution of the stress concentration factor as a function of $\rho=a/b$ for various $G_A/G_B$ and an opening angle of $90^{\circ}$.


\section{V-notch with two circular regions of a different material embracing the tip}

The general theoretical framework described in Sections 2 and 3 can be used for any number of circular domains embracing the notch tip.

\begin{figure}[!ht]
\center
\includegraphics[width=0.48\textwidth]{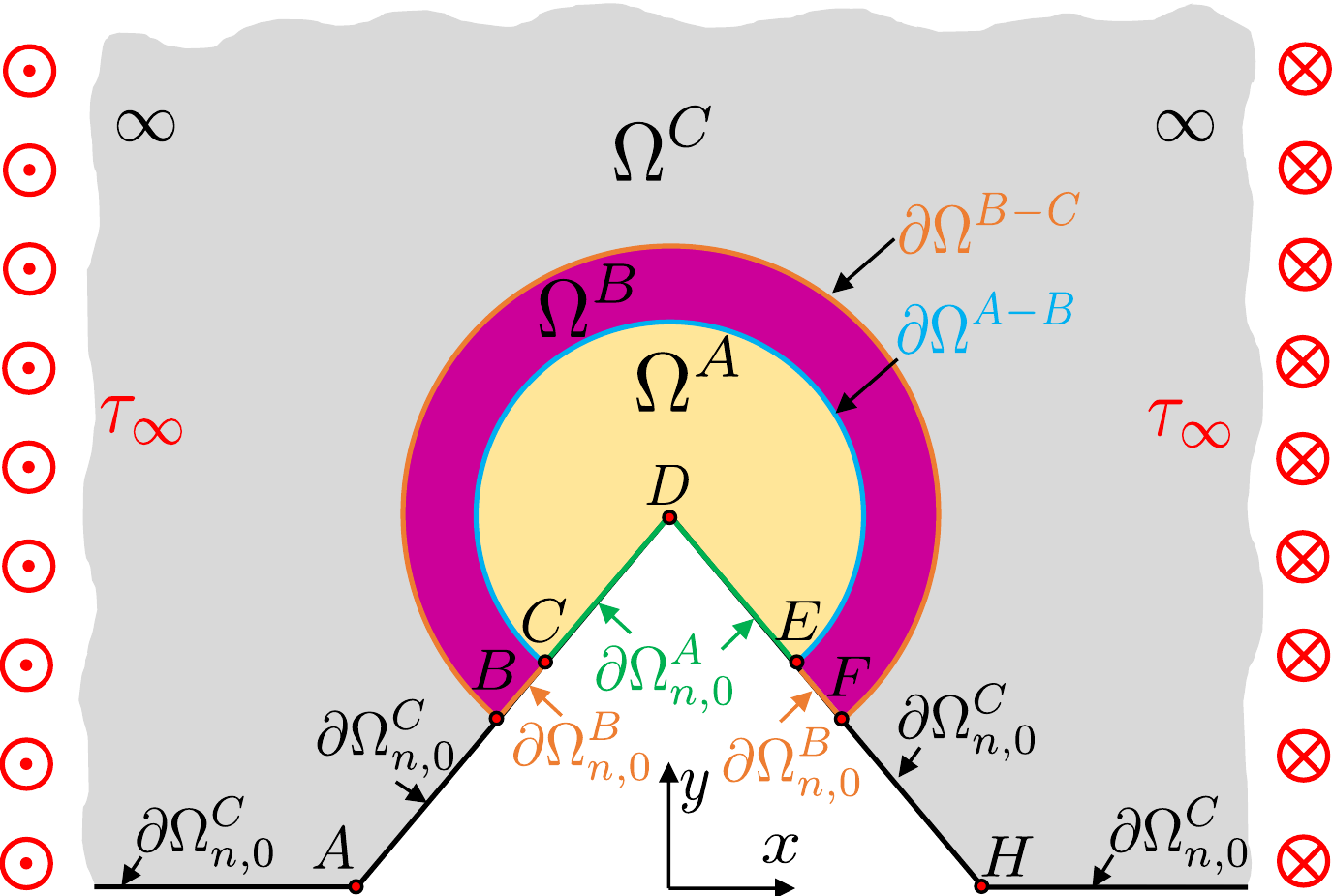}
\caption{Example of semi-infinite domain $\Omega=\Omega^A \cup \Omega^B \cup \Omega^C$ with a finite notch featuring two regions of different materials surrounding the tip. The $\tau_{zn}=0$ condition is applied on $\Omega_{n,0}^A$, $\Omega_{n,0}^B$ and $\Omega_{n,0}^C$ for regions A, B, and C. $\Omega^{A-B}$ represents the interface between regions A and B while $\Omega^{B-C}$ represents the interface between regions B and C.}
\label{multidomain2}
\end{figure}

Let us consider three conformal mappings $z=z\left(\xi_v\right)$, $z=z\left(\xi_1\right)$, and $z=z\left(\xi_2\right)$ with $\xi_v=u_v+iv_v$, $\xi_1=u_1+iv_1$, and $\xi_2=u_2+iv_2$. The three maps are defined so that the condition $v_v=v_{v,0}$ describes $\partial\Omega^A_{n,0}\cup\partial\Omega^B_{n,0}\cup\partial\Omega^C_{n,0}$, the condition $v_1=v_{1,0}$ describes $\partial\Omega^B_{n,0}\cup\partial\Omega^C_{n,0}\cup\partial\Omega^{A-B}$ while $v_2=v_{2,0}$ describes $\partial\Omega^C_{n,0}\cup\partial\Omega^{B-C}$ (see Fig. \ref{multidomain2}). This means that $\xi_v$ represents the curvilinear coordinates of the map for a finite V-notch of opening angle $2\alpha$ and depth $b$ as described in Eq. (\ref{Eq.V}) while $\xi_1$ and $\xi_2$ describe the curvilinear coordinates for a finite V-notch with circular end hole of radii $a_1$ and $a_2$ respectively as described by Eq. (\ref{keyhole_map}).

The relation between the three curvilinear variables is as follows:
\begin{equation}\label{coord3}
    \xi_v=\frac{i}{2}\left(-i\xi_1+\sqrt{4t_1^2-\xi_1^2}\right)=\frac{i}{2}\left(-i\xi_2+\sqrt{4t_2^2-\xi_2^2}\right)
\end{equation}
where $t_1$ and $t_2$ are parameters that depend on the ratio between the radius of the circular regions and the depth of the notch. They can both be calculated leveraging Eqs. (\ref{tsmall}) and (\ref{tap}).

Similar to what was done in Section 2, one can impose the equilibrium equations in each region of the multimaterial domain along with the related compatibility and equilibrium conditions on the interfaces. This leads to the following system of equations: 

\begin{numcases}{}
  \frac{\partial^2 w_A}{\partial u_v^2}+\frac{\partial^2 w_A}{\partial v_v^2} = 0, & $\mbox{for } \xi_v \in \left(-\infty,\infty\right)\times\left[v_{v,0},\infty\right)$\label{primatri}\\
  \frac{\partial w_A}{\partial v_v} = 0, & \mbox{for } $v_{v}=v_{v,0}$\label{secondatri}\\
  \frac{\partial^2 w_B}{\partial u^2}+\frac{\partial^2 w_B}{\partial v^2} = 0, & $\mbox{for } \xi \in \left(-\infty,\infty\right)\times\left[v_{1,0},\infty\right)$\label{terzatri}\\
  \frac{\partial w_B}{\partial v} = 0, & \mbox{for } $v=v_{1,0}$\label{quartatri}\\
  \frac{\partial^2 w_C}{\partial u^2}+\frac{\partial^2 w_C}{\partial v^2} = 0, & $\mbox{for } \xi \in \left(-\infty,\infty\right)\times\left[v_{2,0},\infty\right)$\label{terzatric}\\
  \frac{\partial w_C}{\partial v} = 0, & $\mbox{for } v=v_{2,0}$\label{quartatric}\\
G_A\frac{\partial \omega_A}{\partial v_1} = G_B\frac{\partial w_B}{\partial v_1}, &$ \mbox{for } \xi \in \partial \Omega^{A-B}$\label{equitri}\\
G_B\frac{\partial \omega_B}{\partial v_2} = G_C\frac{\partial w_C}{\partial v_2}, & $\mbox{for } \xi \in \partial \Omega^{B-C}$\label{equitridue}\\
  \omega_A = w_B, & $\mbox{for } \xi \in \partial \Omega^{A-B}\label{ultimatri}$\\
  \omega_B = w_C, & $\mbox{for } \xi \in \partial \Omega^{B-C}\label{ultimatridue}$
\end{numcases}
where $\omega_A\left(u_1,v_1\right)=w_A\left[u_v\left(u_1,v_1\right),v_v\left(u_1,v_1\right)\right]$ represents the displacement in region $\Omega^A$ written as a function of the curvilinear coordinates $u_1,v_1$. On a similar note, $\omega_B\left(u_2,v_2\right)=w_B\left[u_1\left(u_2,v_2\right),v_1\left(u_2,v_2\right)\right]$ represents the displacement in region $\Omega^B$ written as a function of the curvilinear coordinates $u_2,v_2$. The shear moduli in regions $\Omega^A$, $\Omega^B$, and $\Omega^C$ are $G_A$, $G_B$, and $G_C$. In the foregoing system, Eqs (\ref{primatri}), (\ref{terzatri}), and (\ref{terzatric}) represent the equilibrium equations in $\Omega^A$, $\Omega^B$, and $\Omega^C$. Eqs. (\ref{secondatri}), (\ref{quartatri}), and (\ref{quartatric}) are the equilibrium conditions on the stress-free boundary. Eqs. (\ref{equitri}) and (\ref{equitridue}) describe the equilibrium condition on the interfaces $\partial\Omega^{A-B}$ and $\partial\Omega^{B-C}$ respectively. Finally, Eqs. (\ref{ultimatri}) and (\ref{ultimatridue}) represent the compatibility condition on the interfaces.

Following the same theoretical framework described in Section 2 and 3, the solution of Eqs. (\ref{primatri})-(\ref{ultimatridue}) can be found leveraging the following displacement function:
\begin{numcases}{w=}
     w_A\left(u_v,v_v\right) = Cu_v,\quad  &\quad~\quad$\mbox{for }\xi_v\in\Omega^A\quad~\quad$ \label{displacementtri}\\ 
     w_B\left(u_1,v_1\right) = Au_v\left(u_1,v_1\right)+Bu_1,\quad &\quad~\quad$\mbox{for }\xi_1\in\Omega^B\quad~\quad$ \\
     w_C\left(u_2,v_2\right) = Mu_1\left(u_2,v_2\right)+Gu_2,\quad &\quad~\quad$\mbox{for }\xi_2\in\Omega^C\quad~\quad$ \label{displacementtrilast}
\end{numcases}
where $A,B,C,M,G$ are real constants to be found by imposing equilibrium and compatibility conditions on the interfaces and the remote stress condition. 

At this point, it is useful to point out that, as discussed in the foregoing sections, $u_1=2u_v$ along the interface $\Omega^{A-B}$. Along the interface $\Omega^{B-C}$, on the other hand, $v_2=0$. Combining this result with Eq. (\ref{coord3}) and extracting only the real part of the equation provides the following simple relation between the curvilinear coordinates $u_1$ and $u_2$:
\begin{equation}
    u_1=\frac{1}{2}\left[1+\left(\frac{t_1}{t_2}\right)^2\right]u_2
\end{equation}
Substituting this expression into Eqs. (\ref{equitri})-(\ref{ultimatridue}) and imposing the remote stress condition:
\begin{equation}\label{remotegtri}
    \tau_{\infty}=\lim_{\xi \rightarrow \infty}\frac{G_C}{h}\frac{\partial w_C}{\partial u}=G_C\left(M+G\right)
\end{equation}
one obtains the following system of equations:
\begin{numcases}{}
     C = A+2B \\
     A+\frac{B}{2}\left[1+\left(\frac{t_1}{t_2}\right)^2\right] = \frac{M}{2}\left[1+\left(\frac{t_1}{t_2}\right)^2\right] + 2G \\
     G_AC = G_B A \\
     G_B\left\{A+B\left[1-\left(\frac{t_1}{t_2}\right)^2\right]\right\} = G_C\left\{M\left[1-\left(\frac{t_1}{t_2}\right)^2\right]\right\} \\
     \tau_{\infty} = G_C\left(M+G\right)
\end{numcases}
After a few algebraic manipulations, the system allows the calculation of the desired constants:
\begin{equation}
    A=\frac{4\zeta_{AC}t_2^2\tau_{\infty}}{\left(\zeta_{AC}-\zeta_{BC}\right)\left(\zeta_{BC}-1\right)t_1^2+\left(\zeta_{AC}+\zeta_{BC}\right)\left(\zeta_{BC}+1\right)t_2^2}\label{Atri}
\end{equation}
\begin{equation}
    B=\frac{2\left(\zeta_{BC}-\zeta_{AC}\right)t_2^2\tau_{\infty}}{\left(\zeta_{AC}-\zeta_{BC}\right)\left(\zeta_{BC}-1\right)t_1^2+\left(\zeta_{AC}+\zeta_{BC}\right)\left(\zeta_{BC}+1\right)t_2^2}\label{Btri}
\end{equation}
\begin{equation}
    C=\frac{4\zeta_{BC}t_2^2\tau_{\infty}}{G_C\left[\left(\zeta_{AC}-\zeta_{BC}\right)\left(\zeta_{BC}-1\right)t_1^2+\left(\zeta_{AC}+\zeta_{BC}\right)\left(\zeta_{BC}+1\right)t_2^2\right]}\label{Ctri}
\end{equation}
\begin{equation}
    G=\frac{\left[\left(\zeta_{BC}-\zeta_{AC}\right)\left(\zeta_{BC}-1\right)t_1^4-2\zeta_{AC}\left(\zeta_{BC}+1\right)t_1^2t_2^2-\left(\zeta_{AC}+\zeta_{BC}\right)\left(\zeta_{BC}-1\right)t_2^4\right]\tau_{\infty}}{G_C\left[\left(\zeta_{BC}-\zeta_{AC}\right)\left(\zeta_{BC}-1\right)t_1^4-2\left(\zeta_{BC}^2+\zeta_{AC}\right)t_1^2t_2^2+\left(\zeta_{AC}+\zeta_{BC}\right)\left(\zeta_{BC}+1\right)t_2^4\right]}\label{Gtri}
\end{equation}
\begin{equation}
    M=\frac{2\zeta_{BC}t_2^2\tau_{\infty}\left[\left(\zeta_{AC}-\zeta_{BC}\right)t_1^2+\left(\zeta_{AC}+\zeta_{BC}\right)t_2^2\right]}{G_C\left[\left(\zeta_{BC}-\zeta_{AC}\right)\left(\zeta_{BC}-1\right)t_1^4-2\left(\zeta_{BC}^2+\zeta_{AC}\right)t_1^2t_2^2+\left(\zeta_{AC}+\zeta_{BC}\right)\left(\zeta_{BC}+1\right)t_2^4\right]}\label{Mtri}
\end{equation}
where $\zeta_{AC}=G_A/G_C$ and $\zeta_{BC}=G_B/G_C$. Substituting the foregoing equations into the displacement equations, Eqs. (\ref{displacementtri})-(\ref{displacementtrilast}), it is possible to fully characterize the displacement field except for rigid body motions. The final equations are not shown for conciseness. As can be noted from Eqs. (\ref{Atri})-(\ref{Mtri}), the displacement constants depend on the ratios between the radii of the circular regions and the notch depth the notch depth through the parameters $t_1$ and $t_2$, the elastic moduli of the various regions through the dimensionless parameters $\zeta_{AC}$ and $\zeta_{BC}$, and the remote stress $\tau_{\infty}$. On the other hand, the effect of the notch opening angle, $2\alpha$ on the displacement is automatically included by having written $w$ as a function of the curvilinear coordinates $\xi_v$, $\xi_1$, and $\xi_2$ which properly describe the border of the notch.

After deriving the equations for the displacement, the stresses in region $\Omega^A$ can be calculated using the following relation (see Eq. \ref{cartstresses}):
\begin{equation}\label{v_stressestri}
    \tau_{zx}-i\tau_{zy}=G_AC\frac{\mbox{d}\xi_v\left(z\right)}{\mbox{d}z}=G_AC\frac{\left(\xi_v^2-1\right)^{1/2-\alpha/\pi}}{\xi_v^{1-2\alpha/\pi}}
\end{equation}
where $C$ is given by Eq. (\ref{Ctri}). By extracting the real and imaginary parts of the equation, the stresses can be written as follows:
\begin{equation}\label{stressv1bitri}
    \tau_{zx}\left(u_v,v_v\right)=G_AC\frac{\left[\left(u_v^2-v_v^2-1\right)^2+4u_v^2v_v^2\right]^{1/4-\alpha/2\pi}}{\left(u_v^2+v_v^2\right)^{1/2-\alpha/\pi}}\cos{\eta}
\end{equation}
\begin{equation}\label{stressv2bitri}
    \tau_{zy}\left(u_v,v_v\right)=G_AC\frac{\left[\left(u_v^2-v_v^2-1\right)^2+4u_v^2v_v^2\right]^{1/4-\alpha/2\pi}}{\left(u_v^2+v_v^2\right)^{1/2-\alpha/\pi}}\sin{\eta}
\end{equation}
where $\eta=\left(\alpha/\pi-1/2\right)\arg{\left(u_v^2-v_v^2-1+2iu_vv_v\right)}+\left(1-2\alpha/\pi\right)\arg{\left(u_v+iv_v\right)}$.

For region $\Omega^B$, the stresses can be calculated starting from the following equation:
\begin{equation}
    \tau_{zx}-i\tau_{zy}=G_B\left[A\frac{\mbox{d}\xi_v\left(z\right)}{\mbox{d}z}+B\frac{\mbox{d}\xi_1\left(z\right)}{\mbox{d}z}\right]=G_B\left[A\left(\frac{\mbox{d}Z\left(\xi_v\right)}{\mbox{d}\xi_v}\right)^{-1}+B\left(\frac{\mbox{d}Z\left(\xi_1\right)}{\mbox{d}\xi_1}\right)^{-1}\right]
\end{equation}
which can be rewritten in a more compact form as follows:
\begin{equation}\label{keybtri}
    \tau_{zx}-i\tau_{zy}=AG_Bf\left(\xi_v\right)+BG_Bg\left(\xi_1,t_1\right)
\end{equation}
where $A$ and $B$ can be calculated by means of Eqs. (\ref{Atri}) and (\ref{Btri}) and $g\left(\xi_1,t_1\right)$ and $f\left(\xi_v\right)$ were defined in Eqs. (\ref{g_function}) and (\ref{f_function}) respectively. By taking the real and imaginary parts of the equation, it is possible to extract the stress components as a function of the curvilinear coordinates $\xi_v$ and $\xi_1$. 

The stresses in region $\Omega^C$ can be calculated in a similar way by starting from the following equation:
\begin{equation}\label{coco117}
    \tau_{zx}-i\tau_{zy}=G_C\left[M\frac{\mbox{d}\xi_1\left(z\right)}{\mbox{d}z}+G\frac{\mbox{d}\xi_2\left(z\right)}{\mbox{d}z}\right]=G_C\left[M\left(\frac{\mbox{d}Z\left(\xi_1\right)}{\mbox{d}\xi_1}\right)^{-1}+G\left(\frac{\mbox{d}Z\left(\xi_2\right)}{\mbox{d}\xi_2}\right)^{-1}\right]
\end{equation}
This equation can be rewritten as follows:
\begin{equation}\label{keybtricd}
    \tau_{zx}-i\tau_{zy}=MG_Cg\left(\xi_1,t_1\right)+GG_Cg\left(\xi_2,t_2\right)
\end{equation}
where the constants $G$ and $M$ are computed via Eqs. (\ref{Gtri}) and (\ref{Mtri}) respectively. By taking the real and imaginary parts of the equation, it is possible to extract the stress components as a function of the curvilinear coordinates $\xi_1$ and $\xi_2$. 

\begin{figure}[!ht]
\center
\includegraphics[width=1\textwidth]{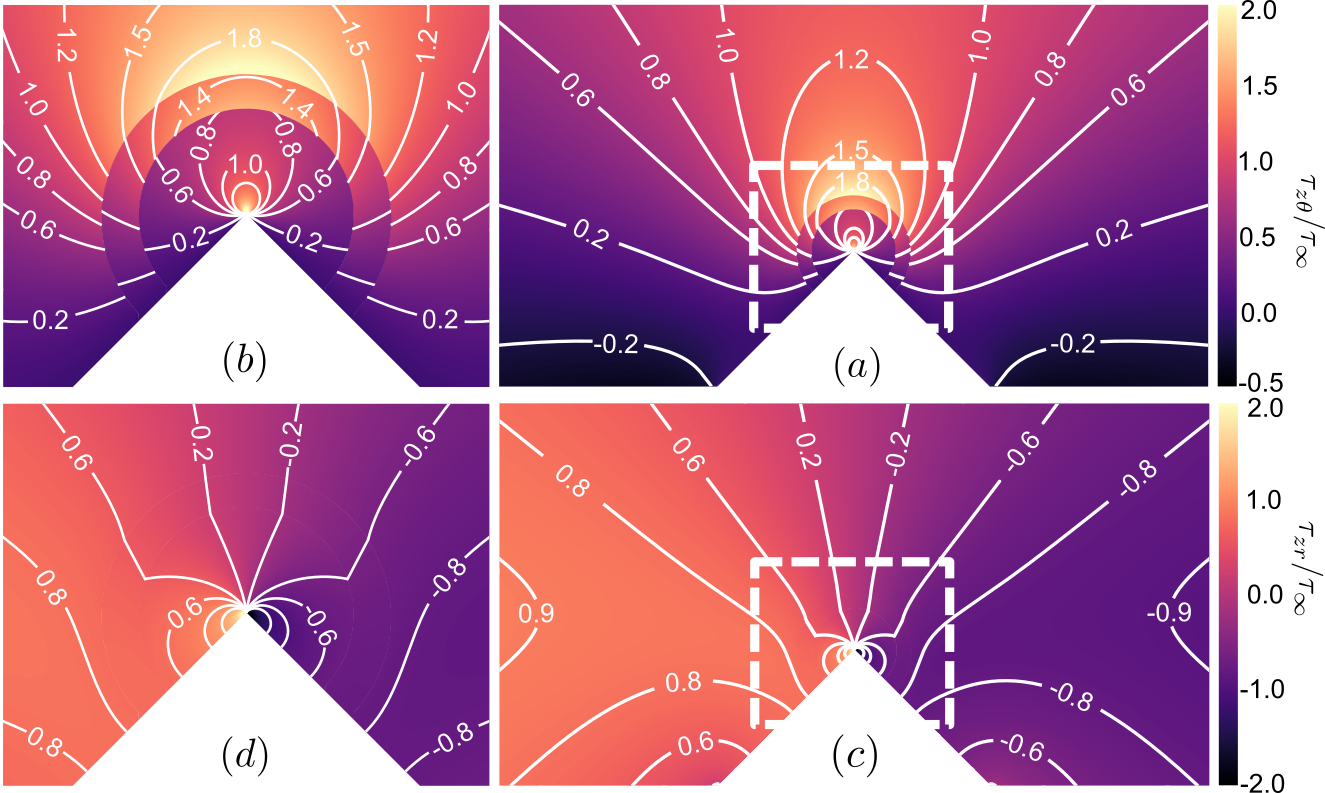}
\caption{Normalized stress distribution for a finite V-notch of depth $b=5$ mm and opening angle $2\alpha=90^{\circ}$ with circular regions made of different material of radii $a_1=0.3b$ and $a_2=0.4b$ embracing the tip. The notch is subjected to a remote stress $\lim_{z\rightarrow \infty}\tau_{zx}=-\tau_{\infty}$. The material inside the circular region $\Omega^A$ features a shear modulus $G_A=1500$ MPa while the material inside the circular region $\Omega^B$ features a shear modulus $G_B=3000$ MPa. The outer region, $\Omega^C$, has a shear modulus $G_C=4500$ MPa: (a) $\tau_{z\theta}/\tau_{\infty}$, (b) magnified view, (c) $\tau_{zr}/\tau_{\infty}$, and (d) magnified view. Note the discontinuity on $\tau_{z\theta}/\tau_{\infty}$ on the interface $\Omega^A$--$\Omega^B$ and $\Omega^B$--$\Omega^C$ while $\tau_{zr}/\tau_{\infty}$ is perfectly continuous for equilibrium.}
\label{triplot}
\end{figure}

Contour plots of the shear stresses calculated by means of Eqs. (\ref{stressv1bitri}), (\ref{stressv2bitri}), (\ref{keybtri}), and (\ref{keybtricd}) for $b=5$ mm, $a_1=0.3b$,, $a_2=0.4b$, $2\alpha=90^{\circ}$, $G_A=1500$ MPa, $G_B=3000$ MPa, and $G_C=4500$ MPa are shown in Figs. (\ref{triplot}a-d). As can be noted, the difference in material properties leads to a significant discontinuity in the tangential component of the stress (Fig. \ref{triplot}a,b). On the other hand, the radial component of the stress is continuous throughout the entire domain for equilibrium as shown by Figs (\ref{triplot}c,d). 

\subsection{Near-tip stress field}
The stress field near the tip of the V-notch can be calculated leveraging the equation for the near-tip stress field of the homogeneous case, Eq. (\ref{vcartstress}). Noting that for the multi-material case the stresses in region $\Omega^A$ differ from the homogeneous case only for the multiplying factor $C$, one can easily obtain the following expression:
\begin{equation}
   \tau_{zx}-i\tau_{zy}=\frac{4\zeta_{BC}\zeta_{AC}t_2^2\tau_{\infty}\exp\left[\frac{\pi\left(q-1\right)}{2q}i\right]\left(\frac{A_v}{q}\right)^{1/q-1}}{\left(\zeta_{AC}-\zeta_{BC}\right)\left(\zeta_{BC}-1\right)t_1^2+\left(\zeta_{AC}+\zeta_{BC}\right)\left(\zeta_{BC}+1\right)t_2^2}z^{1/q-1}
\end{equation}
Based on the foregoing equation, the NSIF \citep{Gross72} can be easily determined leveraging its definition:
\begin{equation}
\begin{split}
   K_3=\sqrt{2\pi}\lim_{y\rightarrow 0}\tau_{zx}y^{1-1/q}&=\frac{\sqrt{2\pi}4\zeta_{BC}\zeta_{AC}t_2^2\tau_{\infty}\left(\frac{A_v}{q}\right)^{1/q-1}}{\left(\zeta_{AC}-\zeta_{BC}\right)\left(\zeta_{BC}-1\right)t_1^2+\left(\zeta_{AC}+\zeta_{BC}\right)\left(\zeta_{BC}+1\right)t_2^2} \\
   & = \frac{4\sqrt{2\pi}\zeta_{BC}\zeta_{AC}t_2^2\tau_{\infty}\left[\frac{b\sqrt{\pi}/\cos\alpha}{q \Gamma\left(1-\alpha/\pi\right)\Gamma\left(1/2+\alpha/\pi\right)}\right]^{1-1/q}}{\left(\zeta_{AC}-\zeta_{BC}\right)\left(\zeta_{BC}-1\right)t_1^2+\left(\zeta_{AC}+\zeta_{BC}\right)\left(\zeta_{BC}+1\right)t_2^2}
   \end{split}
\end{equation}
or, in a more compact form:
\begin{equation}\label{NSIFtri}
   K_3=\tau_{\infty} b^{1-1/q} k_3^{(1)}\left(\alpha\right)k_3^{(2)}\left( a_1/b,a_2/b,\zeta_{AC},\zeta_{BC}\right)
   \end{equation}
where:
\begin{equation}\label{k3trialpha}
   k_3^{(1)}\left(\alpha\right)=\sqrt{2\pi}\left[\frac{\sqrt{\pi}/\cos\alpha}{q \Gamma\left(1-\alpha/\pi\right)\Gamma\left(1/2+\alpha/\pi\right)}\right]^{1-1/q}
   \end{equation}
\begin{equation}\label{k3tricic}
   k_3^{(2)}\left(a_1/b,a_2/b,\zeta_{AC},\zeta_{BC}\right)=\frac{4\zeta_{BC}\zeta_{AC}t_2^2}{\left(\zeta_{AC}-\zeta_{BC}\right)\left(\zeta_{BC}-1\right)t_1^2+\left(\zeta_{AC}+\zeta_{BC}\right)\left(\zeta_{BC}+1\right)t_2^2}
   \end{equation}
are dimensionless functions. It is worth noting that $k_3^{(1)}$ is a purely geometrical function that accounts for the effect of the opening angle $2\alpha$ and takes the same form as the homogeneous case, Eq. (\ref{kappa}). On the other hand, the other dimensionless function $k_3^{(2)}$ accounts for the effect of the size of the circular regions and their elastic properties on the NSIF. It is interesting to note that this function has the following asymptotic values:
\begin{numcases}{}
     k_3^{\left(2\right)} = \frac{2\zeta_{BC}}{1+\zeta_{BC}}\quad  &\quad$\mbox{for }\zeta_{AC}\rightarrow\zeta_{BC}$ \label{coco1} \\ 
     k_3^{\left(2\right)} = \frac{4\zeta_{BC}t_2^2}{\left(1+\zeta_{BC}\right)^2t_2^2-\left(1-\zeta_{BC}\right)^2t_1^2}\quad  &\quad$\mbox{for }\zeta_{AC}\rightarrow 1$ \label{coco4}\\ 
     k_3^{\left(2\right)} = \frac{4\zeta_{BC}t_2^2}{\left(1+\zeta_{BC}\right)t_2^2-\left(1-\zeta_{BC}\right)t_1^2}\quad  &\quad$\mbox{for }\zeta_{AC}\rightarrow \infty$ \label{coco5}\\ 
     k_3^{\left(2\right)} = 0\quad  &\quad$\mbox{for }\zeta_{AC}\rightarrow 0$ \label{fortunata1} \\ 
     k_3^{\left(2\right)} = 0\quad  &\quad$\mbox{for }\zeta_{BC}\rightarrow 0$ \label{fortunata2} \\ 
     k_3^{\left(2\right)} = \frac{2\zeta_{AC}}{1+\zeta_{AC}}\quad  &\quad$\mbox{for }\zeta_{BC}\rightarrow 1$ \label{coco2} \\ 
     k_3^{\left(2\right)} = 0\quad  &\quad$\mbox{for }\zeta_{BC}\rightarrow \infty$ \label{fortunata3}\\ 
     k_3^{\left(2\right)} = \frac{2\zeta_{AC}}{1+\zeta_{AC}}\quad  &$\quad\mbox{for }t_2\rightarrow t_1$ \label{coco3}\\ 
     k_3^{\left(2\right)} = \frac{4\zeta_{AC}\zeta_{BC}}{\left(\zeta_{AC}+\zeta_{BC}\right)\left(\zeta_{BC}+1\right)}\quad  &\quad$\mbox{for }t_2\rightarrow \infty$
\end{numcases}
Equations (\ref{fortunata1}) and (\ref{fortunata2}) show that if any of the circular regions has elastic moduli tending to zero then the dimensionless NSIF will also tend to zero. On the other hand, Eq. (\ref{fortunata3}) indicates that if the outer shell, region $\Omega^B$, is very stiff compared to the outer region then it will shield the tip of the notch from the stresses and the NSIF will also tend to zero. For cases in which the modulus of the first circular region coincides with the modulus of the second region $\Omega^B$ or the modulus of the second region coincides with the one of the outer region $\Omega^C$, Eqs. (\ref{coco1}) and (\ref{coco2}) show that the dimensionless function will tend to $2\zeta_{BC}/(1+\zeta_{BC})$ or $2\zeta_{AC}/(1+\zeta_{AC})$ which combined with $k_3^{(1)}$ give exactly the dimensionless function for the bimaterial case, Eq. (\ref{k3bi}). As Eq. (\ref{coco3}) shows, a similar result is obtained when the radius of the second circular region tends to the radius of the first region, $t_1\rightarrow t_2$. Finally, if the first circular region features the same modulus of region $\Omega^C$ or if its modulus is significantly larger then the dimensionless function depends on the radii of $\Omega^A$ and $\Omega^B$ and the ratio $\zeta_{BC}=G_B/G_C$ according to Eqs. (\ref{coco4}) and (\ref{coco5}).

\begin{figure}[!ht]
\center
\includegraphics[width=0.5\textwidth]{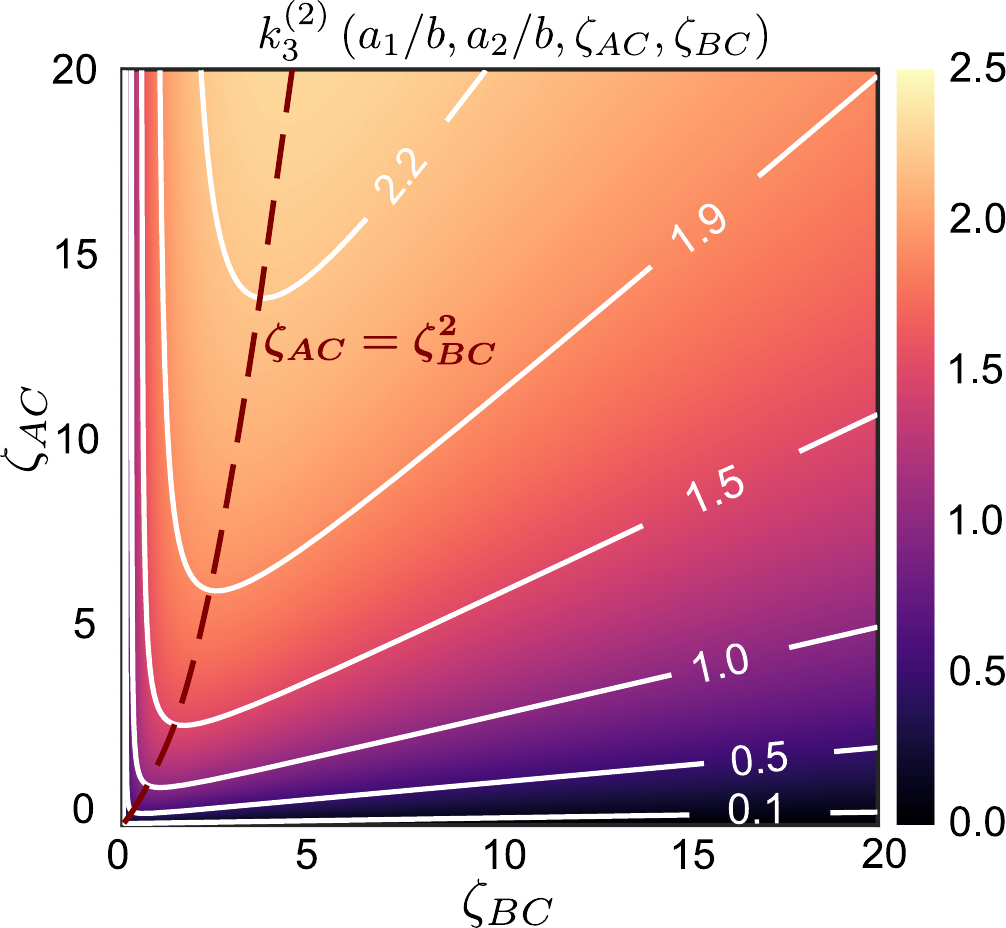}
\caption{Contour plot of the dimensionless $k_3^2\left(a_1/b,a_2/b,\zeta_{AC},\zeta_{BC}\right)$ as a function of the elastic property ratios $\zeta_{AC}=G_A/G_C$ and $\zeta_{BC}=G_B/G_C$ for $2\alpha=90^{\circ}$, $b=5$ mm, $a_1/b=0.3$, and $a_2/b=0.4$.}
\label{k3tri}
\end{figure}

The foregoing equations enable a thorough investigation of the evolution of the NSIF as a function of the geometrical and material parameters of the problem. Figure (\ref{k3tri}) presents a contour plot of $k_3^{(2)}\left(a_1/b,a_2/b,\zeta_{AC},\zeta_{BC}\right)$ as a function of the elastic parameters of the circular regions, $\zeta_{AC}=G_A/G_C$ and $\zeta_{BC}=G_B/G_C$, for an opening angle $2\alpha=90^{\circ}$, a notch depth $b=5$ mm, $a_1/b=0.3$, and $a_2/b=0.4$. As can be noted, the relation between $k_3^{(2)}$ and the material parameters is quite interesting. For a given ratio between the shear modulus of region $\Omega^B$ and region $\Omega^C$, decreasing the shear modulus of region $\Omega^A$ always leads to a reduction of the dimensionless function and hence the NSIF. This shows that utilizing softer materials in the inner region embracing the notch flanks is an effective way to mitigate the intensity of the stress field close to the tip. On the other hand, for a given ratio $\zeta_{AC}$ between the shear modulus of the inner circular region $\Omega^A$ and $\Omega^C$, the dependence of the dimensionless function is more complex. For $\zeta_{BC}\le \sqrt{\zeta_{AC}}$, the $k_3^{(2)}$ decreases smoothly with increasing values of $\zeta_{BC}$. In other words, for a constant $\zeta_{AC}$, increasing the stiffness of the outer circular region $\Omega^B$ can reduce the NSIF down to a minimum corresponding to the case in which $\zeta_{BC}=\sqrt{\zeta_{AC}}$. For $\zeta_{BC}> \sqrt{\zeta_{AC}}$, increasing the stiffness of the outer circular region can only increase the NSIF. These observations along with Eqs. (\ref{NSIFtri})-(\ref{k3tricic}) provide significant guidance for the effective design of multimaterial notches with low NSIF and potentially higher damage tolerance compared to traditional designs. 

\begin{figure}[!ht]
\center
\includegraphics[width=1\textwidth]{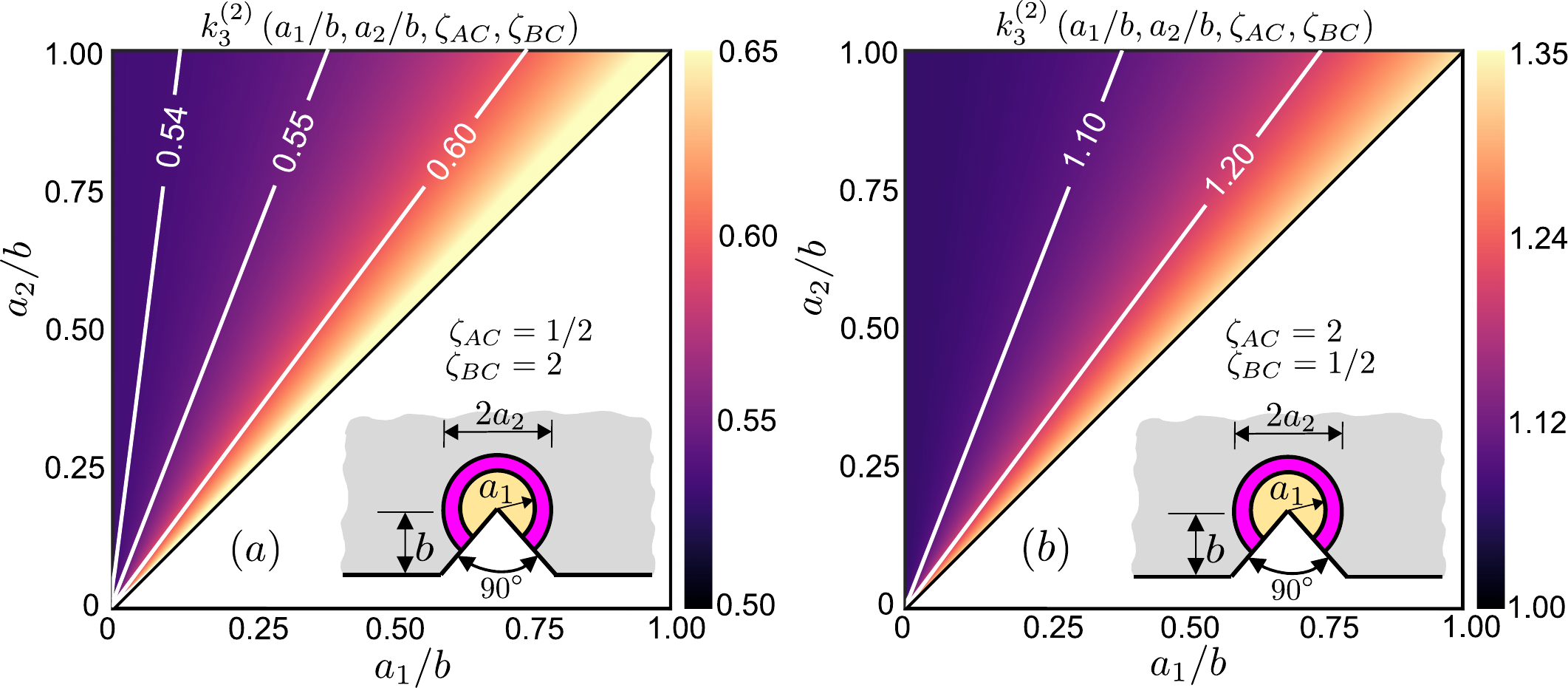}
\caption{Contour plots of the dimensionless $k_3^2\left(a_1/b,a_2/b,\zeta_{AC},\zeta_{BC}\right)$ as a function of the radii of the circular regions embracing the tip for two combinations of elastic properties: (a) $\Omega^A$ is softer than $\Omega^C$ while $\Omega^B$ is stiffer than $\Omega^C$, (b) $\Omega^A$ is stiffer than $\Omega^C$ while $\Omega^B$ is softer than $\Omega^C$. For all the cases, $b=5$ mm and $2\alpha=90^{\circ}$.}
\label{abasoft}
\end{figure}

A contour plot of $k_3^{(2)}\left(a_1/b,a_2/b,\zeta_{AC},\zeta_{BC}\right)$ as a function of the sizes of the multimaterial circular regions as expressed by the ratios $a_1/b$ and $a_2/b$ is presented in Figs. (\ref{abasoft}a,b) for $2\alpha=90^{\circ}$ and $b=5$ mm. 

Fig. (\ref{abasoft}a) considers the case in which the inner circular region is softer than $\Omega^C$ while the outer circular region is stiffer: $\zeta_{AC}=1/2$ and $\zeta_{BC}=2$. It is interesting to note that, for a given size of the outer circular region $a_2/b$, increasing the radius of the inner region $a_1/b$ leads to an increase in the NSIF. This is explained by the fact that the stiffer outer region tends to partially shield the inner region from the remote stress. By increasing the radius of the inner region, the volume fraction of the outer region decreases leading to a slight increase of the stress intensity close to the tip. For a given $a_2/b$ the maximum increase in the NSIF is obtained when the inner radius tends to the critical value: $a_1/b\rightarrow a_2/b$. On the other hand, increasing $a_2/b$ for a given inner radius $a_1/b$ leads to a reduction of the dimensionless function $k_3^2$, and hence the NSIF. This due to the increase in shielding effect of region $\Omega^B$ by increasing its extent. 

Fig. (\ref{abasoft}b) considers the case in which the inner circular region is stiffer than $\Omega^C$ while the outer circular region is softer: $\zeta_{AC}=2$ and $\zeta_{BC}=1/2$. The first observation that can be made by comparing (\ref{abasoft}b) to (\ref{abasoft}a) is that for any combination of $a_1/b$ and $a_2/b$, the values of $k_3^2$ are always large then in the previous case. This confirms that the most effective way to reduce the NSIF is to have softer materials in the region close to the tip. Similar to what was observed before, increasing the size of the outer region while keeping $a_1/b$ constant leads to a reduction of the NSIF. This is again due to an increase in shielding of $\Omega^A$ by the outer circular region $\Omega^B$. On the other hand, increasing the size of the inner region while keeping $a_2/b$ constant leads to an increase of the NSIF. These results suggest that it is possible to design the multimaterial domain to mitigate the Notch Stress Intensity Factor (NSIF) without excessively load the other outer regions by finding a suitable combination of $a_1/b$ and $a_2/b$.

Finally, it is important to note that, thanks to Eqs. (\ref{NSIFtri})-(\ref{k3tricic}), it is possible to calculate the Notch Stress Intensity Factor (NSIF) as a function of the material and geometrical parameters of the problem. Once the NSIF is calculated, the Cartesian stresses near the tip can be calculated taking advantage of Eqs. (\ref{tvbi1}) and (\ref{tvbi2}) while the polar stress components can be calculated using Eqs. (\ref{neartiptheta1}) and (\ref{neartiptheta}).

\subsection{Maximum stress in $\Omega^B$}\label{maxbsection}
The analysis performed in Section 3 revealed that in addition to the NSIF, also the maximum stress in the other region depends on the material and geometrical parameters of the problem. Soft materials in $\Omega^A$ can lead to lower NSIFs but also higher stress concentrations in $\Omega^B$. This can potentially undermine the benefits of reducing the NSIF if the failure is initiated at the point of maximum stress concentration instead of the notch tip. Accordingly, it is important to investigate how the maximum stress in $\Omega^B$ is affected by the material and geometrical configurations now that there are three regions made of different materials instead of only two. Towards this goal, it is useful to note that $v_1=u_1=u_v=0$ at the point of maximum stress. Considering Eq. (\ref{v-key}), this means that $v_v=t_1$ in the location of maximum stress. Inserting this result into Eq. (\ref{keybtri}) and extracting the real part of the equation leads to the following expression for the maximum stress:
\begin{equation}
    \tau_{zx,B}^{\max}=\frac{4\tau_{\infty}\zeta_{BC}^2t_2^2}{\left(\zeta_{AC}-\zeta_{BC}\right)\left(\zeta_{BC}-1\right)t_1^2+\left(\zeta_{AC}+\zeta_{BC}\right)\left(\zeta_{BC}+1\right)t_2^2}\left(\frac{\sqrt{t_1^2+1}}{t_1}\right)^{1-\frac{2\alpha}{\pi}}
\end{equation}
It is interesting to note that the foregoing equation differs from the homogeneous case, Eq. (\ref{t_max}), only for a multiplying factor. As expected, when $\zeta_{AC} \rightarrow 0$ and $\zeta_{BC} \rightarrow 1$  the stress concentration tends to the homogeneous case of V-notch with a circular end hole of radius $a_1$.

\begin{figure}[!ht]
\center
\includegraphics[width=0.5\textwidth]{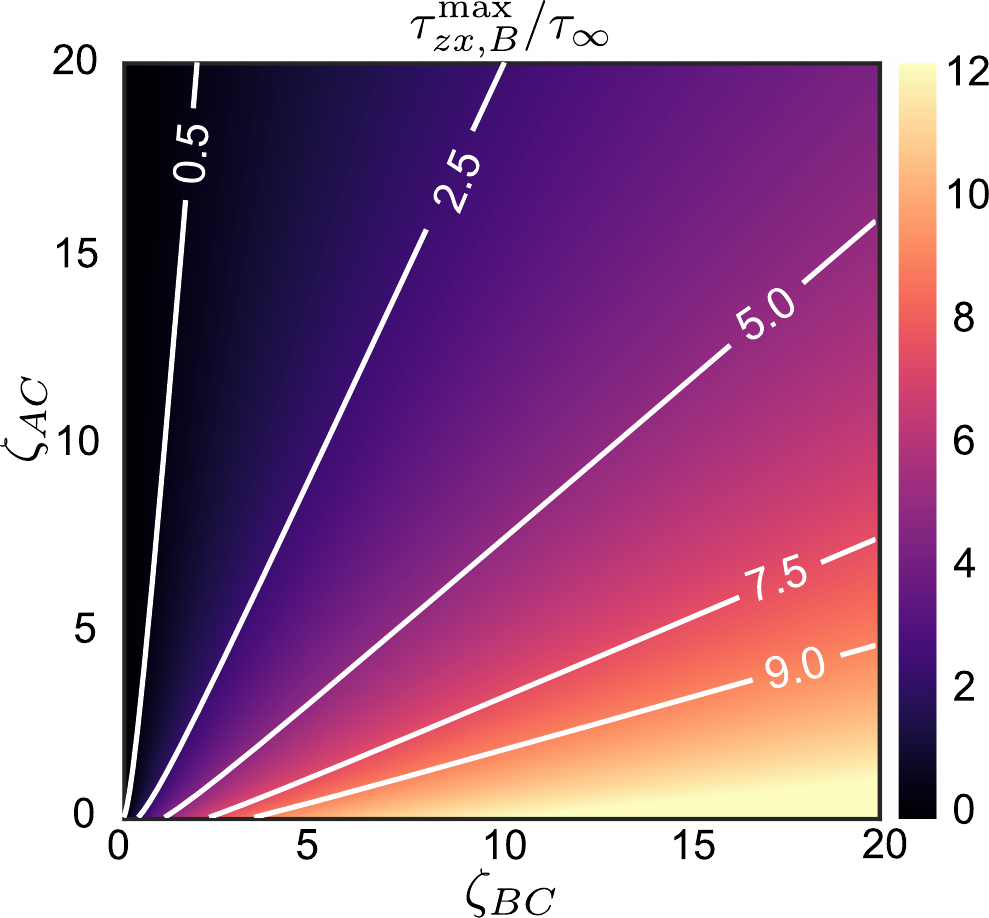}
\caption{Contour plot of the stress concentration in region $\Omega^B$ as a function of the elastic property ratios $\zeta_{AC}=G_A/G_C$ and $\zeta_{BC}=G_B/G_C$ for $2\alpha=90^{\circ}$, $b=5$ mm, $a_1/b=0.3$, and $a_2/b=0.4$.}
\label{maxstress}
\end{figure}

Figure (\ref{maxstress}) provides a contour plot of the stress concentration factor, $\tau_{zx,B}^{\max}/\tau_{\infty}$, as a function of the elastic property ratios $\zeta_{AC}=G_A/C_C$ and $\zeta_{BC}=G_B/C_C$ for a given opening angle $2\alpha=90^{\circ}$, $b=5$ mm, $a_1/b=0.3$, and $a_1/b=0.4$. As can be noted, there is a strong dependence of the stress concentration on the elastic parameters. In particular, for a given $\zeta_{BC}$, increasing the ratio between the shear modulus of the inner circular region $\Omega^A$ and the shear modulus of $\Omega^C$ always leads to a reduction of the stress concentration. This result can be explained by considering that increasing the stiffness of $\Omega^A$ allows this region to take more load hence relieving the stresses in $\Omega^B$. On the other hand, increasing the stiffness of $\Omega^B$ while keeping $\zeta_{AC}$ constant always leads to an increase of the stress concentration. This result is particularly important because it was shown in the previous section that one of the most effective ways to reduce the NSIF is to increase the shear modulus of $\Omega^B$. However, Fig. (\ref{maxstress}) clearly shows that this leads to an increase of the stress concentration in the region. A conclusion that can be drawn here is that, in order to increase the overall capacity of the structure, the combination of material properties of the regions embracing the notch will need to be selected carefully to make sure that by reducing the NSIF one does not trigger an unwanted failure in $\Omega^B$.

\begin{figure}[!ht]
\center
\includegraphics[width=1\textwidth]{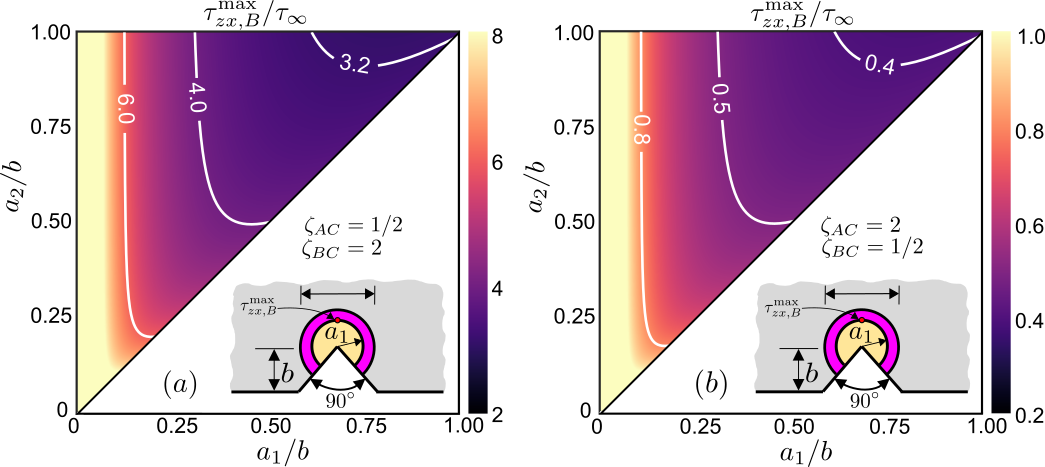}
\caption{Contour plots of the stress concentration in region $\Omega^B$ as a function of the radii of the circular regions embracing the tip for two combinations of elastic properties: (a) $\Omega^A$ is softer than $\Omega^C$ while $\Omega^B$ is stiffer than $\Omega^C$, (b) $\Omega^A$ is stiffer than $\Omega^C$ while $\Omega^B$ is softer than $\Omega^C$. For all the cases, $b=5$ mm and $2\alpha=90^{\circ}$.}
\label{maxab}
\end{figure}

Figures (\ref{maxab}a,b) show the effects on the stress concentration in $\Omega^B$ of the radii of the circular regions $\Omega^A$ and $\Omega^B$ for an opening angle $2\alpha=90^{\circ}$, $b=5$ mm, and two combinations of elastic properties. Fig. (\ref{maxab}a) considers the case in which the inner circular region $\Omega^A$ is softer than $\Omega^C$ while the outer circular region $\Omega^B$ is stiffer: $\zeta_{AC}=1/2$ and $\zeta_{BC}=2$. It can be noted that for a given $a_1/b$ increasing the radius of the outer circular region always leads to a decrease of the stress concentration in $\Omega^B$. On the other hand, for a given $a_2/b$ increasing the inner radius generally leads to a decrease of the stress concentration except for cases in which both $a_1/b$ and $a_2/b$ are very close to $1$ (upper right portion of the contour plot). Intuitively, this can be explained by considering the extreme case in which $\zeta_{AC}\rightarrow 0$. This would correspond the the case of a V-notch with an end hole of radius $a_1$ and a circular shell of shear modulus $G_B$ and radius $a_2$ surrounding it. Increasing the radius $a_1$ would reduce the sharpness of the notch decreasing the stress concentration. Although the case represented in Fig. (\ref{maxab}a) is not as extreme and $\zeta_{AC}=1/2$, the same trend can be noted. Similar conclusions can be drawn from Fig. (\ref{maxab}b) which describes the case in which the inner region $\Omega^A$ is stiffer than $\Omega^C$ while the outer circular region $\Omega^B$ is softer: $\zeta_{AC}=2$ and $\zeta_{BC}=1/2$. The main difference compared to the previous case is that, for given $a_1/b$ and $a_2/b$, the use of a soft material in region $\Omega^B$ in lieu of a stiff material reduces significantly the stress concentration. 

\subsection{Maximum stress in $\Omega^C$}
The calculation of the maximum stress in $\Omega^C$ can be done following the same procedure outlined in Section $\ref{maxbsection}$. This time, it is useful to note that in the position of maximum stress $u_2=v_2=0$ while $u_1=0$. Leveraging Eq. (\ref{coord3}) it is easy to show that this means that:
\begin{equation}
  v_1=\frac{t_2^2-t_1^2}{t_2}  
\end{equation}
at the location of maximum stress. Substituting this result into Eq. (\ref{coco117}), introducing Eqs. (\ref{Gtri}) and (\ref{Mtri}), and taking the real part leads to the following equation for the maximum stress:
\begin{equation}
\begin{split}
   \tau_{zx}^{\max}&=\left(M\frac{t_1^2+t_2^2}{t_2^2}+2G\right)\left(\frac{\sqrt{t_2^2+1}}{t_2}\right)^{1-\frac{2\alpha}{\pi}} \\
   & =\frac{2\tau_{\infty}\left[\left(\zeta_{BC}-\zeta_{AC}\right)t_1^2+\left(\zeta_{BC}+\zeta_{AC}\right)t_2^2\right]}{\left(\zeta_{AC}-\zeta_{BC}\right)\left(\zeta_{BC}-1\right)t_1^2+\left(\zeta_{BC}+\zeta_{AC}\right)\left(\zeta_{BC}+1\right)t_2^2}\left(\frac{\sqrt{t_2^2+1}}{t_2}\right)^{1-\frac{2\alpha}{\pi}}
   \end{split}
\end{equation}
The foregoing equation is very similar to the homogeneous case except for a multiplying factor that depends on the interplay between the radii of the circular regions and their elastic shear moduli. It is worth noting that for $\zeta_{AC}\rightarrow \zeta_{BC}\rightarrow 0$ the stress concentration tends to the one of a V-notch with an end hole of radius $a_2$. On the other hand, the case $\zeta_{AC}\rightarrow \zeta_{BC}$ leads to the same stress concentration of the bimaterial case discussed in Section 3.

\begin{figure}[!ht]
\center
\includegraphics[width=0.5\textwidth]{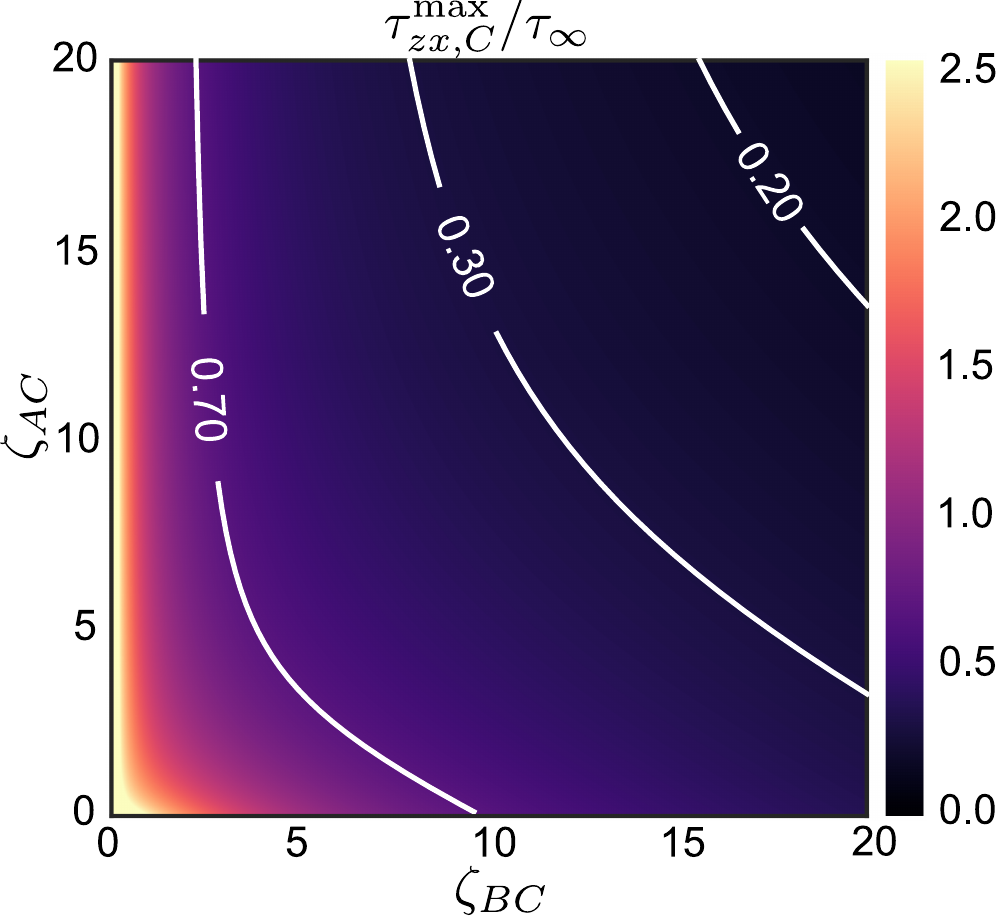}
\caption{Contour plot of the stress concentration in region $\Omega^C$ as a function of the elastic property ratios $\zeta_{AC}=G_A/G_C$ and $\zeta_{BC}=G_B/G_C$ for $2\alpha=90^{\circ}$, $b=5$ mm, $a_1/b=0.3$, and $a_2/b=0.4$. }
\label{ccmax}
\end{figure}

To investigate more thoroughly the effects of the mechanical and geometrical parameters of the problem on the stress concentration in $\Omega^C$, Fig. (\ref{ccmax}) shows $\tau_{zx,C}^{\max}/\tau_{\infty}$ as a function of the elastic parameters $\zeta_{AC}$ and $\zeta_{BC}$ for an opening angle $2\alpha=90^{\circ}$, a depth $b=5$ mm, $a_1/b=0.3$, and $a_2/b=0.4$. As can be noted, the elastic parameters have a very significant effect on the stress concentration. In particular, it is worth noting that for a given $\zeta_{BC}$, increasing the elastic modulus of the inner circular region $\Omega^A$ tends to reduce the stress concentration in region $\Omega^C$. This is because, for a given remote stress $\tau_{\infty}$, increasing the modulus of $\Omega^A$ increases the load taken by such region thus relieving part of the stress concentration in $\Omega^C$. Similarly, for a given $\zeta_{BC}$, increasing the modulus of the outer circular region $\Omega^B$ always reduces the stress concentration in $\Omega^C$. This is again due to the increased load taken by $\Omega^B$ which leads to a decrease of the stresses in the other two regions.

\begin{figure}[!ht]
\center
\includegraphics[width=1\textwidth]{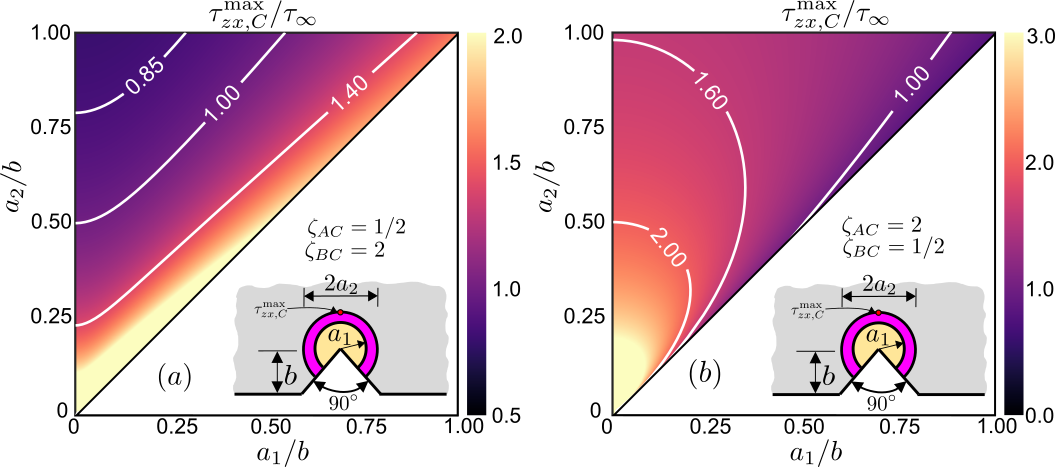}
\caption{Contour plots of the stress concentration in region $\Omega^C$ as a function of the radii of the circular regions embracing the tip for two combinations of elastic properties: (a) $\Omega^A$ is softer than $\Omega^C$ while $\Omega^B$ is stiffer than $\Omega^C$, (b) $\Omega^A$ is stiffer than $\Omega^C$ while $\Omega^B$ is softer than $\Omega^C$. For all the cases, $b=5$ mm and $2\alpha=90^{\circ}$.}
\label{maxCab}
\end{figure}

Figures (\ref{maxCab}a,b) show the stress concentration $\tau_{zx,C}^{\max}/\tau_{\infty}$ as a function of $a_1/b$ and $a_2/b$ for an opening angle $2\alpha=90^{\circ}$, a notch depth $b=5$ mm, and two combinations of elastic parameters. In particular, Fig. (\ref{maxCab}a) shows the case in which the inner circular region is softer than $\Omega^C$ while the outer circular region $\Omega^B$ is stiffer: $\zeta_{AC}=1/2$ and $\zeta_{BC}=2$. From the figure, it can be noted that, for a given $a_1/b$, increasing the size of the outer circular region $\Omega^B$ always leads to a reduction of the stress concentration in $\Omega^C$. This is because the volume fraction of material with a high shear modulus increases. This more load can be taken by $\omega^B$ relaxing part of the stresses in $\Omega^C$. On the other hand, for a given $a_2/b$, increasing the radius of the inner circular region always increases the stress concentration. This is because, increasing the radius of $\Omega^A$, the volume fraction of softer material increases. This means that more load has to be taken by the other two region with the result that the stress concentration in $\Omega^C$ increases.

Figure (\ref{maxCab}b) shows the case in which the inner circular region is stiffer than $\Omega^C$ while the outer circular region $\Omega^B$ is softer: $\zeta_{AC}=2$ and $\zeta_{BC}=1/2$. Similar to the previous case, it can be noted that increasing the size of the outer circular region $\Omega^B$ for a given $a_1/b$ tends to reduce the stress concentration. On the other hand, for a given $a_2/b$, larger radii of the inner circular region $\Omega^A$ lead to a decrease of the stress concentration. This trends, opposite to the one presented in (\ref{maxCab}a), is explained by the fact that the inner region is stiff in this case. Hence, increasing its radius increases the volume fraction of stiff material that can take more load and relieve region $\Omega^C$ from some of the stress.

\section{Nonlinear Computational Modeling}
\begin{figure}[!ht]
\center
\includegraphics[width=0.35\textwidth]{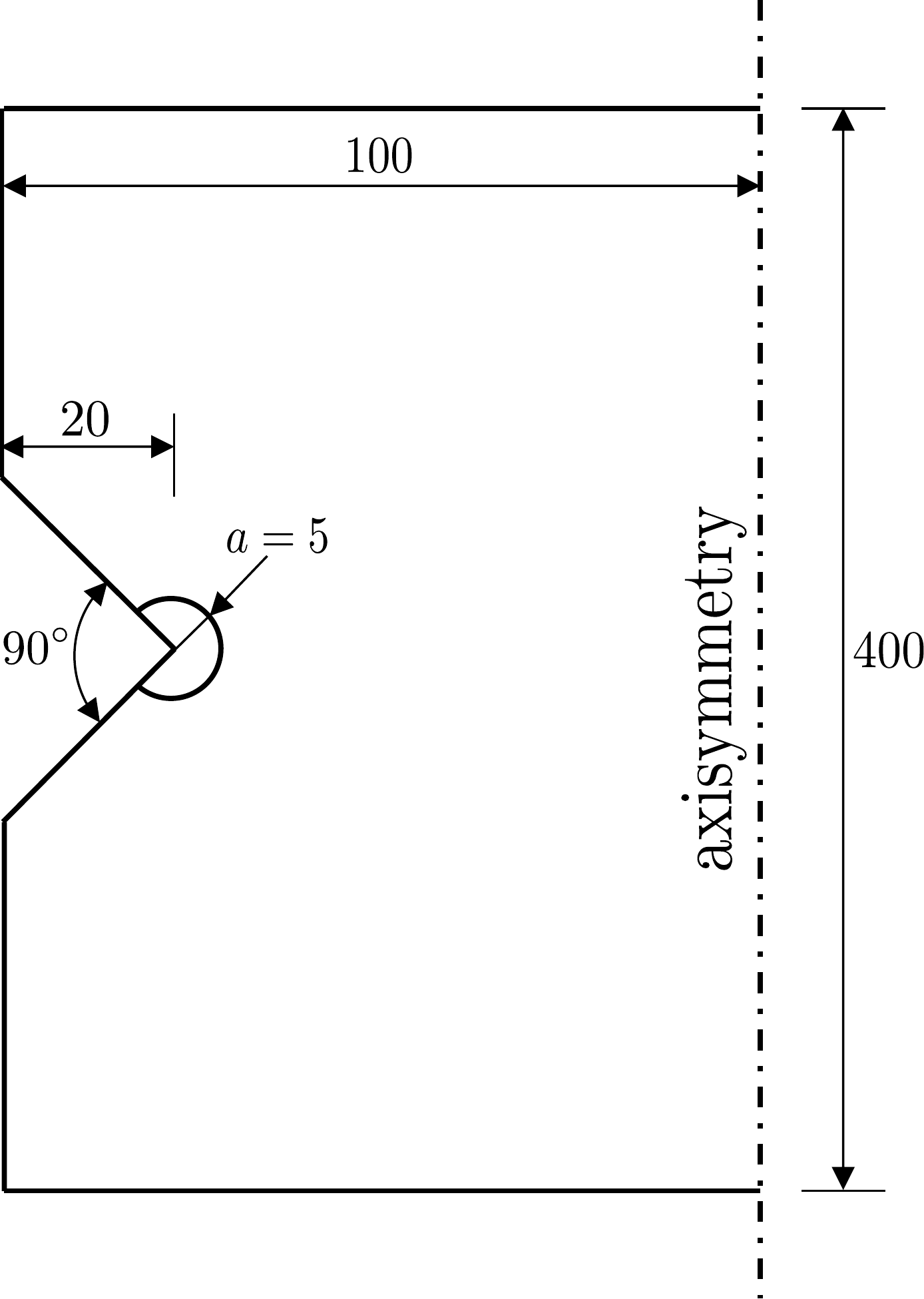}
\caption{Circular shaft weakened by a finite V-notch under torsion investigated in this work. All the dimensions are in mm.}
\label{schema}
\end{figure}

Sections 3 and 4 investigated the stress distribution and NSIFs in antiplane shear and torsion assuming a linear elastic behavior. Based on the closed-form solution, it was found that it is possible to design a combination of materials and radii for the circular regions embracing the notch to reduce the stress intensity and ultimately obtain a higher structural capacity compared to the homogeneous case. However, it is well known that even materials that exhibit a relatively brittle behavior in mode I and II loading conditions might be more prone to plastic deformations in the presence of significant antiplane shear. \cite{berto2012fracture}, for instance, reported significant plastic deformations in torsion tests at room temperature on notched specimens made of polymethyl methacrylate (PMMA), which is generally considered a rather brittle material. Accordingly, while the conclusions drawn from Sections 3 and 4 are valid for extremely brittle materials, it is important to demonstrate that the increase in structural capacity can be obtained also in the presence of inelastic deformations that are likely to occur in antiplane shear and torsion. Towards this goal, this section focuses on the nonlinear computational modeling of the proposed multimaterial system. As a case study, a circular shaft featuring a finite V-notch under torsion was investigated. For simplicity, only two configurations where analyzed. The first configuration is the homogeneous case, where the system is made of an epoxy polymer, a material commonly used in additive manufacturing. This is used as a benchmark to evaluate the performance of the second configuration which is a bimaterial system like the one shown in Fig. (\ref{multidomain}) where a circular region of radius $a$ is made of vulcanized rubber while the rest of the structure is made of epoxy. As can be noted from Fig. (\ref{schema}), which provides all the geometrical details of the problem, the circular shaft has a radius $100$ mm and a length of $400$ mm and it features a V-notch of opening angle $2\alpha=90^{\circ}$ and depth $b=20$ mm. A circular partition of radius $a=5$ mm centered at the tip of the notch and embracing it was created to allow the assignment of a different material behavior. 

It is worth mentioning that this section serves the only purpose of validating the performance of the proposed multimaterial system in the presence of realistic inelastic deformation. No optimization studies where performed to identify the ideal combination of radii and material properties. Furthermore, only a system composed of a soft and an hard phase was investigated. It is likely that thorough optimization studies can lead to more significant differences between the performance of homogeneous and the bimaterial system. However, this is beyond the scope of the present study.

\subsection{Elasto-plastic-damage model}
The structure investigated in this work is assumed to be made of epoxy, a thermoset polymer that can exhibit significant plastic behavior followed by strain softening. Several seminal studies which investigated the behavior of epoxy polymers under significant biaxial and triaxial stress conditions have shown that yielding strongly depends on the interplay between hydrostatic and deviatoric components of the stress field (see e.g. \citeauthor{asp1995}, \citeyear{asp1995}, \citeauthor{asp1996}, \citeyear{asp1996}). This is confirmed by the fact that the yielding stress in uniaxial tension is generally lower than the yielding stress in uniaxial compression \citep{werner2014, poulain2014, hu2003deformation, xia2003deformation}. Recently, it was also shown that the hydrostatic stress strongly affects the strain at failure \citep{Qiaositu}. Furthermore, the hydrostatic stress has been shown to affect the ultimate stress. This has led to the development of a number of different failure criteria that account for the effect of the volumetric stress including e.g. Christensen's failure criterion for glassy polymers \citep{christensen2013theory}, the paraboloidal yielding/failure criterion \citep{Parabola1, Parabola2} or the adaptation of other models such as Drucker-Prager model \citep{canal2009failure} and Mohr-Coulomb \citep{gonzalez2007mechanical} to simulate polymers. For large dilatational stresses, it was shown that the failure of the polymer occurs by microcavitation and the Dilatational Energy Density (DED) criterion provides a superior prediction of the failure behavior \citep{asp1996}.  For an interesting recent comparison between some of the foregoing failure criteria to predict the failure initiation in the matrix of fiber composites, the reader is referred to \citep{kumagai2020multiscale}.

\begin{figure}[!ht]
\center
\includegraphics[width=0.5\textwidth]{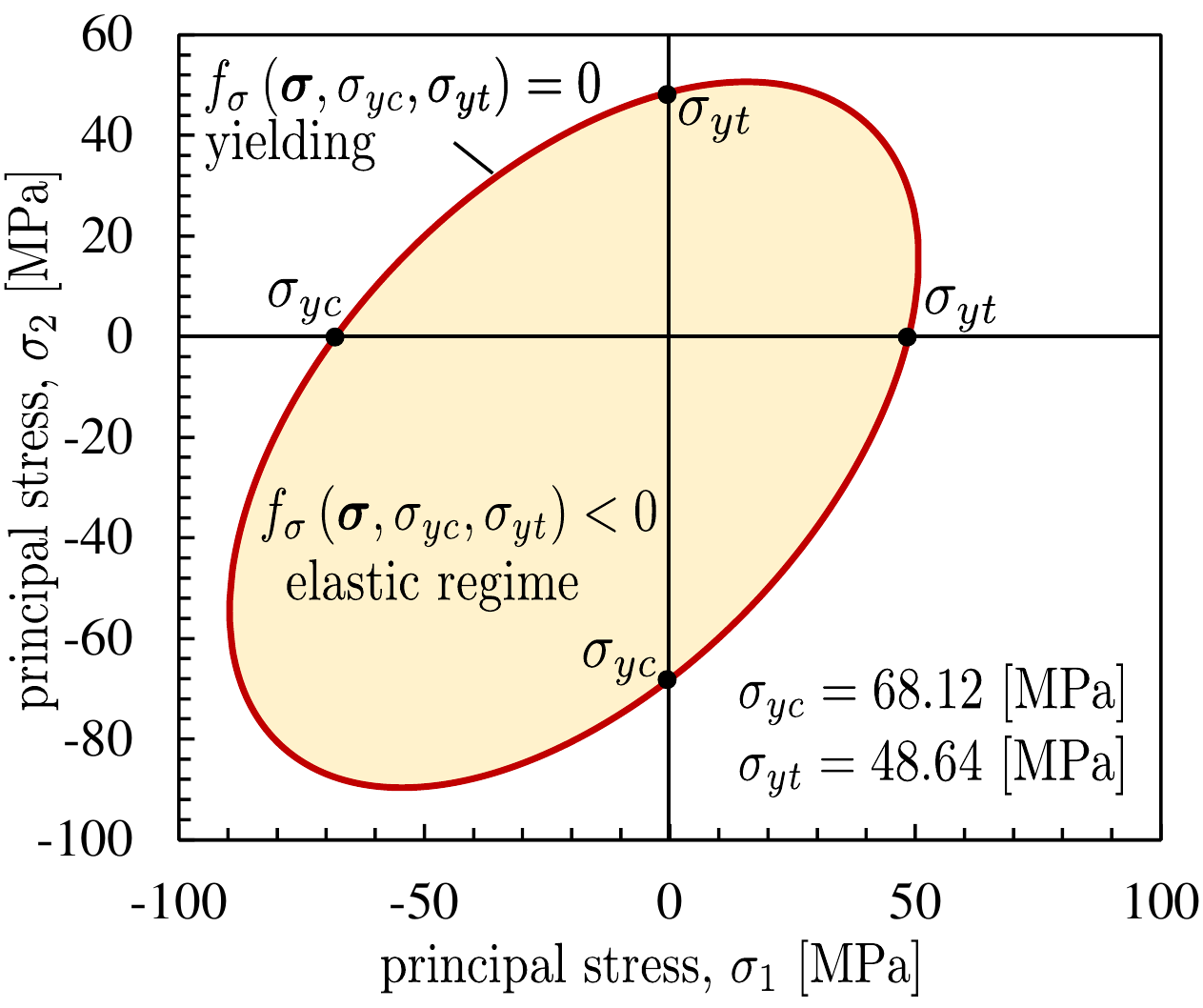}
\caption{Example of biaxial yield surface, $\left(\sigma_3=0\right)$, predicted by the paraboloidal criterion \citep{Parabola1, Parabola2}. Yielding stresses taken from \citep{poulain2014}.}
\label{yieldingplot}
\end{figure}

For the loading conditions investigated in this work the dilationational stress is not significant and failure by microcavitation is very unlikely. Accordingly, the yield surface of the polymer is assumed to follow the paraboloidal criterion \citep{Parabola1, Parabola2}: 
\begin{equation}\label{yielding}
 f_{\sigma}\left(\boldsymbol{\sigma},\sigma_{yc},\sigma_{yt}\right)=6J_2+2\left(\sigma_{yc}-\sigma_{yt}\right)I_1-2\sigma_{yc}\sigma_{yt}   
\end{equation}
where $\sigma_{yt}$ and $\sigma_{yc}$ are the absolute values of the tensile and compressive yield stresses, $J_2=\frac{1}{2}s_{ij}s_{ij}$ is the second invariant of the deviatoric stress tensor $\boldsymbol{s}$ (note that repeated indexes imply summation), and $I_1=\mbox{Tr}\left(\boldsymbol{\sigma}\right)$ is the first stress invariant. Figure \ref{yieldingplot} exemplifies the biaxial yield surface for $\sigma_{yt}=48.64$ MPa and $\sigma_{yc}=68.12$ MPa which are typical values of the initial yield stresses in tension and compression for epoxy \citep{poulain2014}. This model has been shown to capture the multiaxial behavior of thermosets extremely well except for cases of large dilatational stresses \citep{asp1996}.

To avoid positive volumetric strain under hydrostatic pressure, a non-associative flow rule characterized by the following plastic potential is used in this work:
\begin{equation}\label{flow rule}
 g_{\sigma}\left(\boldsymbol{\sigma}\right)=\sigma_{vM}^2+\alpha p^2   
\end{equation}
with $\sigma_{vM}=\sqrt{3J_2}=$ von Mises equivalent stress, $p=1/3I_1=$ hydrostatic pressure, and $\alpha=9/2\left(1-2\nu_p\right)/\left(1+\nu_p\right)$ with $\nu_p=$ plastic Poisson's ratio. Then, the non-associative flow rule reads \citep{jirasek2001inelastic}:
\begin{equation}\label{plastic_strain}
    \Delta\boldsymbol{\varepsilon}^p=\Delta\lambda \frac{\partial g_{\sigma}}{\partial \boldsymbol{\sigma}}
\end{equation}
where $\Delta\lambda$ is the increment of the plastic multiplier, subjected to the Kuhn-Tucker consistency conditions and to be updated via the return mapping algorithm. After a few algebraic manipulations, it is easy to show that:
\begin{equation}
    \frac{\partial g_{\sigma}}{\partial \boldsymbol{\sigma}}=3\boldsymbol{s}+\frac{2}{3}\alpha p \boldsymbol{\delta}
\end{equation}
Substituting the foregoing equation in the non-associative flow rule, it is possible to find the following direct relationship for the plastic strain increment:
\begin{equation}
    \Delta \boldsymbol{\varepsilon}_p=\Delta \lambda \left(3\boldsymbol{s}+\frac{2}{3}\alpha p \boldsymbol{\delta}\right)
\end{equation}

\begin{figure}[!ht]
\center
\includegraphics[width=1\textwidth]{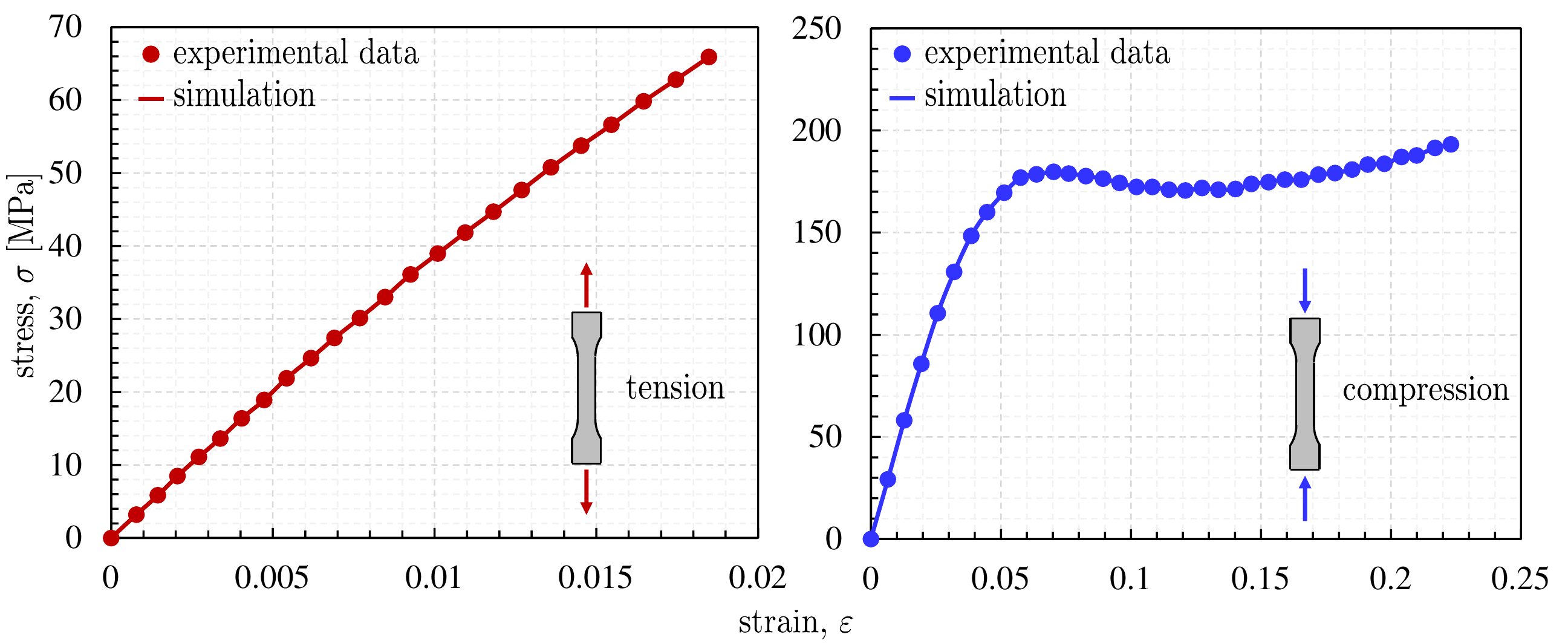}
\caption{Calibration of the elasto-plastic model in tension and compression for Epoxy 3501-6. Experimental data taken from \citep{werner2014}.}
\label{Plastic}
\end{figure}

From the plastic strain increment, it is possible to define an equivalent plastic strain as follows:
\begin{equation}
\dot \varepsilon^e_p=\sqrt{\frac{1}{1+2\nu_p^2} \boldsymbol{\dot\varepsilon}_p: \boldsymbol{\dot\varepsilon}_p}
\end{equation}
which is used to link the hardening behavior of the polymer as characterized from uniaxial tests to the hardening in the presence of a general multiaxial stress state \citep{jirasek2001inelastic}. In this work, isotropic hardening is assumed with hardening equations for both tension and compression written in terms of the equivalent plastic strain $\sigma_{yt}=\sigma_{yt}\left(\varepsilon_p^e\right)$, $\sigma_{yc}=\sigma_{yc}\left(\varepsilon_p^e\right)$. Figs. (\ref{Plastic}a,b) show the uniaxial tension and compression curves utilized in this work to obtain the hardening laws. The symbols represent the experimental data provided in \citep{werner2014} while the solid lines represent the fitting obtained by the proposed elasto-plastic model.

The model is implemented within an elastic predictor-plastic corrector algorithm \citep{belytschko2014nonlinear}. At the beginning of the $(n+1)-$th time increment, an initial trial stress is calculated assuming an elastic behavior:
\begin{equation}
    \boldsymbol{\sigma}_{n+1}^{tr}=\boldsymbol{\sigma}_{n}+\boldsymbol{D}^{e}:\Delta\boldsymbol{\varepsilon}
\end{equation}
from which it is possible to extract the deviatoric and volumetric parts:
\begin{numcases}{}
         \boldsymbol{s}_{n+1}^{tr}=\boldsymbol{s}_{n}+2G\Delta \boldsymbol{e} \\
         p_{n+1}^{tr}=p_{n}+K\Delta \varepsilon_v 
\end{numcases}
where $\varepsilon_v=$volumetric strain, $\boldsymbol{e}=$ deviatoric strain tensor, $G=$ shear modulus, and $K=$ bulk modulus. 

After calculating the trial stress, the yielding condition is checked using the current tensile and compressive yield stresses. If $f_{\sigma}\left(\boldsymbol{\sigma},\sigma_{yc},\sigma_{yt}\right)< 0$ the material is still in the elastic regime and the trial stress can be considered as the correct stress: $\boldsymbol{\sigma}_{n+1}=\boldsymbol{\sigma}_{n+1}^{tr}$. On the other hand, if $f_{\sigma}\left(\boldsymbol{\sigma},\sigma_{yc},\sigma_{yt}\right)\ge 0$, the material has entered the plastic regime and the stress is not admissible. Stresses and plastic strain must be properly recalculated to guarantee the yielding condition $f_{\sigma}\left(\boldsymbol{\sigma},\sigma_{yc},\sigma_{yt}\right)=0$. Towards this goal, one can note that:
\begin{equation}
    \boldsymbol{\sigma}_{n+1}=\boldsymbol{\sigma}_{n+1}^{tr}-\boldsymbol{D}^e: \Delta \boldsymbol{\varepsilon}_p
\end{equation}
from which the following simple relationships can be derived:
\begin{equation}\label{s}
 \boldsymbol{s}_{n+1}= \frac{\boldsymbol{s}_{n+1}^{tr}}{1+6\Delta \lambda G}
\end{equation}
\begin{equation}\label{p}
 p_{n+1}= \frac{p_{n+1}^{tr}}{1+2\Delta \lambda \alpha K}
\end{equation}
Thanks to the foregoing expressions, the equivalent plastic strain increment can be calculated as a function of the trial stresses as follows:
\begin{equation}
\Delta \boldsymbol{\varepsilon}_p=\frac{\Delta \lambda}{\sqrt{1+2\nu^2}}\sqrt{9\frac{\boldsymbol{s}_{n+1}^{tr}:\boldsymbol{s}_{n+1}^{tr}}{(1+6\Delta\lambda G)^2}+\frac{4\alpha^2 p_{n+1}^2}{3(1+2K\alpha\Delta \lambda)^2}}
\end{equation}
where the only remaining unknown is the plastic multiplier increment $\Delta \lambda$. This can be found by imposing the yielding condition:
\begin{equation}
    f_{\sigma}\left[\boldsymbol{\sigma}_{n+1}\left(\Delta \lambda\right), \sigma_{yt}\left(\Delta \lambda\right), \sigma_{yc}\left(\Delta \lambda\right)\right]=0
\end{equation}
which is solved numerically using the Newton-Raphson algorithm. After $\Delta \lambda$ is found, the new stress can be calculated using Eqs. (\ref{s}) and (\ref{p}). 

In addition to the curves shown in Fig. (\ref{Plastic}a,b), the other data used for the calibration of the model is $E=4600$ GPa, $\nu=$ 0.35. These properties were obtained from \citep{werner2014}.

\subsubsection{Damage model}
The combination of the damage model with the elasto-plastic model can be described by means of the following equation:
\begin{equation}\label{damage}
    \boldsymbol{\sigma}=\left(1-\omega\right)\overline{\boldsymbol{\sigma}}=\left(1-\omega\right)\boldsymbol{D}_e:\left(\boldsymbol{\varepsilon}-\boldsymbol{\varepsilon}_p\right)
\end{equation}
where $\omega$ is a scalar describing the amount of isotropic damage, $\boldsymbol{D}_e$ is the elastic stiffness, $\boldsymbol{\varepsilon}$ is the total strain, $\boldsymbol{\varepsilon}_p$ is the plastic strain, $\overline{\boldsymbol{\sigma}}$ is the effective stress and $\boldsymbol{\sigma}$ is the nominal stress. Eq. (\ref{damage}) underlies two different ways to combine stress-based plasticity with strain-based scalar damage. One possible approach is to have the plastic part expressed in terms of the effective stress (i.e. in the undamaged space). The other approach consists in having the plasticity part written in terms of the nominal stress (in the damage space). As was shown by Grassl and Jir\'asek in an excellent study \citep{grassl2006damage}, only the first class of models provides local uniqueness without any restrictions on the model parameters. Accordingly, in this work, it was decided to have the plastic component of the model written in terms of the effective stress. 

The evolution of the damage variable $\omega$ was calculated as a function of the equivalent plastic strain, $\varepsilon_p^e$, assuming a linear strain softening leading to the following expression:
\begin{equation}
    \omega=1-\frac{\langle\varepsilon_p^e-\varepsilon_0^e\rangle}{\varepsilon_f^e-\varepsilon_0^e}
\end{equation}
In the foregoing equation, $\varepsilon_0^e$ is the equivalent plastic strain at damage initiation and $\varepsilon_f^e$ is the equivalent plastic strain when the element is fully damaged. In the context of the crack band model \citep{BazOh83,Baz97,Sal21a}, the strain at complete failure must be adjusted as a function of the material fracture energy and the element size to guarantee a mesh-independent energy dissipation upon fracture:
\begin{equation}
    \varepsilon_f^e=\varepsilon_0^e+\frac{2G_f}{\sigma_y\left(\varepsilon_0^e\right)h_e}-\frac{\sigma_y\left(\varepsilon_0^e\right)}{E}
\end{equation}
where $\sigma_y\left(\varepsilon_0^e\right)$ is the yield stress at damage initiation, $G_f$ is the fracture energy, and $h_e$ is the characteristic length of the element. The crack band model has been shown to accurately and objectively capture strain localization phenomena in a variety of materials including e.g. concrete \citep{BazOh83, bavzant2000microplane, caner2013microplane, caner2013microplane2}, composites \citep{salviato2016experimental,salviato2016spectral}, and polymers \citep{qiao2019strength,Qiao19Sal,Qiaositu}.
For the simulations performed in this work, the fracture energy was assumed to be $G_f=1$ N/mm which is a typical value for epoxy polymers \citep{qiao2019strength}. The elasto-plastic-damage model was implemented as a VUMAT subroutine in ABAQUS/Explicit.
\subsection{Hyperelastic-damage model}
In this work, the material in $\Omega^A$ is assumed to be vulcanized rubber. Considering the large deformations that can occur at the tip of the notch, a nearly-incompressible hyperleastic model was implemented with the strain energy density following a 3-term Ogden model \citep{ogden1972large}: 
\begin{equation}\label{ogden}
    W=\sum_{i=1}^3\frac{2\mu_i}{\alpha_i^2}\left(\overline{\lambda}_1^{\alpha_i}+\overline{\lambda}_2^{\alpha_i}+\overline{\lambda}_3^{\alpha_i}-3\right)+\frac{K}{2}\left(J-1\right)^2
\end{equation}
In the foregoing expression $\overline{\lambda}_i=\lambda_i/J^{1/3}$, $\lambda_i$ with $i=1...3$ are the stretch ratios, $J=\det\left(\boldsymbol{F}\right)$ is the Jacobian, $K$ is the bulk modulus for small deformations, and $\mu_i$ and $\alpha_i$ are material properties to be calibrated via experimental data. Starting from the strain energy density, the Cauchy stress tensor can be easily calculated as follows \citep{bower2009applied}:
\begin{equation}\label{hyperstress1}
\boldsymbol{\sigma}=\frac{1}{J}\boldsymbol{F}\frac{\partial W}{\partial \boldsymbol{F}}
\end{equation}
where $\boldsymbol{F}$ is the deformation gradient.

\begin{figure}[!ht]
\center
\includegraphics[width=0.5\textwidth]{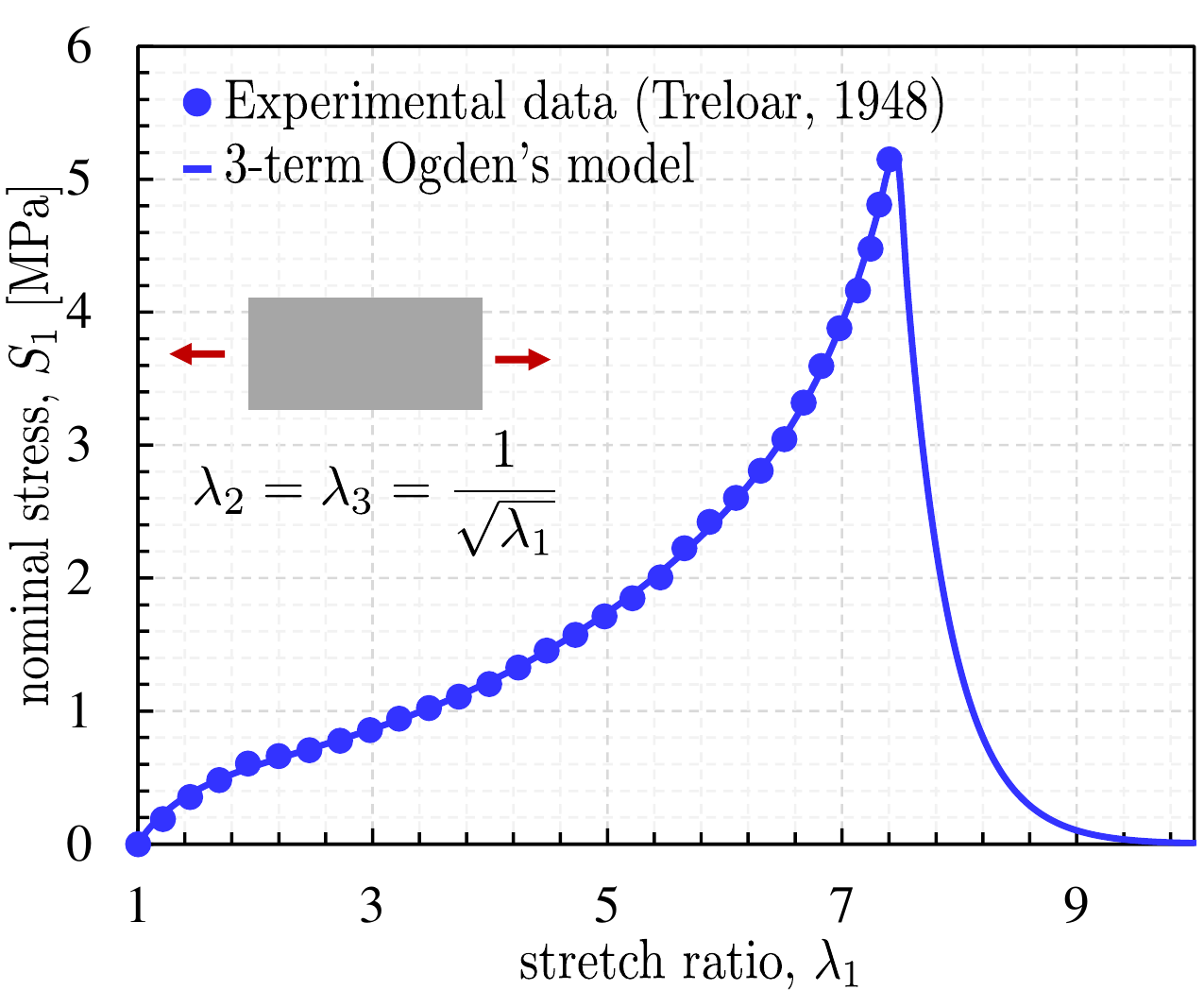}
\caption{Uniaxial tensile behavior of the hyperelastic material considered in region $\Omega^A$. The symbols represent data on vulcanized rubber taken from \citep{treloar1948stresses}. The solid line shows the fitting by the 3-term Ogden's model \citep{ogden1972large}. Note that the model has been modified to include strain softening using an improved version of the framework presented in \citep{volokh2007hyperelasticity}. Shown is the regularized softening curve for an element size of $0.3$ mm using the crack band model \citep{BazOh83}. }
\label{rubber}
\end{figure}

The strain energy density obtained from Eq. (\ref{ogden}) increases indefinitely with increasing stretch ratios. Accordingly, it is not suitable to describe damage and fracture. To enable to capture strain softening a modified version of the strain energy function proposed in \citep{volokh2007hyperelasticity} is used in this work: 
\begin{numcases}{\Psi\left(W\right)=}
         W\quad  &\quad\mbox{for }$W\le W_0$ \label{sed1}\\
         W_0+\Phi\left[1-\exp\left(-\frac{W-W_0}{\Phi}\right)\right]\quad  &\quad\mbox{for }$W > W_0$ \label{sed2}
\end{numcases}
where $W$ is calculated by means of Eq. (\ref{ogden}), and $W_0$ and $\Phi$ are parameters that describe the softening behavior of the material. $W_0$ controls the damage initiation while $\Phi$ controls the initial slope of the softening curve. Following the crack band model \citep{BazOh83, Baz97, Sal21a}, these parameters must be adjusted as a function of the total fracture energy, $G_f$, and the element characteristic length, $h_e$. 

Substituting Eqs. (\ref{sed1}) and (\ref{sed2}) into (\ref{hyperstress1}), the Cauchy stress tensor now reads:
\begin{equation}\label{hyperstress2}
\boldsymbol{\sigma}=\frac{1}{J}\boldsymbol{F}\frac{\partial W}{\partial \boldsymbol{F}}\exp\left(-\frac{\langle W-W_0 \rangle}{\Phi}\right)
\end{equation}
where $\langle x \rangle=\max\left(x,0\right)$. Figure (\ref{rubber}) shows the calibration of Eq. (\ref{hyperstress2}) using uniaxial tensile experimental data from \citep{treloar1948stresses} using the Levenberg-Marquardt algorithm \citep{levenberg1944method,marquardt1963algorithm}. As can be noted, Eq. (\ref{hyperstress2}) can accurately capture the experimental data with $\alpha_1=0.46$, $\alpha_2=15.81$, $\alpha_3=3.52$, $\mu_1=1.79$ MPa, $\mu_2=1.75\times 10^{-13}$ MPa, $\mu_3=2.13\times 10^{-2}$ MPa and $K=2$ GPa. The figure also shows the exponential strain softening predicted by the hyperelastic model for an element size of $0.3$ mm for which $W_0=1.22\times10^{-2}$ J/mm$^3$ and $\Phi=1.00\times10^{-2}$ J/mm$^3$. For this work, the fracture energy was assumed to be $G_f=3.7$ N/mm which is a typical value for vulcanized rubber. 

The foregoing model was implemented as a VUMAT subroutine in ABAQUS/Explicit.

\begin{figure}[!ht]
\center
\includegraphics[width=1\textwidth]{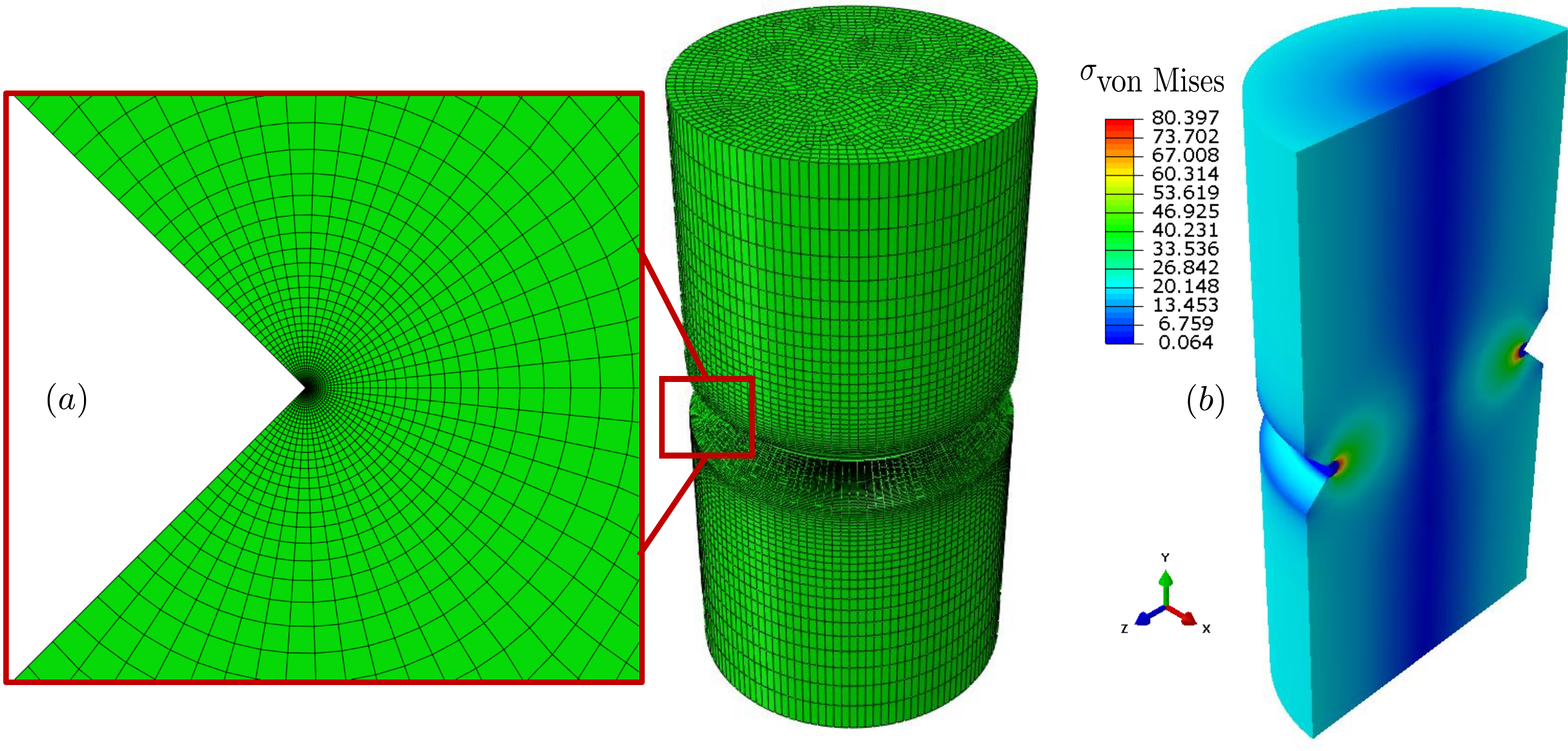}
\caption{FE model of the circular shaft weakened by a finite V-notch under torsion: (a) FE mesh with $926,223$ hexahedral elements (the insert shows a magnification of the mesh at the notch tip), (b) von Mises stress distribution for the bimaterial system at peak load.}
\label{FEM}
\end{figure}

\subsection{Finite Element model}
For the nonlinear simulation of the torsion problem a finite element model was created in ABAQUS/Explicit. In order to allow the fracture pattern to be non-axisymmetric, a full three-dimensional problem was solved in ABAQUS/Explicit as shown in Figs. (\ref{FEM}a,b). A mesh of $926,223$ hexahedral elements with reduced integration was used to accurately resolve the stress and strain fields. As the insert in Fig. (\ref{FEM}a) shows, particular care was devoted to getting a very regular and fine mesh close to the notch tip. A thorough convergence study confirmed that this allowed an objective description of the stress and strain fields before localization. On the other hand, as described in the foregoing sections, mesh objective results in the presence of localization where guaranteed by the use of the crack band model at the constitutive level \citep{BazOh83,Baz97, Sal21a}.

For the application of the boundary conditions, two reference points were assigned at the center of the top and bottom surfaces respectively. A tie constrain to the reference point at the top was assigned to all the nodes on the top surface while the nodes of the bottom surface were tied to the reference point at the bottom. To simulate a torsional load, the reference point at the bottom was completely fixed. The simulation was performed in rotation-control by applying a linearly increasing rotation along the axis, $\theta$, on the top reference point and measuring the reaction torque, $T$. A typical von Mises stress distribution at peak load for the bimaterial case is shown in Fig. (\ref{FEM}b) as an example.

\subsection{Results and discussion}
Thanks to the nonlinear computational models it was possible to investigate the structural behavior of the homogeneous and bimaterial systems under torsion. As mentioned in the foregoing sections, the bimaterial system featured a circular region surrounding the notch tip made of vulcanized rubber while the rest of the structure was made of epoxy. Rubber was chosen since the linear elastic solution for the NSIF showed that the use of soft materials at the tip can help reduce the stress intensity. In addition, rubber features a very large fracture energy compared to epoxy which helps achieving higher ultimate loads. 

\begin{figure}[!ht]
\center
\includegraphics[width=0.52\textwidth]{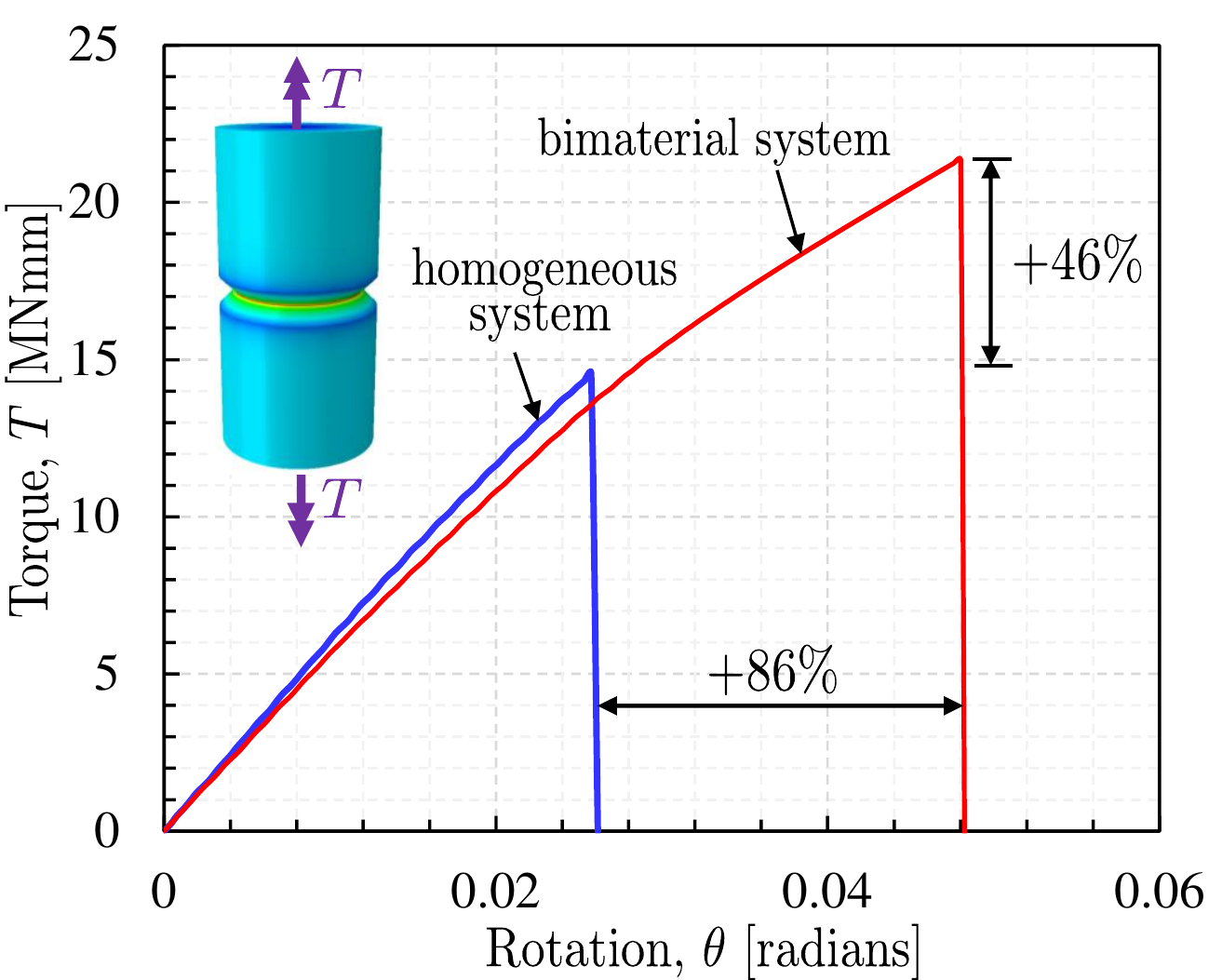}
\caption{Nonlinear simulation of the structural response (Torque vs rotation angle) for the case of a homogeneous structure and a bimaterial notch. Note that the addition of a soft phase in front of the notch increases the structural capacity of $46\%$ while the rotation at failure is increased of $86\%$ compared to the homogeneous case. }
\label{simulations}
\end{figure}

The results of the simulations are summarized in Fig. (\ref{simulations}) which shows the torque as a function of the torsion angle for the two configurations investigated. As can be noted, the addition of a soft phase to mitigate the intensity of the stress field indeed improves the performance of the structure. In fact, it is worth noting that the structural capacity of the bimaterial system made of rubber and epoxy is almost $50\%$ higher of the homogeneous system made of epoxy. At the same time, the blunting of the notch enabled by the soft phase seems to promote large plastic deformations before reaching the peak load. This translates into a larger rotation at failure which is almost $90\%$ higher than the homogeneous case. Another interesting result shown in Fig. (\ref{simulations}) is that, since the soft phase occupies only a very limited region surrounding the notch tip, the overall stiffness of the structure decreases of only $8\%$ compared to the homogeneous case. This is an additional benefit considering that several other approaches to increase the capacity and damage tolerance of structural components often lead to a significant stiffness loss.

Finally, it is worth stating here again that the goal of the nonlinear simulations was to demonstrate the validity of the proposed multimaterial system in increasing the structural capacity also in the presence of realistic inelastic strains. It is expected that different material combinations as well as geometrical configurations can lead to different results. It is also envisioned that the use of multiple circular regions like shown in Fig. (\ref{multidomain2}) instead of just one can lead to higher tunability of the system and enhanced performance. Thorough optimization studies can lead to even larger gains in performance compared to the simple case investigated in this work. However, this was not in the scope of the present work and it will be the subject of future studies.

\section{Conclusions}
This study investigated the addition of circular regions embracing the tip of finite V-notches to reduce the Notch Stress Intensity Factor (NSIF) and increase the capacity of structures subjected to antiplane shear and torsion. Towards this goal, a general framework for finding the closed-form solution for stresses, displacements, and NSIFs was presented for the first time. Moreover, nonlinear simulations using elasto-plastic-damage and hyperelastic-damage models were performed to investigate the increase of structural capacity.

Based on the results presented in this work it is possible to elaborate the following conclusions:
\begin{enumerate}
    \item The stress and displacement distributions for finite V-notches featuring $N$ circular regions of different materials embracing the tip under antiplane shear and torsion can be calculated in closed-form by combining the solutions of $N$ homogeneous sub-problems. The first sub-problem is a homogeneous finite V-notch with the same depth and opening of the original problem. Each of the other $N-1$ sub-problems is a finite V-notch with a circular end hole of radius corresponding to the circular region surrounding the notch in the original problem. Once the $N$ solutions are combined, the remaining parameters can be uniquely identified by imposing the compatibility conditions on the interfaces and the remote stress condition.
    \item For the proposed configuration, the multimaterial regions do not affect the order of the singularity of the stress field close to the tip which, like in the homogeneous case, depends only on the opening angle of the notch: $\sigma_{ij}\propto 1/r^{1-\lambda_3}$ with $r=$ distance from the notch tip, $\lambda_3=\pi/\left(2\pi-2\alpha\right)$, and $2\alpha=$ notch opening angle. This is in stark contrast to the case in which the bimaterial interface does not embrace the notch (see e.g. \citeauthor{bogy1971two} \citeyear{bogy1971two}, \citeauthor{hein1971stress} \citeyear{hein1971stress}, \citeauthor{yu2010scaling} \citeyear{yu2010scaling}, \citeauthor{le2010scaling} \citeyear{le2010scaling}).
    \item The Notch Stress Intensity Factor (NSIF) and the stress concentrations at the material interfaces are shown to be significantly influenced by the combination of material properties and radii of the circular regions. For the bimaterial case, the NSIF decreases with decreasing ratios between the shear modulus of the region surrounding the notch tip and the one of the outer region. In contrast, the stress concentration at the bimaterial interface increases. This shows that it is possible to decrease the intensity of the stress field by using softer materials close to the notch tip. However, particular care must be devoted to avoid excessive stress concentrations which can trigger early failures at the material interfaces. For the multimaterial case, the NSIF and stress concentration factors show a complex dependence on a combination of elastic moduli and region radii.
    \item A significant advantage of the closed-form solution presented in this work compared to numerical approaches is that it allows to explicitly account for the effects of all the material and geometrical parameters of the problem on the NSIFs and the stress concentrations. This makes it a valuable tool for design of multimaterial structures under antiplane shear and torsion.
    \item Advanced hyperelastic-damage and elasto-plastic-damage computational models confirmed the benefits of the proposed multimaterial system also in the presence of nonlinear deformations. Investigating a bimaterial system under torsion made of vulcanized rubber and epoxy showed that the structural capacity can be effortlessly increased of almost $50\%$ while the nominal rotation at failure can be increased of almost $90\%$ compared to the homogeneous case. Since the soft region is limited to the notch tip, the foregoing results are obtained at the expenses of a structural stiffness reduction of only $8\%$. It is expected that even larger benefits can be achieved by performing thorough optimization studies. 
    \item It is expected that a similar approach to increase the structural capacity and damage tolerance can be extended to other loading conditions including Mode I and II. Furthermore, it is expected that the proposed system can be used to significantly increase the fatigue life of notched structural components. Considering the low stresses in high-cycle fatigue and the fact that most of the life is spent in damage initiation, it is probably possible to avoid the use of nonlinear computational models and rely completely on the closed-form solution presented in this work for design. The proposed solution e.g. can be used within the Strain Energy Density (SED) fatigue failure criterion by Lazzarin and co-workers which has been extremely successful with the description of high-cycle fatigue failure for a number of brittle material systems \citep{lazzarin2001finite, lazzarin2008some, berto2009review}.
\end{enumerate}

\section*{Acknowledgments}
The author acknowledges the financial support from the National Science Foundation under Grant No. CMMI-2032539.

\clearpage
\linespread{1}\selectfont
\bibliographystyle{elsarticle-harv}
\bibliography{bib}

\end{document}